\documentclass[onecolumn,secnumarabic,amssymb,amsmath,amsfonts nobibnotes,nofootinbib, aps, prd,preprintnumbers]{revtex4-1}

\usepackage{epsfig,graphicx}
\usepackage{chngcntr}
\usepackage{rotating}
\usepackage{wrapfig}
\usepackage{parskip} 
\usepackage{graphicx}
\usepackage{caption}
\usepackage{subcaption}
\usepackage[linkcolor=blue, citecolor=red, urlcolor=blue]{hyperref}
\begin{document}

\title{Vacuum Stability and  Electroweak Precision   in the Two Higgs Doublet Model with Vector-Like Quarks }%

\author{\sf Kivanc Y. Cingiloglu\footnote{kivanc.cingiloglu@concordia.ca},\; Mariana Frank\footnote{mariana.frank@concordia.ca}}%

\affiliation{ Department of Physics,  
			Concordia University, 7141 Sherbrooke St.West,\\
			Montreal, Quebec, Canada H4B 1R6.\\}
\date{\today}%
			
			\begin{abstract}

We present a comprehensive analysis of the vacuum stability of the Two-Higgs Doublet Model, for both Type-I and Type-II, augmented by vector-like quarks in either singlet, doublet or triplet representations. We review  the model briefly before introducing the extra fermionic states and their interactions, and impose restrictions on the parameters coming both from theoretical considerations and experimental bounds. We then study the renormalization group equation evolution of the parameters of the model in order to isolate the parameter regions that satisfy vacuum stability requirements. We then add the  electroweak precision observables to insure that the resulting parameter space is consistent with the data. We include complete expressions for the renormalization group equations and the $\mathbb{S}$ and $\mathbb{T}$ parameters used. Finally we summarize the effects of various vector-like quark representations on the parameter space. We  indicate the regions constrained, highlighting the differences between representations in Type-I and Type-II, and pinpoint the effects of the interplay between the extended model and the additional fermions. 

\end{abstract}
\maketitle
\bigskip


\section{Introduction}
 \label{sec:intro}

The discovery of the Higgs boson \cite{ATLAS:2012yve,CMS:2012qbp} marked a significant milestone in particle physics, validating the existence of the missing piece of the Standard Model (SM). Yet the data collected supporting the Higgs discovery seem to indicate that principles of stability, renormalizability, and naturalness, which motivated the introduction the Higgs boson in the first place, appear in conflict with  the properties of the Higgs field itself. The
idea of naturalness seems to be in conflict with the surprising degree of fine-tuning of both
parameters in the Higgs field potential \cite{Degrassi:2012ry,Buttazzo:2013uya}. Related to this is the issue of the stability of the electroweak vacuum which  arises from the behaviour of the Higgs potential under renormalization group equations. To address this issue, it became imperative to explore extensions of the SM that could resolve this instability while remaining consistent with experimental observations.  These explorations involve extending the particle content by additional states, and/or extending the symmetry group (which in turn, result in the presence of new particles). 
\\

Such additional particles can be fermions or bosons. While the former are limited, the later appear to have a wider range of applicability.
The issue with additional fermions is the following. In the SM, gauge invariance does not allow for the introduction of bare mass
terms for quarks and leptons, since these terms are not gauge invariant. So quark and lepton
masses only arise from Yukawa interactions, after spontaneous gauge symmetry breaking. Additional fermionic families (quark or lepton) are ruled out by the Higgs data, since both the digluon production cross section (and decay), and diphoton decay channel, agree with the SM predictions and thus are inconsistent with the existence of additional fermions in the loops. 
\\

However, if the fermionic components have vector-like structure, rather than SM-chiral-like, their left- and right-handed components have the same couplings, allowing for bare mass terms which are gauge invariant. The addition of these particles is one of the simplest extensions of the SM. Because of their vector-like nature, they do not contribute to gauge anomalies
and are less restricted than their chiral counterparts by 
current experimental data. They may populate the desert between the SM and the scale of grand-unification, without worsening the hierarchy problem. Vector-like quarks (VLQs), allowed to mix that couple with the third generation
quarks (top and bottom partners), 
 appear in composite
Higgs models with a partially-composite top quark \cite{Agashe:2004rs,Ferretti:2013kya,Kaplan:1991dc,Ferretti:2014qta}.  They are naturally present in theories with extra dimensions \cite{Chang:1999nh,Gherghetta:2000qt,Contino:2003ve,Gopalakrishna:2011ef,Gopalakrishna:2013hua,Couture:2017mbd} and in Little Higgs Models \cite{Arkani-Hamed:2002iiv,Perelstein:2003wd,Schmaltz:2002wx}.  Finally, VLQs can be introduced
in non-minimal supersymmetric models to increase corrections to the Higgs mass
without significantly affecting electroweak precision observables \cite{Martin:2010dc,Martin:2009bg,Babu:2008ge}, and they appear also in  GUT-inspired, supersymmetric models
\cite{Kang:2007ib}.
\\

 Additionally, VLQ may  explain some of the mismatch between the SM predictions and observed data.  For instance the CKM matrix, which encodes couplings for each of the three
generation quarks  is, by construction, unitarity. However, the recent dataset collected after 2018
\cite{FlavourLatticeAveragingGroup:2019iem} disfavours the CKM unitarity of the first row for three generations of quarks to 99.998\% C.L., a problem 
confirmed by the determination of $V_{ud}$ from superallowed beta decays \cite{Bryman:2021teu}. While improved lattice evaluations of decay constants and form factors for kaons and pions, and corrections to the nuclear beta decay have shrunk  the discrepancy to $3 \sigma$, referred to as the Cabibbo Angle Anomaly \cite{Kirk:2023oez}, 
 introducing VLQs seems the most promising avenue, because of they are able to yield right-handed  charged quark currents, which can modify the CKM matrix results \cite{Albergaria:2023vls}. An additional VLQ family could 
also explain quark and lepton mass hierarchies \cite{CarcamoHernandez:2021yev}.
\\

In the context of the SM, vector-like quarks (VLQ) contribute to the stability of the vacuum, due to their strong coupling.  It is well-known that in the SM, the stability of the vacuum is threatened by the strong coupling of the top to the Higgs boson \cite{Altarelli:1994rb}. The simplest cure is to add a scalar singlet field, which mixes with the SM Higgs boson and compensates for the top quark contribution \cite{Ghorbani:2021rgs}. Vector-like quarks, due to their distinct representation under the electroweak group, offer a promising avenue for mitigating the vacuum stability problem. The question remains, how would the vacuum stability  be affected by the addition of VLQs to the particle content. 
\\

In a previous work \cite{Arsenault:2022xty}, we analyzed the effects of all possible representations of vector-like quarks and their implications for maintaining vacuum stability within the  SM augmented by an additional scalar. We have shown that, even with the addition of VLQs, the presence of the additional scalar was still a necessity. We extend this analysis here to the study of the effect of introducing vector-like quarks into the a simple extension of the SM, the Two-Higgs Doublet Model (2HDM). Thus we effectively replace the singlet scalar by scalars in a doublet representation. Our study involves analyzing all anomaly-free representations of vector-like quarks and their implications for maintaining vacuum stability within this model. As several versions of the model exist, we shall concentrate here on Type-I (where the fermions couple to only one Higgs doublet and the other is inert) and Type-II (where up quarks and neutrinos couple to one Higgs doublet, while down quarks and charged leptons couple to the other). The latter is of particular interest as it is consistent with the interaction structure required in supersymmetry.
\\

The Two-Higgs doublet models, seen as one of the simplest extensions of the SM,  have received a great deal of attention in the literature, see for example  \cite{Branco:2011iw,Gunion:2002zf,Coleppa:2013dya,Altmannshofer:2012ar,Ivanov:2006yq, Ginzburg:2005dt,Basler:2017nzu,Accomando:2019jrb,Song:2022xts,Cherchiglia:2017uwv,Haber:2015pua,BhupalDev:2014bir,Baglio:2014nea,Eberhardt:2013uba,Barroso:2013awa,Eriksson:2009ws,WahabElKaffas:2007xd,Maniatis:2006fs,Davidson:2005cw} and references therein. 
There are several motivations for extending the SM to 2HDMs. The best known  is, as alluded to before, supersymmetry.  In supersymmetric theories, the scalars belonging to multiplets of different chiralities cannot couple together in the Lagrangian, and thus a single Higgs doublet cannot give mass to both up- and down-type quarks. In addition, cancellation of anomalies also requires the presence of  an additional doublet. Another motivation for 2HDMs comes from axion models \cite{Kim:1986ax}. It was noted \cite{Peccei:1977hh}  that a possible CP-violating term in the QCD Lagrangian can be rotated away if the Lagrangian contains a global $U(1)$ symmetry, but this is possible only if there are two Higgs doublets.  And yet another motivation for 2HDMs comes from the fact that the SM is unable to generate a sufficiently large baryon asymmetry of the Universe, while 2HDMs can, due to  additional sources of CP violation \cite{Trodden:1998qg}.
\\

In this  paper, we investigate the effects of vector-like quarks in the context of  extending of SM to the 2HDM framework.   By incorporating vector-like quarks into 2HDM, we analyze whether we can overcome the negativity of quartic Higgs boson self couplings by finding a viable parameter space consistent with various  theoretical and experimental constraints in Type-I and Type-II 2HDM scenarios. Furthermore, we delve into the consequences of these extensions on precision electroweak observables. We focus on two separate components: first the oblique parameters originating from purely the 2HDM and second, on the impact of vector-like quark contributions on these observables. These analyses shed light on the potential alterations to electroweak measurements that arise from the inclusion of vector-like quarks in multi-Higgs scenarios.  Through numerical simulations, we demonstrate the significant role that vector-like quarks play in stabilizing the electroweak vacuum while maintaining agreement with precision electroweak measurements. Our aim is to provide  insights into the potential avenues for extending the SM  to address some of its shortcomings and set the theoretical framework  for future explorations and for experimental validations. 
\\

Our work is organized as follows. In Section \ref{sec:VS2HDM} we review the 2HDM. In the following Section \ref{sec:VS2HDM+VLQ} we review vector-like quarks, in singlet, doublet or triplet representations, setting the general Lagrangian responsible for their interaction, as well as reviewing experimental searches and theoretical considerations responsible for restricting their masses. Section \ref{sec:rgef} is dedicated to our exploration of the parameter space of the 2HDM with VLQs which satisfies vacuum stability bounds. Section \ref{sec:electroweakprecision} explores the constraints imposed by electroweak precision observables on the surviving parameter space, looking separately at the restrictions coming from the 2HDM alone, in \ref{subsec:2HDMSTU}, and from the VLQs, in \ref{subsec:VLQSTU}. We summarize our findings and conclude in Section \ref{sec:conclusion}. Finally, in the Appendices \ref{sec:appendix} we gather all RGE formulas for the VLQ representations  used in this work.
\\


\section{The Two Higgs Doublet Model}
\label{sec:VS2HDM}
In what follows, we present brief summary of the Two Higgs Doublet Model (2HDM). Extensive reviews of the 2HDMs of Type-I and Type-II are in {\it e.g.}\cite{Branco:2011iw,Gunion:2002zf}. The most general scalar
potential contains 14 parameters and can have CP-conserving, CP-violating, and charge violating
minima.  We make
several simplifying assumptions: that CP is conserved in the Higgs
sector, allowing one to distinguish between scalars and pseudoscalars, that CP is
not spontaneously broken, and that discrete symmetries eliminate from the potential all
quartic terms odd in either of the doublets.
\\

\begin{figure}[htbp]
	\hspace{-2.5cm}
	\begin{subfigure}{.33\textwidth}
		\includegraphics[height=1.8in]{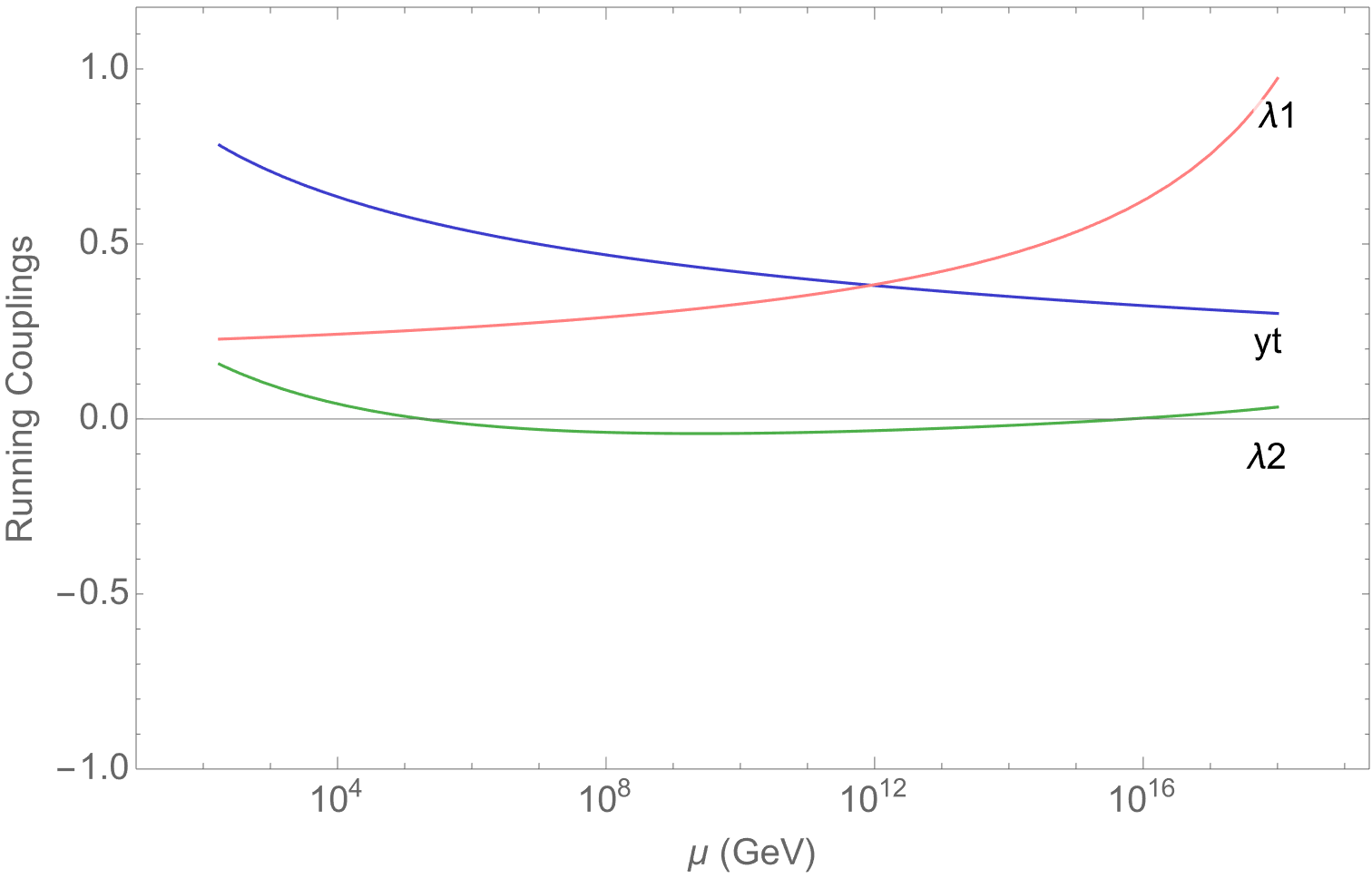}
		\caption{}
	\end{subfigure}
	\begin{subfigure}{.33\textwidth}
		\includegraphics[height=1.8in]{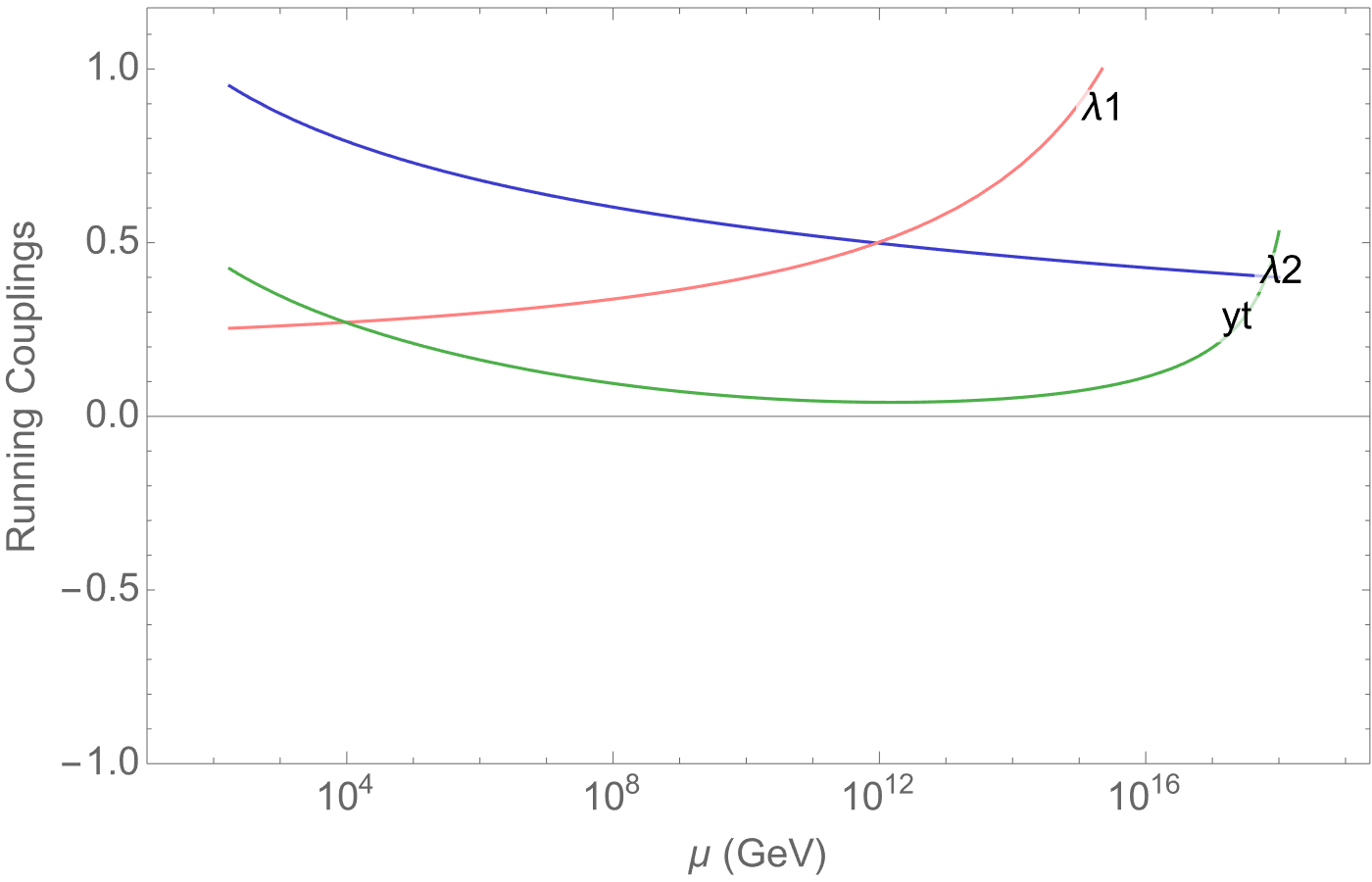}
		\caption{}
	\end{subfigure}
	\begin{subfigure}{.3\textwidth}
		\includegraphics[height=1.8in]{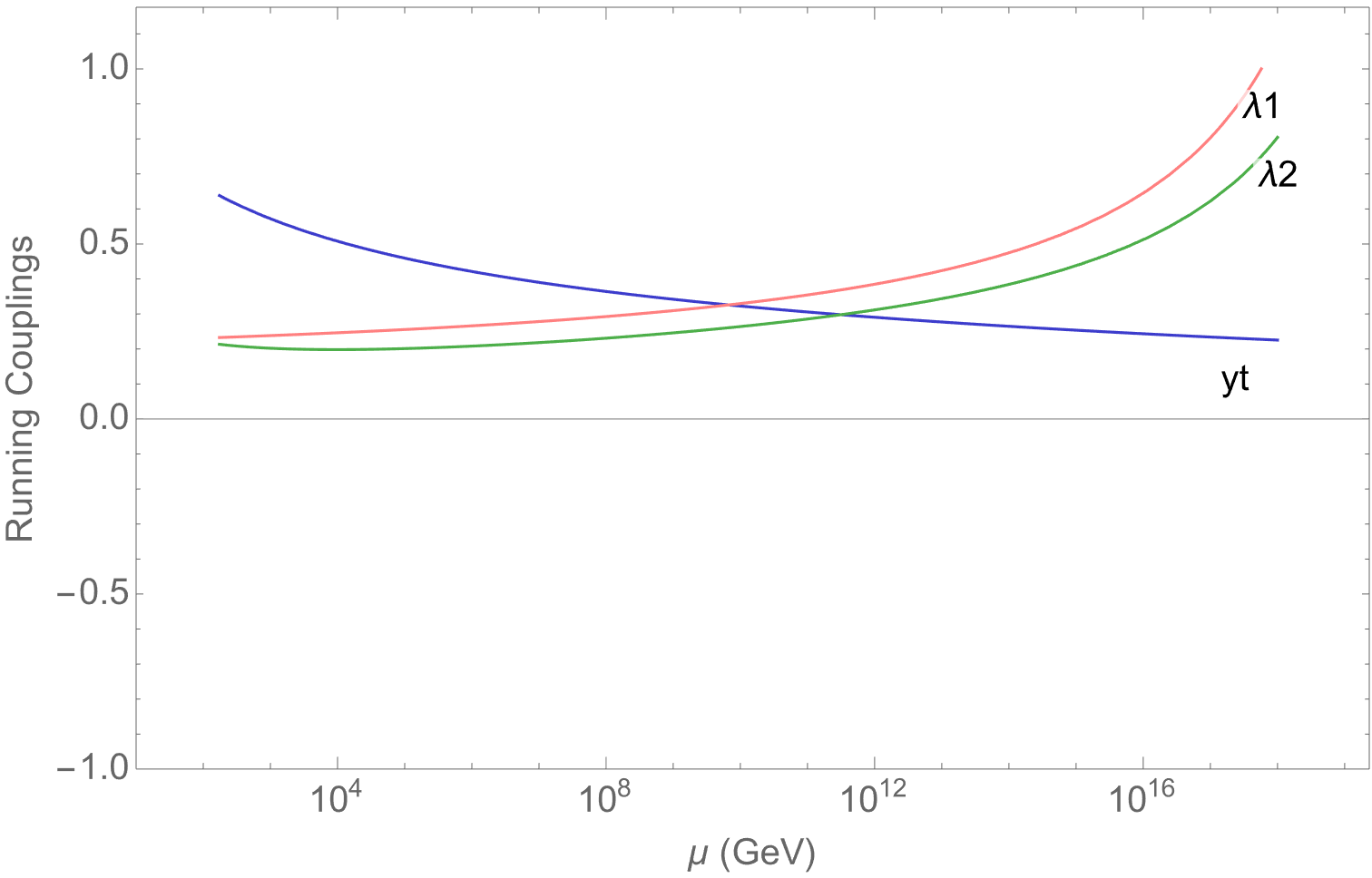}
		\caption{}
		\end{subfigure}
  \caption{The RGE running of the top Yukawa and scalar couplings $\lambda_1$ and $\lambda_2$ in 2HDM fixed at $\tan\beta=6$ for (a) $M_H=450$ GeV, (b) $M_H=600$ GeV, (c) $M_H=700$ GeV }
  \label{fig:2HDMrunning}
\end{figure}

The 2HDM scalar potential for the two doublet fields with hypercharge $Y=$1, which is invariant under the gauge symmetry of the SM, $SU(3)_C \otimes SU(2)_L \otimes U(1)_Y$ and satisfy a discrete ${\cal Z}_2$ symmetry is given by \cite{Branco:2011iw}
\begin{eqnarray}
\label{eq:2HDMpot}
V(\Phi_1,\Phi_2)&=&m_{11}^2\Phi_1^{\dagger}\Phi_1+m_{22}^2\Phi_2^{\dagger}\Phi_2-m_{12}(\Phi_1^{\dagger}\Phi_2+\Phi_2^{\dagger}\Phi_1)+\frac{\lambda_1}{2}(\Phi_1^{\dagger}\Phi_1)^2+\frac{\lambda_2}{2}(\Phi_2^{\dagger}\Phi_2)^2\, \nonumber\\
&+&\lambda_3(\Phi_1^{\dagger}\Phi_1)(\Phi_2^{\dagger}\Phi_2)+\lambda_4(\Phi_1^{\dagger}\Phi_2)(\Phi_2^{\dagger}\Phi_1)+\frac{\lambda_5}{2}\left[(\Phi_1^{\dagger}\Phi_2)^2+(\Phi_2^{\dagger}\Phi_1)^2\right]\,,
\end{eqnarray}
where the complex doublets are perturbed around their minimums $v_i$ as
\begin{equation}
\Phi_i=\left(\begin{matrix} w_i^+ \\ \frac{v_i+\rho_i+i\eta_i}{\sqrt{2}}\end{matrix}\right),\qquad (i=1,2)
\end{equation}
with $\sqrt{v_1^2+v_2^2}=v=246$ GeV, and the $m^2_{12}$ term softly breaks the ${\cal Z}_2$ symmetry.  The reason for introducing ${\cal Z}_2$ symmetry is to avoid tree-level flavour-changing neutral currents. Minimizing the 2HDM potential Eq. \ref{eq:2HDMpot} breaks electroweak symmetry and  allows the scalar potential   be fully described in terms of seven independent parameters. The scalar couplings at $\mu_0$ can be expressed in terms of the physical masses of the two CP-even scalars, $h$ and $H$, as:
\begin{eqnarray}
\label{eq:intscalarcouplings}
\lambda_{1}&=&\frac{M_H^2 \cos^2 \alpha + M_h^2 \sin^2 \alpha}{v^2\cos^{2}\beta} \, , \nonumber
 \\
\lambda_{2}&=&\frac{M_h^2 \cos^2 \alpha + M_H^2 \sin^2 \alpha}{v^2\sin^{2}\beta}\, , \nonumber 
\\
\lambda_{3}&=&\frac{\sin2\alpha}{v^{2}\sin2\beta}(M_H^2-M_h^2) + \frac{2M_{H^{\pm}}^{2}}{v^{2}}\, , \nonumber
\\
\lambda_{4}&=&\frac{M_{A}^{2}-2M^{2}_{H^{\pm}}}{v^{2}}\, , \nonumber
\\
\lambda_{5}&=&-\frac{M_{A}^{2}}{v^{2}}
\end{eqnarray}
along with $\tan\beta=v_2/v_1$ and $\alpha$ the mixing angle between the two CP-even scalars. In contrast to the SM vacuum which conserves CP symmetry but breaks $SU(2)_L\otimes U(1)_Y$ symmetry, there are four possible vacuum states in 2HDM. Charge-breaking vacuum occurs when the charged component of either of scalars acquires a non-zero VEV. $U(1)$ symmetry is spontaneously broken, and the photon gets a non-zero mass
\begin{equation}
\langle \Phi_1\rangle_{CB}=\frac{1}{\sqrt{2}}\left(\begin{matrix} 0 \\ c_1\end{matrix}\right), \qquad \langle \Phi_2\rangle_{CB}=\frac{1}{\sqrt{2}}\left(\begin{matrix} c_2 \\ c_3\end{matrix}\right),
\end{equation}
CP-breaking vacuum occurs when there is a relative phase difference between the vacuum expectation values (VEVs) of the neutral components of the scalar doublets
\begin{equation}
\langle \Phi_1\rangle_{CP}=\frac{1}{\sqrt{2}}\left(\begin{matrix} 0 \\ v_1\end{matrix}\right), \qquad \langle \Phi_2\rangle_{CP}=\frac{1}{\sqrt{2}}\left(\begin{matrix} 0 \\ v_1e^{i\eta}\end{matrix}\right)\, .
\end{equation}
The inert vacuum state happens when either one of the scalar fields acquire a non-zero VEV.
\begin{equation}
\langle \Phi_1\rangle_{IN}=\frac{1}{\sqrt{2}}\left(\begin{matrix} 0 \\ v\end{matrix}\right), \qquad \langle \Phi_2\rangle_{IN}=\frac{1}{\sqrt{2}}\left(\begin{matrix} 0 \\ 0 \end{matrix}\right),
\end{equation}
while mixed(normal) vacuum occurs when both of the neutral components of the scalar doublets have non-zero and positive VEVs
\begin{equation}
\langle \Phi_1\rangle_{N}=\frac{1}{\sqrt{2}}\left(\begin{matrix} 0 \\ v_1\end{matrix}\right), \qquad \langle \Phi_2\rangle_{N}=\frac{1}{\sqrt{2}}\left(\begin{matrix} 0 \\ v_2\end{matrix}\right)
\end{equation}
If all different vacua could have existed simultaneously in 2HDM potential, then one can undoubtedly think that the probability of transition between these states is non-zero. However, this was proven to be impossible once a certain type of vacuum was chosen \cite{Barroso:2013awa}. Furthermore, tree level vacuum stability is insured if the following conditions for the potential parameters in softly broken ${\cal Z}_2$ symmetry and CP conserving case are  satisfied:
\begin{equation}
\label{eq:stabilitytree1}
\lambda_1>0, \qquad \lambda_2>0, \qquad \lambda_3+\sqrt{\lambda_1\lambda_2}>0,\qquad \lambda_3+\lambda_4-\lvert\lambda_5\rvert>-\sqrt{\lambda_1\lambda_2}\, .
\end{equation}
These conditions are further put into a form which guarantees that the vacuum remains stable globally only if 
\begin{equation}
\label{eq:stabilitytree2}
-m_{12}\left(m_{11}^2-\frac{\lambda_1}{\lambda_2}m_{22}^2 \right)\left(\tan\beta-\left[\frac{\lambda_1}{\lambda_2}\right]^{1/4} \right)>0
\end{equation}
We consider the case where $M_h < M_H$ (with $h$ the SM-like Higgs boson), the light Higgs masses scenario and mixed vacuum in this study. Based on how ${\cal Z}_2$ symmetry is imposed on the 2HDM Lagrangian, four types of Yukawa interactions arise. Here we consider only two versions of the model:
\begin{itemize}
\item Type-I: All fermions couple to the $\Phi_2$ doublet, and the discrete symmetry is described as $\Phi_2\to-\Phi_2$.
\item Type-II: All charged leptons and down type quarks couple to $\Phi_1$ and all up type quarks couple to $\Phi_2$.
\end{itemize}
Although the conditions Eq. \ref{eq:stabilitytree1}-\ref{eq:stabilitytree2} are necessary, they are not sufficient to guarantee absolute stability of the electroweak vacuum at next-to leading order (NLO). In fact, the renormalization group equations (RGE) running of quartic couplings $\lambda_{1,2}$ in Type-I and Type-II are severely affected by negative corrections of top and bottom Yukawa couplings. 
\begin{eqnarray}
\label{eq:rge2HDM}
\frac{d \lambda_2^{I}}{d \ln \mu^2}&=& \frac{1}{16 \pi^2} \left[12\lambda_2^2+4\lambda_3^2+4\lambda_3\lambda_4+2\lambda_4^2+2\lambda_5^2-3\lambda_1(-4y_t^2+g_1^2+3g_2^2)-12y_t^4-12y_b^4+... \right ] \, , \nonumber \\
\frac{d \lambda_1^{II}}{d \ln \mu^2}&=& \frac{1}{16 \pi^2} \left[12\lambda_1^2+4\lambda_3^2+4\lambda_3\lambda_4+2\lambda_4^2+2\lambda_5^2-3\lambda_1(g_1^2+3g_2^2)-12y_b^4+... \right ] \, , \nonumber \\
\frac{d \lambda_2^{II}}{d \ln \mu^2}&=& \frac{1}{16 \pi^2} \left[12\lambda_2^2+4\lambda_3^2+4\lambda_3\lambda_4+2\lambda_4^2+2\lambda_5^2-3\lambda_2(-4y_t^2+g_1^2+3g_2^2)-12y_t^4+...\right ] \, ,
 \end{eqnarray}
where gauge portal terms are not shown here due to their positive contributions. In Fig. \ref{fig:2HDMrunning}, we present the running couplings of the quartic couplings $\lambda_1$ and $\lambda_2$ in 2HDM Type-II. Moreover, positivity of the quartic couplings are not directly guaranteed at loop level by simply imposing a mixing between CP-even scalars but it also relies on the other free parameters of 2HDM\footnote{A similar analysis for the Higgs singlet model (HSM) has been performed in Ref. \cite{Ghorbani:2021rgs}.}.  While $M_H=450$ GeV is insufficient to preserve the stability as $\lambda_2<0$ around $\mu\sim10^6$ GeV, increasing the scale to $M_H=600$ GeV and $M_H=700$ GeV secured the positivity of quartic couplings up to $\Lambda_{Pl}$.
We further use the radiative decay constant on the scalar initial conditions at RGE level \cite{Sirlin:1985ux}
\begin{equation}
\delta_\lambda(\mu)= \frac{G_F M_Z^2}{8 \sqrt{2} \pi^2} \left[\xi f_1(\xi, \mu)+f_0(\xi, \mu) +\xi^{-1} f_{-1}(\xi, \mu)\right],
\end{equation}
with $\xi= M_H^2/M_Z^2$, $G_F=1.16635 \times 10^{-5}$ GeV$^{-2}$ and
\begin{eqnarray}
f_1(\xi, \mu)&=&6 \ln \frac{\mu^2}{M_H^2}+\frac32 \ln \xi-\frac12 Z\left(\frac{1}{\xi}\right)- Z\left(\frac{c_W^2}{\xi}\right)-\ln c_W^2+\frac92 \left(\frac{25}{9}-
\sqrt{\frac13} \pi \right)\, \nonumber \\
f_0(\xi, \mu)&=&6 \ln \frac{\mu^2}{M_Z^2}\left [1+2c_W^2-2 \frac{m_t^2}{M_Z^2} \right]+\frac{3c_W^2 \xi}{\xi-c_W^2}+2Z\left(\frac{1}{\xi}\right) +4c^2 Z\left(\frac{c_W^2}{\xi}\right) +\frac{3c_W^2 \ln c_W^2}{s_W^2} \nonumber \\
&+&12c_W^2 \ln c_W^2-\frac{15}{2}(1+2c_W^2) -3 \frac{m_t^2}{M_Z^2} \left [ 2 Z\left(\frac{m_t^2}{\xi M_Z^2}\right)+4 \ln \frac{m_t^2}{M_Z^2} -5 \right], \nonumber \\
f_{-1}(\xi, \mu)&=&6 \ln \frac{\mu^2}{M_Z^2}\left [1+2c_W^4-24\frac{m_t^2}{M_Z^2} \right] -6Z\left(\frac{1}{\xi}\right)- 12 c_W^4 Z\left(\frac{c_W^2}{\xi}\right)-12 c_W^4 \ln c_W^2  \nonumber \\
&+&8(1+2c_W^4) + \left [  Z\left(\frac{m_t^2}{\xi M_Z^2}\right)+ \ln \frac{m_t^2}{M_Z^2} -2 \right]\,
\end{eqnarray}
with 
\begin{eqnarray}
Z(z)&=&\left \{ \begin{array}{lr} 2A\tan^{-1}(1/A), &z>1/4\\
A \ln [(1+A)/(1-A)], &z<1/4,
\end{array}\right.  \\
A&=&|1-4z|^1/2
\end{eqnarray} 
Note that the radiative correction is multiplicative only for $\lambda_2$ in 2HDM Type-I. 
We discard the contributions of all Yukawa couplings except for those of the third generation quarks, and also, we include electroweak radiative correction terms for increased accuracy \cite{Hempfling:1994ar}
\begin{equation}
\Delta_t(\mu_0)= \Delta_W(\mu_0) +\Delta_{QED}(\mu_0) +\Delta_{QCD}(\mu_0) \, ,
\end{equation}
with 
\begin{eqnarray}
\Delta_W(\mu_0)&=&\frac{G_Fm_t^2}{16 \sqrt{2} \pi^2}\left (-9 \ln\frac{m_t^2}{\mu_0^2}-4 \pi \frac{M_H}{m_t}+11 \right )\, , \nonumber \\
\Delta_{QED}(\mu_0)&=&\frac{\alpha}{9 \pi} \left (3 \ln \frac{m_t^2}{\mu_0^2}-4 \right)\, , \nonumber \\
\Delta_{QCD}(\mu_0)&=&\frac{\alpha_s}{9 \pi} \left (3 \ln \frac{m_t^2}{\mu_0^2}-4 \right)\, . 
\end{eqnarray}
Therefore, the initial condition for the top Yukawa coupling becomes 
\begin{equation}
y_t=\frac{\sqrt{2} m_t}{v} [1+\Delta_t(\mu_0)] \, .
\end{equation}
We now proceed with adding the contributions of VLQs into the model.
\\
\section{ 2HDM with vector-like quarks}
\label{sec:VS2HDM+VLQ}
\subsection{Theoretical Considerations}
\label{subsec:VLQtheory}
Using 2HDM potential in Section \ref{sec:VS2HDM}, we investigate the effect of introducing vector-like quarks on the stability of electroweak vacuum. Unlike SM-like (chiral) fermions whose left-handed and right-handed components transform differently under $SU(3)_C\otimes SU(2)_L \otimes U(1)_Y$, vector-like fermions have the same interactions regardless of chirality. However when we consider  incorporating them into the SM framework, it becomes necessary to introduce a new scalar boson into the Lagrangian. The additional scalar boson plays a crucial role in maintaining the stability of the 2HDM potential up to Planck scale. The rationale behind this requirement stems from the fact that the inclusion of extra fermions leads to a decrease in the effective self-coupling of the Higgs boson. Consequently, this extension could potentially exacerbate the negative evolution of the Higgs quartic coupling when compared to the scenario within the SM without additional particles. The presence of the new scalar boson serves as a remedy to this situation. 
\\

The question remains how would VLQs affect models with different scalar representations, such as 2HDMs. Throughout our work, we uphold the condition that the potential of 2HDM must remain positive up to the Planck scale. The main motivation for our study, is to establish the limitations that constrain the masses of vector-like quarks and the mixing angles with the SM quarks. Establishing these constraints is essential  in preserving the stability of the electroweak vacuum.
\\

The new VLQs states interact with the Higgs bosons through Yukawa interactions. The allowed anomaly-free multiplet states for the  vector-like quarks, together with their nomenclature, are listed in Table \ref{tab:VQrepresentations} \cite{Aguilar-Saavedra:2013qpa,Aguilar-Saavedra:2013wba,Ellis:2014dza,Carmi:2012yp}. The first two representations are $U$-like and $D$-like singlets \cite{Kang:2007ib,Fajfer:2013wca}, the next three are doublets (one SM-like, two non-SM like), and the last two are triplets. Note that the latter allow for quarks with exotic charges, $Q_X=5/3$ and $Q_Y=-4/3$. The various representations are distinguished by their $SU(2)_L$ and hypercharge numbers.
\begin{table}[htbp]
\caption{\label{tab:VQrepresentations}\sl\small Representations of Vector-Like Quarks, with quantum numbers under $SU(2)_L \times U(1)_Y$.}
  \begin{center}
 \small
 \begin{tabular*}{0.99\textwidth}{@{\extracolsep{\fill}} c| ccccccc}
 \hline\hline
	Name &${\cal U}_1$ &${\cal D}_1$ &${\cal D}_2$ &${\cal D}_X$ &${\cal D}_Y$ &${\cal T}_X$ 
	&${\cal T}_Y$\\
  Type&Singlet &Singlet &Doublet&Doublet &Doublet &Triplet 
	&Triplet\\
	 \hline
	   &$T$ &$B$ &$\left ( \begin{array}{c} T \\ B \end{array} \right ) $ &$ \left (\begin{array}{c}  X \\ T  \end{array}\right)$ &$\left ( \begin{array}{c} B \\ Y\end{array} \right ) $ &$ \left (\begin{array}{c} X\\T \\ B \end{array} \right ) $
	  & $\left ( \begin{array}{c} T \\ B\\Y \end{array} \right )$\\
  \hline
  $SU(2)_L$ &1 &1 &2 &2 & 2 &3 &3 \\
  \hline
  $Y$ &$ 2/3$ &$ -1/3$ &$1/6 $ &$ 7/6$ &$-5/6$ &$2/3$ &$-1/3$ \\
      \hline
    \hline
   \end{tabular*}
\end{center}
 \end{table}

The Yukawa and other relevant interaction terms between the vector-like quarks and SM quarks are, in the bare ($\Phi_1,\Phi_2$) basis for Type-I:
\begin{eqnarray}
\label{eq:lagtype1}
{\cal L}^{I}_{SM}&=& -y_u {\bar q}_L \Phi_2^c u_R -y_d{\bar q}_L \Phi_2 d_R \nonumber \\
{\cal L}^{I}_{{\cal U}_1, {\cal D}_1}&=& -y_T{\bar q}_L \Phi_2^c U_{1_R} -y_B{\bar q}_L \Phi_2 D_{1_R}-y_M ({\bar U}_{1_L} \Phi_2 U_{1_R} +{\bar D}_{1_L} \Phi_2 D_{1_R})-M_U {\bar U}_L U_R-M_D {\bar D}_L D_R, \nonumber \\
{\cal L}^{I}_{{\cal D}_2}~~&=& -y_T {\bar D}_{2_L} \Phi_2^c u_{R} -y_B{\bar D}_{2_L} \Phi_2 d_{R}-y_M ({\bar D}_{2_L} \Phi_2^c D_{2_R}+y_B{\bar D}_{2_L} \Phi_2 D_{2_R})-M_D {\bar D}_{2_L}  D_{2_R}, \nonumber \\
{\cal L}^{I}_{{\cal D}_X, {\cal D}_Y}&=& -y_T {\bar D}_{X_L}\Phi_2 u_{R} -y_B{\bar D}_{Y_L} \Phi_2^c d_{R}-y_M( {\bar D}_{X_L}\Phi_2 D_{X_R} +y_B{\bar D}_{Y_L} \Phi_2^c D_{Y_R})-M_X {\bar D}^{I}_{X_L} D_{X_R}-M_Y {\bar D}_{Y_L} D_{Y_R}, \nonumber \\
{\cal L}^{I}_{{\cal T}_X, {\cal T}_Y}&=& -y_T {\bar q}_{L}\tau^a \Phi_2^c  {\cal T}^a_{X_R} -y_B{\bar q}_{L} \tau^a \Phi_2 {\cal T}^a_{Y_R}-y_M ({\bar {\cal T} }_{X_L}\tau^a \Phi_2^c  {\cal T}^a_{X_R} +y_B{\bar {\cal T}}_{Y_L} \tau^a \Phi_2 {\cal T}^a_{Y_R})-M_X {\bar  {\cal T}}_{X_L}  {\cal T}_{X_R}-M_Y {\bar  {\cal T}}_{Y_L}  {\cal T}_{Y_R}\nonumber \\ 
\end{eqnarray}
and for Type-II:
\begin{eqnarray}
\label{eq:lagtype2}
{\cal L}^{II}_{SM}&=& -y_u {\bar q}_L \Phi_2^c u_R -y_d{\bar q}_L \Phi_1 d_R \nonumber \\
{\cal L}^{II}_{{\cal U}_1, {\cal D}_1}&=& -y_T{\bar q}_L \Phi_2^c U_{1_R} -y_B{\bar q}_L \Phi_1 D_{1_R}-y_M ({\bar U}_{1_L} \Phi_2 U_{1_R} +{\bar D}_{1_L} \Phi_1 D_{1_R})-M_U {\bar U}_L U_R-M_D {\bar D}_L D_R, \nonumber \\
{\cal L}^{II}_{{\cal D}_2}~~&=& -y_T {\bar D}_{2_L} \Phi_2^c u_{R} -y_B{\bar D}_{2_L} \Phi_1 d_{R}-y_M ({\bar D}_{2_L} \Phi_2^c D_{2_R}+y_B{\bar D}_{2_L} \Phi_1 D_{2_R})-M_D {\bar D}_{2_L}  D_{2_R}, \nonumber \\
{\cal L}^{II}_{{\cal D}_X, {\cal D}_Y}&=& -y_T {\bar D}_{X_L} \Phi_2 u_{R} -y_B{\bar D}_{Y_L} \Phi_1^c d_{R}-y_M( {\bar D}_{X_L} \Phi_2 D_{X_R} +y_B{\bar D}_{Y_L} \Phi_1^c D_{Y_R})-M_X {\bar D}_{X_L} D_{X_R}-M_Y {\bar D}_{Y_L} D_{Y_R}, \nonumber \\
{\cal L}^{II}_{{\cal T}_X, {\cal T}_Y}&=& -y_T {\bar q}_{L}\tau^a \Phi_2^c  {\cal T}^a_{X_R} -y_B{\bar q}_{L} \tau^a \Phi_1 {\cal T}^a_{Y_R}-y_M ({\bar {\cal T} }_{X_L}\tau^a \Phi_2^c  {\cal T}^a_{X_R} +y_B{\bar {\cal T}}_{Y_L} \tau^a \Phi_1 {\cal T}^a_{Y_R})-M_X {\bar  {\cal T}}_{X_L}  {\cal T}_{X_R}-M_Y {\bar  {\cal T}}_{Y_L}  {\cal T}_{Y_R}\nonumber \\ .
\end{eqnarray}
where $\Phi_i^c=i \sigma^2 \Phi_i^\star$, (i=1,2), $y_u$, $y_d$, $y_T$, and $y_B$ are the Yukawa couplings of the scalar fields $\Phi_{1,2}$ to vector-like and to SM quarks, while $y_M$ is the Yukawa coupling of the scalar fields to only vector-like quarks.
\\\\
The gauge eigenstate fermion fields resulting from the mixing can be written in general as,
\begin{eqnarray}
{\cal T}_{L,R}=&\left(\begin{matrix}
t\\T\end{matrix}\right)_{L,R}\ \qquad 
{\cal B}_{L,R}=&\left(\begin{matrix} b\\B\end{matrix}\right)_{L,R}\ 
\,
\label{eq:gauge_eigenst_fields}
\end{eqnarray}
The mass eigenstate fields are denoted as $(t,_1, t_2)$ and $(b_1, b_2)$ and they are found through bi-unitary transformations,
\begin{eqnarray}
{\mathbf T}_{L,R}&=&\left(\begin{matrix} t_1 \\t_2\end{matrix}\right)_{L,R}=V_{L,R}^t \left(\begin{matrix} t\\T\end{matrix}\right)_{L,R}\nonumber \\
{\mathbf B}_{L,R}&=&\left(\begin{matrix} b_1\\b_2\end{matrix}\right)_{L,R}=V_{L,R}^b \left(\begin{matrix} b\\B\end{matrix}\right)_{L,R}
\, ,
\end{eqnarray}
where
\begin{equation}
V_{L,R}^{t}=\left(
\begin{matrix}
\cos\theta^t& -\sin\theta^t\\
\sin\theta^t &\cos\theta^t\end{matrix} 
\right)_{L,R}\, , \qquad
V_{L,R}^{b}=\left(
\begin{matrix}
\cos\theta^b & -\sin\theta^b\\
\sin\theta^b &\cos\theta^b \end{matrix}
\right)_{L,R}\, ,
\label{eq:rotation_matrices}
\end{equation}
In the following we abbreviate $\cos\theta_L^t\equiv c_L^t$,.... 
Through these rotations we obtain the diagonal mass matrices 
\begin{equation}
\label{eq:massmatrices}
M^t_{diag}=V_L^t M^t (V_R^t)^\dagger=\left(\begin{matrix}m_{t_1} & 0 \\0 &m_{t_2}\end{matrix}\right)
\quad , \quad
M^b_{diag}=V_L^b M^b (V_R^b)^\dagger=\left(\begin{matrix}m_{b_1} & 0 \\0 &m_{b_2}\end{matrix}\right)\, .
\end{equation}
By using the gauge eigenstate fields, the mass matrices in the top and bottom sector are given, after spontaneous symmetry breaking
\begin{eqnarray}
-{\cal L}^{t}_{Yuk}&=&\left(\begin{matrix} t_L & T_L \end{matrix}\right) \left(
\begin{matrix}
y_t\frac{v}{\sqrt{2}}& y_T\frac{v}{\sqrt{2}}\\
y_T\frac{v}{\sqrt{2}}&y_M\frac{v}{\sqrt{2}}+M_T\end{matrix} 
\right)\left(\begin{matrix} t_R \\ T_R\end{matrix}\right)\,, \nonumber \\
-{\cal L}^{b}_{Yuk}&=&\left(\begin{matrix} b_L & B_L \end{matrix}\right) \left(
\begin{matrix}
y_b\frac{v}{\sqrt{2}}& y_B\frac{v}{\sqrt{2}}\\
y_B\frac{v}{\sqrt{2}}&y_M\frac{v}{\sqrt{2}}+M_B\end{matrix} 
\right)\left(\begin{matrix} b_R \\ B_R\end{matrix}\right)
\end{eqnarray}
The mass eigenvalues for top partners in Type-I, II+VLQ model are
\begin{eqnarray}
 m_{t_1,t_2}^2=\frac{1}{4}\left [( y_t^2+y_T^2+y_M^2)v^2 \right] \left[ 1\pm \sqrt{1-\left( \frac{2y_t y_M }{ (y_t^2+y_T^2+y_M^2)}\right )^2} \right]
 \label{eq:eigenvecTt}
 \end{eqnarray}
 with eigenvectors
 \begin{equation}
\left(  \begin{array}{c} t_1 \\ t_2  \end{array} \right )_{L,R}=  \left(  \begin{array}{cc} \cos \theta^t_{L,R} & \sin \theta^t_{L,R} \\ -\sin \theta^t_{L,R}  & \cos \theta^t_{L,R} \end{array}  \right)
\left( \begin{array}{c} t \\T  \end{array} \right )_{L,R}
\, .
 \end{equation}
Diagonalization of the mass matrices Eq. \ref{eq:massmatrices} is useful for expressing the mixing angles for top and bottom sector in terms of the free parameters of the model
\begin{eqnarray}
 \label{eq:mixing}
 \tan (2\theta^t_L)&=&\frac{2y_T y_M}{ y_M^2-y_t^2-y_T^2}\nonumber \\
\tan (2\theta^t_R)&=& \frac{2y_t y_T}{y_M^2+y_t^2-y_T^2}\, ,
\end{eqnarray}
\\

Charge assignments of the non-SM-like quarks do not allow the $X$ and $Y$ fields to mix with the other fermions. Therefore, these vector-like quarks are also mass eigenstates. Bottom sector mixing angle can be obtained with the replacement $t\to b$ and $\theta^t\to\theta^b$. And solving Eq. \ref{eq:mixing} for the Yukawa couplings we end up with the relations between mass eigenvalues and mixing angles:
\begin{equation}
\frac{y_T}{y_t}=s_L^tc_L^t\frac{m_t^2\frac{\tan\theta^{t^{2}}_L}{\tan\theta^{t^{2}}_R}-m_T^2\frac{\tan\theta^{t^{2}}_R}{\tan\theta^{t^{2}}_L}}{m_Tm_t}
\end{equation} 
Connection between mass eigenstates and mixing angles of the SM quarks to VLQs in anomaly-free states is unique for each representation \cite{Aguilar-Saavedra:2013qpa}
 \begin{eqnarray}
 \label{eq:mixingrelations}
{\rm For~ doublets:} 
&(XT): &m_X^2=m_T^2 (\cos \theta^t_R)^{2}+m_t^2 (\sin \theta^t_R)^{2}\nonumber \\
&(TB): &m_T^2 (\cos \theta^t_R)^{2}+m_t^2 (\sin \theta^t_R)^{2}=m_B^2 (\cos \theta^b_R)^{2}+m_b^2 (\sin \theta^b_R)^{2}
\nonumber \\
&(BY):&m_Y^2=m_B^2 (\cos \theta^b_R)^{2}+m_b^2 (\sin \theta^b_R)^{2}\nonumber \\
\nonumber \\
{\rm For ~triplets:}& (XTB): &m_X^2=m_T^2 (\cos \theta^t_L)^{2}+m_t^2 (\sin \theta^t_L)^{2}\nonumber \\
 && \phantom{m_X^2} =m_B^2 (\cos \theta^b_L)^{2} + m_b^2(\sin \theta^b_L)^{2} \, ,\nonumber \\
  &&{\rm where}~\sin(2\theta^b_L)= \sqrt{2}{m_T^2-m_t^2\over  (m_B^2-m_b^2)}\sin(2\theta^t_L)\, .\nonumber \\
  &(TBY):& m_Y^2=m_B ^2 (\cos \theta^b_L)^2+m_b^2 (\sin \theta^b_L)^{2}\nonumber \\
  &&\phantom{m_Y^2}=m_T^2 (\cos \theta^t_L)^{2}+m_t^2 (\sin \theta^t_L)^{2} \, , \nonumber\\
 &&{\rm where}~\sin(2\theta^b_L)= {m_T^2-m_t^2\over \sqrt{2} (m_B^2-m_b^2)}\sin(2\theta^t_L)\, , 
 \label{relations}
\end{eqnarray}
and where 
\begin{eqnarray}
m_{T,B} (\tan \theta^{t,b}_R)&=m_{t,b}(\tan \theta^{t,b}_L)\qquad &{\hbox{for~singlets,~triplets}}\nonumber \\
m_{T,B} (\tan \theta^{t,b}_L)&=m_{t,b} (\tan \theta^{t,b}_R)\qquad &{\hbox{for~doublets}} \, .
\label{angles}
\end{eqnarray}
We defined here $m_T=M_U+y_T v \sin \beta/\sqrt{2}$, $m_B=M_D+y_B v \cos \beta/\sqrt{2}$, $m_t=y_t  v \sin \beta/\sqrt{2}$, $m_b=y_b  v \cos \beta/\sqrt{2}$ for Type-II  models (for Type-I, replace $v\cos \beta$ and $v \sin \beta$ by $v$) , while $m_X=M_X$ and $m_Y=M_Y$.
Initial conditions for all Yukawa couplings are modified with mixing relations. For Type-I+VLQ, all fermions acquire mass by interacting with VEV of $\Phi_2$
\begin{eqnarray}
y^I_t(\mu_0)&=&\frac{\sqrt{2} m_t}{v}\frac{1}{\sqrt{\cos^2 \theta_L+x_t^2 \sin^2 \theta_L}}\, , \nonumber \\
y^I_T(\mu_0)&=&\frac{\sqrt{2} m_T}{v}\frac{\sin \theta_L \cos \theta_L (1-x_t^2)}{\sqrt{\cos^2 \theta_L+x_t^2 \sin^2 \theta_L}}\, , \nonumber \\
y^I_B(\mu_0)&=&\frac{\sqrt{2} m_B}{v}\frac{\sin \theta_L \cos \theta_L (1-x_b^2)}{\sqrt{\cos^2 \theta_L+x_b^2 \sin^2 \theta_L}}\, , \nonumber \\
y^I_M(\mu_0)&=&\sum_{i=X,T,B,Y}\frac{C_R m_i}{v}\sqrt{\cos^2 \theta_L+x_t^2 \sin^2 \theta_L}\, , \nonumber \\
\end{eqnarray}
whereas in Type-II+VLQ, $\tan\beta$, which is the ratio of VEVs, modifies the initial conditions to read 
\begin{eqnarray}
\label{eq:type2IC}
y^{II}_t(\mu_0)&=&\frac{\sqrt{2} m_t}{v\sin\beta}\frac{1}{\sqrt{\cos^2 \theta_L+x_t^2 \sin^2 \theta_L}}\, , \nonumber \\
y^{II}_T(\mu_0)&=&\frac{\sqrt{2} m_T}{v\sin\beta}\frac{\sin \theta_L \cos \theta_L (1-x_t^2)}{\sqrt{\cos^2 \theta_L+x_t^2 \sin^2 \theta_L}}\, , \nonumber \\
y^{II}_B(\mu_0)&=&\frac{\sqrt{2} m_B}{v\cos\beta}\frac{\sin \theta_L \cos \theta_L (1-x_b^2)}{\sqrt{\cos^2 \theta_L+x_b^2 \sin^2 \theta_L}}\, , \nonumber \\
y^{II}_M(\mu_0)&=&\sum_{i=X,T,B,Y}\frac{C_R m_i}{v}\sqrt{\cos^2 \theta_L+x_t^2 \sin^2 \theta_L}\, , \nonumber \\
\end{eqnarray}
where $C_R=(\sqrt{2},\frac{1}{\sqrt{2}},\frac{\sqrt{2}}{3}$) is the representation dependent weight factor with $x_b=m_b/m_B$, and as before $x_t=m_t/m_T$. Since X and Y fields do not mix with other fermions of the model, their low-energy Yukawa couplings are not altered by mixing relations. However, $y_X$ and $y_Y$ have indirect effects on the coupled RGEs, as seen from Eq. \ref{sec:apprge1}-\ref{sec:apprge2} for Type-I and Type-II  analysed in this work. Furthermore, the initial conditions on the VLQ Yukawa couplings in Type-II have different $\beta$ dependences in Eq. \ref{eq:type2IC} based on which field is an up- or down-type member of the multiplets.
\\
\subsection{Restrictions on VLQ masses}
\label{subsec:VLQlimits}
Bounds on masses of VLQs were established by the direct searches at the LHC by ATLAS \cite{ATLAS:2015uhg,ATLAS:2016scx,ATLAS:2015ktd,ATLAS:2015vzd, ATLAS:2017lvm} and by CMS \cite{CMS:2017ked,CMS:2017gsh,CMS:2015hyy,CMS:2015hyy,CMS:2015jwh,CMS:2013wkd,CMS:2013hwy} collaborations, obtained from specific mechanisms such as single production \cite{ATLAS:2023bfh} and pair production \cite{ATLAS:2022hnn,ATLAS:2017vdo} at $s=\sqrt{13}$ TeV.  The constraints are sensitively dependent on the assumed decay channels of the light VLQs, which are allowed by kinematics to decay into a SM quark. If VLQs decay only to the third generation quarks, then following channels could be observed:\footnote{For VLQs that carry non-SM-like hypercharges, the following CC and NC channels are also allowed and searched for accordingly: $X\to tW^+$, $Y\to bW^-$, $T(B)\to X(Y)W^-(W^+)$. See also \cite{Aguilar-Saavedra:2013qpa}.} $T(B)\to W^+(W^-)t(b)$, $T(B)\to Zt(b)$,  $T(B)\to Ht(b)$, hence the bounds become relatively stronger due to the final states. The constraints $m_T>1.27$ TeV, $m_B>$1.2 TeV are obtained for  singlets, whereas doublets require slightly higher mass limits $m_T>1.46$ TeV, $m_B>$1.32 TeV through pair production. Nonetheless, the lower limits on the VLQ masses in the range of $[800 , 1400]$ GeV and $\sin\theta<0.18$ from Run 2 \cite{ATLAS:2018cye} are still compatible with the data \cite{ATLAS:2023pja}. It should be noted that these limits are decreased if the first and the second generation SM quarks are also included. However, since the Yukawa couplings play an essential role due their direct relations to masses, these models are commonly unfavored. As our work concerns 2HDMs, we consider a lowest limit on $m_T$ of 800 GeV, to allow for the consideration of the largest parameter space for the electroweak vacuum stability and electroweak observables (EWPOs).
\\

Corrections to the mass of $W$-boson are calculated  using the oblique parameters. To this end, precision experiments carried out at the Tevatron \cite{CDF:2022hxs} that signal any type of shift in $\Delta M_W$ are used to describe effects from new physics (NP). Since both the scalar and the fermion sector contribute to EWPO, the combined corrections significantly rely on scalar extensions in addition to vector-like fermions. Singlet (HSM) \cite{Lopez-Val:2014jva,Xiao:2014kba} and triplet  (HTM) scalar models \cite{Bahrami:2014ska,Shimizu:2023rvi} have already been studied. However, for 2HDM+VLQ, we are only interested in constraints coming from the $\chi^2(S,T)$[VLQ+2HDM] analysis in order to generate a viable space for the electroweak vacuum stability requirements. 
\\

There are alternative ways for corrections to Higgs self-energies which would manifest themselves, especially when the new particles carry SM-like colour and electroweak quantum numbers. In these scenarios, for every diagrammatic contribution to the self-energies, one could replace one of the Higgs bosons by its vacuum expectation value and attach two SM gauge bosons to the loop. From there, one can obtain a corresponding diagrammatic contribution to the Higgs decays to SM gauge bosons. A rough estimation of possible deviations from precision electroweak measurements, which pushed new physics to  $\Lambda_{NP}\sim$ 1 TeV, is based on the estimate of the size of Higgs oblique
corrections  roughly given by ${\cal O} (v^2/\Lambda^2) \sim 5\%$. If VLQs enter the loop diagrams, new fermions or charged bosons contribute to the loop-induced diphoton decay and/or gluon fusion channels of the Higgs bosons. Effects of VLQs on Higgs couplings have been explored in studies for singlet \cite{Aguilar-Saavedra:2006uim,Xiao:2014kba}, doublet \cite{Dawson:2012di} and triplet models \cite{Benbrik:2015fyz}. The $T$- singlet VLQ model established  an upper bound $\sin\theta_L<0.4$ from the combined $H\to gg$ and $H\to\gamma\gamma$ cross section and branching ratio, respectively, while in the doublet ($TB$) representation an upper limit  $\sin\theta_L<0.115$ was obtained only from contribution to gluon fusion cross section $\mu_{\gamma\gamma}\leq 1.03$, while the triplet ($XTB$) model contribution is  $\mu_{\gamma\gamma}\leq1.18$ around $m_{VLQ}\simeq 1$ TeV. Consequently, all these studies have shown VLQ corrections which match  the earlier correction scale from NP models. 
\\

Another constraint comes from $B$-physics, namely from $b\to s\gamma$. This is by far the most significant constraint throughout the work carried here since $\tan\beta$ alone varies in a significantly large interval for the  2HDM models without VLQs. Thus we assumed here that the scalar parameters that are fine-tuned with respect to VLQ signals are extremely sensitive to new physics. There are studies in literature which extend 2HDM with singlet \cite{Arhrib:2016rlj} and doublet \cite{Benbrik:2022kpo} VLQs, providing solid constraints to work around $\tan\beta\leq$12, $M_{H^{\pm}}=[80,1000]$ GeV, and $m_{VLQ}>$1 TeV along with small mixings $\sin\theta_L<$0.08.
\\

The pertubativity of the Yukawa couplings $y_i$ is one of the weakest constraints at tree level, extending the upper bound of the mixing angle as $\sin\theta^t_L=[0.77 , 0.31]$ versus $m_T=[0.8, 2]$ TeV \cite{Aguilar-Saavedra:2013qpa}. Unitarity entails the S-matrix for scalar scattering to be unitary at high energy \cite{Horejsi:2005da}, imposing upper limits as $M_{\Phi^0}<\sqrt{\frac{4\pi}{\sqrt{2}G_F}}= 870$ GeV for scalar-scalar scattering and $M_{\Phi^0}<\sqrt{\frac{8\pi}{3\sqrt{2}G_F}}= 712$ GeV for gauge boson-scalar scattering in 2HDM.  

\section{RGE Analysis of the Parameter Space of 2HDM with VLQs}
\label{sec:rgef}
The effect of fermions on the stability of the electroweak vacuum  without extending scalar sector beyond the SM Higgs field is to drive the Higgs  self-coupling negative at larger scales, so the potential becomes unbounded from below, and there is no resulting stability. Theoretical considerations indicate that if the validity of the SM is extended to $\Lambda_{Pl}$, a second, deeper minimum is located near the Planck scale such that
the electroweak vacuum is metastable \cite{Degrassi:2012ry,Buttazzo:2013uya}. The additional scalar bosons  maintain positivity of the Higgs self coupling while the renormalization flow tends to decrease it further at higher energy scales \cite{Altarelli:1994rb}. Moreover, a common feature of both observed and exotic fermions  is that Yukawa couplings generally further lower scalar couplings since Yukawa couplings are negatively affected by NLO contributions. However, this is not always the case, and it  depends on how the structure of gauge interactions have been affected by new fields. Through the possibility of various interaction portals, vector-like fermions open new ways to remedy  stabilizing the electroweak vacuum.
\\ 

A straightforward  approach would be to extend the gauge sector of the SM as the gauge beta-functions have positive  effects on quartic coupling RGEs \cite{Tang:2013bz}. However, additional gauge symmetries might also come short of  being able to express the current SM interactions as they have relatively small contributions compared to other remedies. Nonetheless, these corrections, $\Delta\beta_1=\frac{8}{3}n_FG_2G_3y_f^2$ and $\Delta\beta_{2,3}=\frac{8}{3}n_Fd_{2,3}S_2(G_{2,3})$\footnote{Here $S_2(G_i)$ are Dynkin indices for the groups $G_2$ and $G_3$.}, are multiplicative with respect to new fermion families and these contributions are already manifest at the RGE level, as we shall see in Section \ref{sec:appendix}. Yukawa and scalar portals have shown promising results, providing non-critical surfaces of  electroweak vacuum stability \cite{Hiller:2022rla}. As shown in Section \ref{subsec:VLQtheory}, Yukawa portals lead to mixing between vector-like quarks and the SM quarks. Due to mixing constraints, for an energy scale less than the mass of $m_{VLQ}$, decoupling occurs and VLQs contribute to RGE running as if they were massless. Furthermore, in the presence of VLQs, beyond-SM gauge couplings have larger values compared to the SM ones, thus reducing the corrections to Yukawa couplings running at energy scale $\mu\geq m_{VLQ}$. This could be shown, for instance, for the top quark, where the beta function 
\begin{equation}
\beta_t\supset y_t[C_f y_f-C_1g_1-C_2g_2-C_3g_3]
\end{equation} 
which in turn shows that $y_{t}(\mu)<y^{SM}_{t}(\mu)$. Moreover, gauge and Yukawa couplings have opposite sign contributions in scalar RGEs Eq. \ref{sec:apprge1}-\ref{sec:apprge2} when fermions are allowed to interact with the scalars of the model. This characteristic can be seen from all scalar RGEs except the one governs $\lambda_1^I$, which is not allowed to interact with fermions through Yukawa couplings due to ${\cal Z}_2$ symmetry. Thus,  from RGE structure, the gauge and Yukawa couplings could lead to upward shifts in the Higgs quartic couplings though the condition  $\lambda_{1,2}>\lambda^{SM}_{1,2}$ in the presence of VLQ. We note that, in this context, vector-like quarks have been studied with only the SM Higgs field \cite{Gopalakrishna:2018uxn} and within the additional Higgs singlet model \cite{Arsenault:2022xty}.
\\

\begin{figure}[htbp]
	\centering
	\begin{subfigure}{.5\textwidth}\hspace{-1.5cm}
		\includegraphics[height=1.8in]{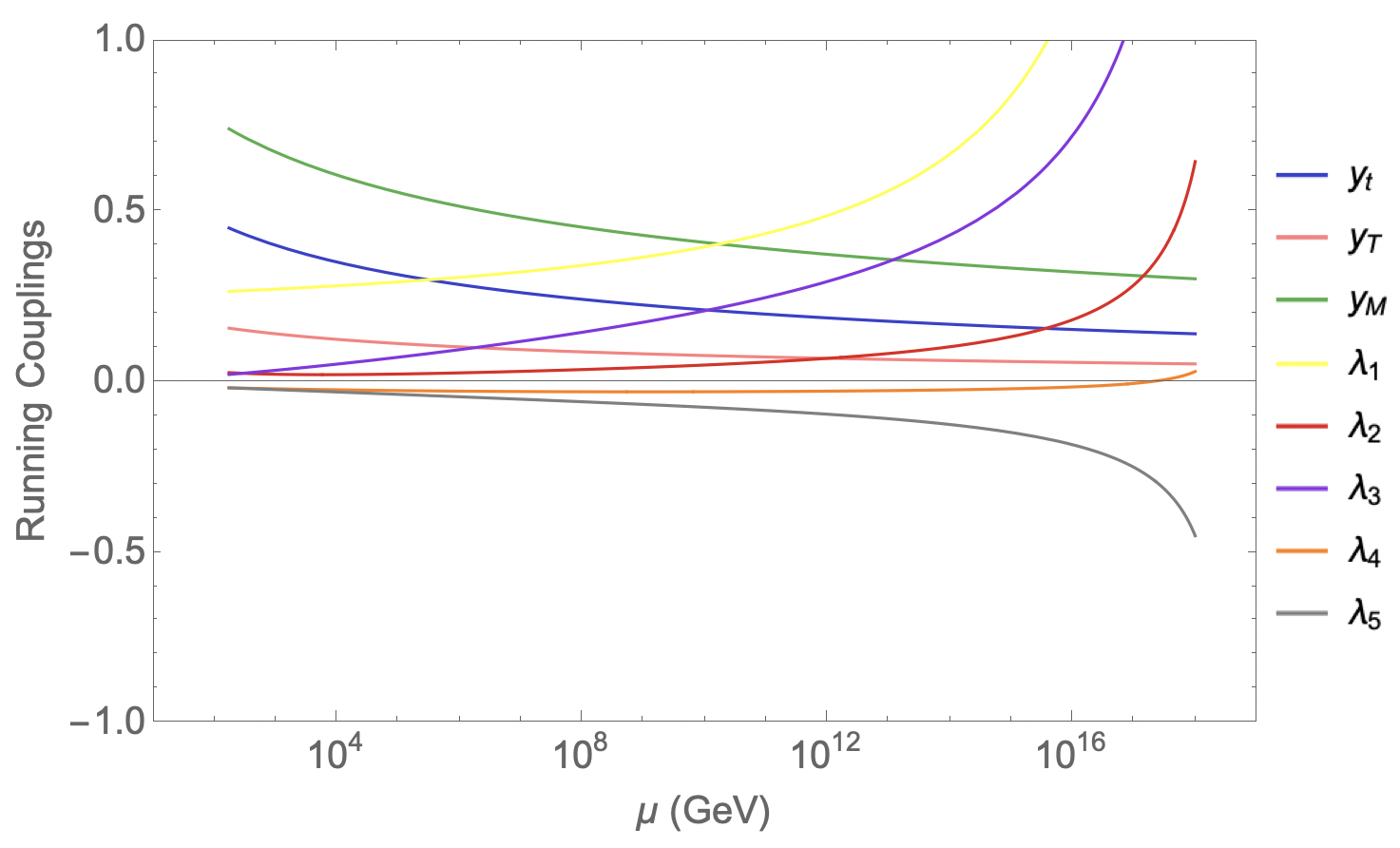}
		\caption{$\mathcal{U}_{1}$ (T) - Type I}
	\end{subfigure}\hspace{-0.5cm}
	\begin{subfigure}{.5\textwidth}
		\includegraphics[height=1.8in]{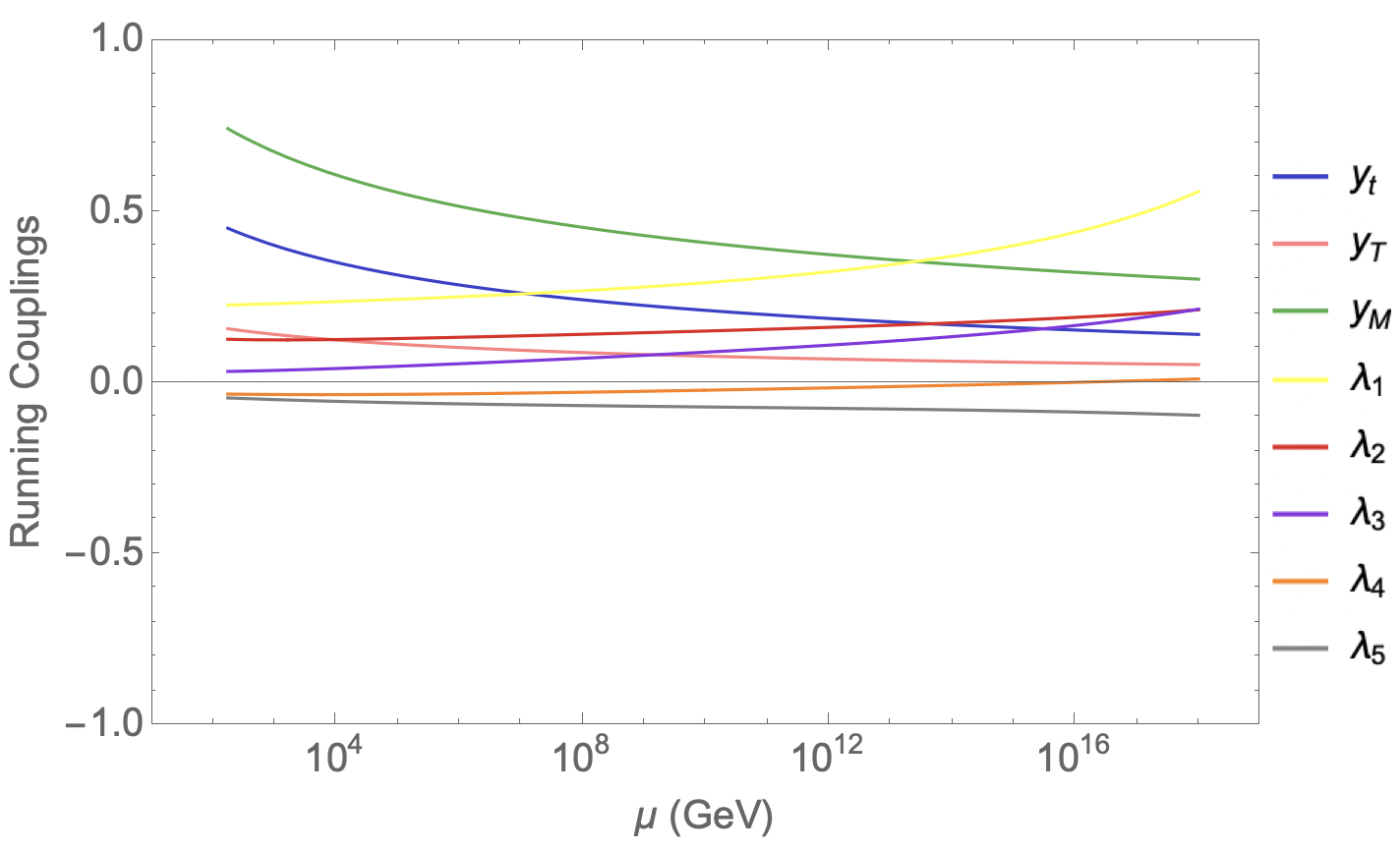}
		\caption{$\mathcal{U}_{1}$ (T) - Type II}
	\end{subfigure}\\
\begin{subfigure}{.5\textwidth}\hspace{-1.5cm}
		\includegraphics[height=1.8in]{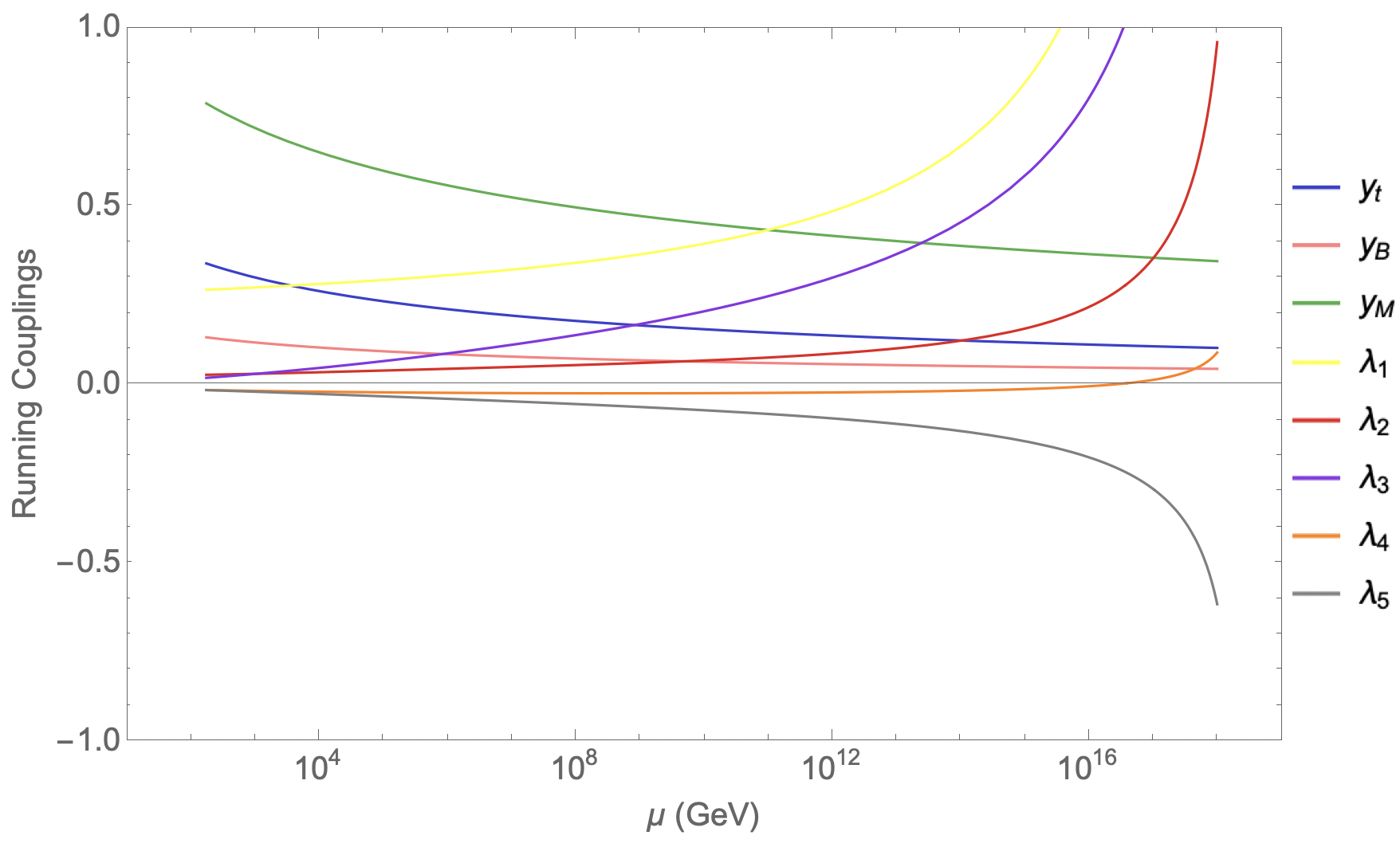}
		\caption{$\mathcal{D}_{1}$ (B) - Type I}
	\end{subfigure}\hspace{-0.5cm}
	\begin{subfigure}{.5\textwidth}
		\includegraphics[height=1.8in]{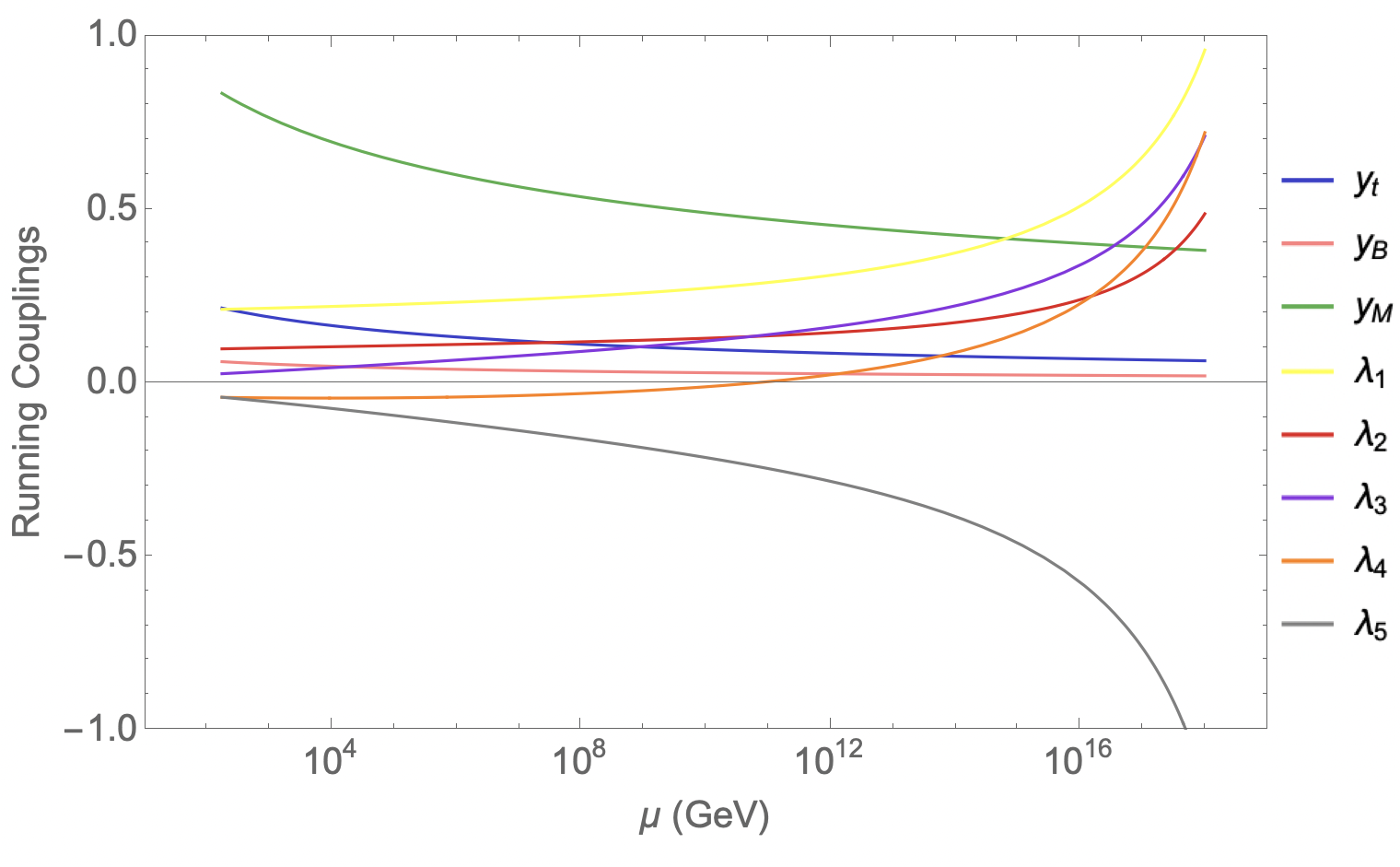}
		\caption{$\mathcal{D}_{1}$ (B) - Type II}
	\end{subfigure}	
  \caption{The RGE running of the Yukawa and scalar couplings for models with vector-like quarks. We plot results for Type-I on the left column and for Type -II on the right column. Top panel: singlet vector-like representation, ${\cal U}_1$. Bottom panel: singlet vector-like representation, ${\cal D}_1$. For singlet models, we have set $m_T=0.8$ TeV, $m_B=0.85$ TeV, $M_H=450$ GeV, $M_{H^{\pm}}=M_A=200$ GeV $\mu_0=m_t$, $\tan\beta=6$, and mixing angles  $\sin \alpha=0.1$ and $\sin \theta_L=0.08$.}
  \label{fig:rgesinglets}
\end{figure}

\begin{figure}[htbp]
	\centering
	\begin{subfigure}{.5\textwidth}\hspace{-1.5cm}
		\includegraphics[height=1.8in]{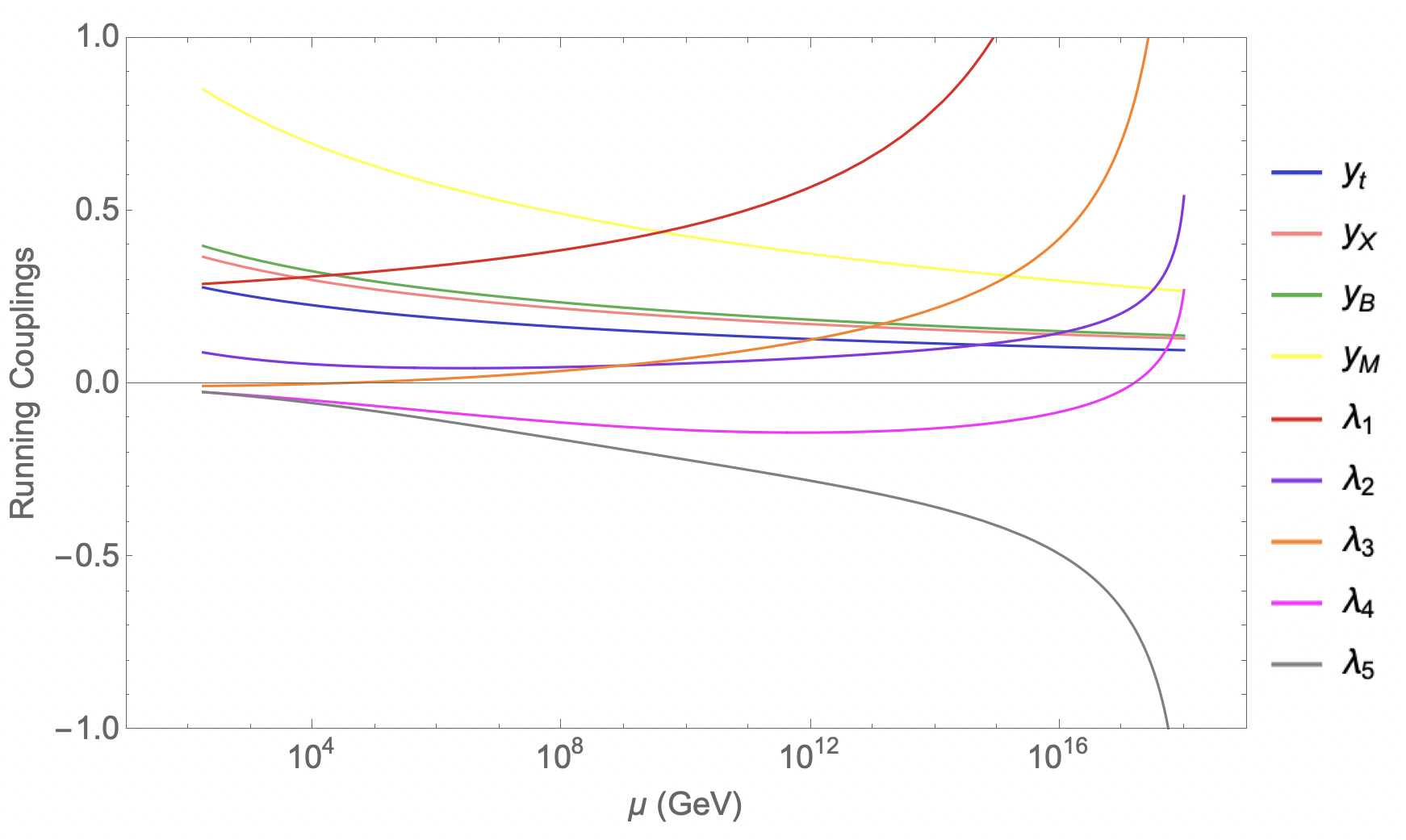}
		\caption{$\mathcal{D}_{X}$ (XT) - Type I}
	\end{subfigure}\hspace{-0.5cm}
	\begin{subfigure}{.5\textwidth}
		\includegraphics[height=1.8in]{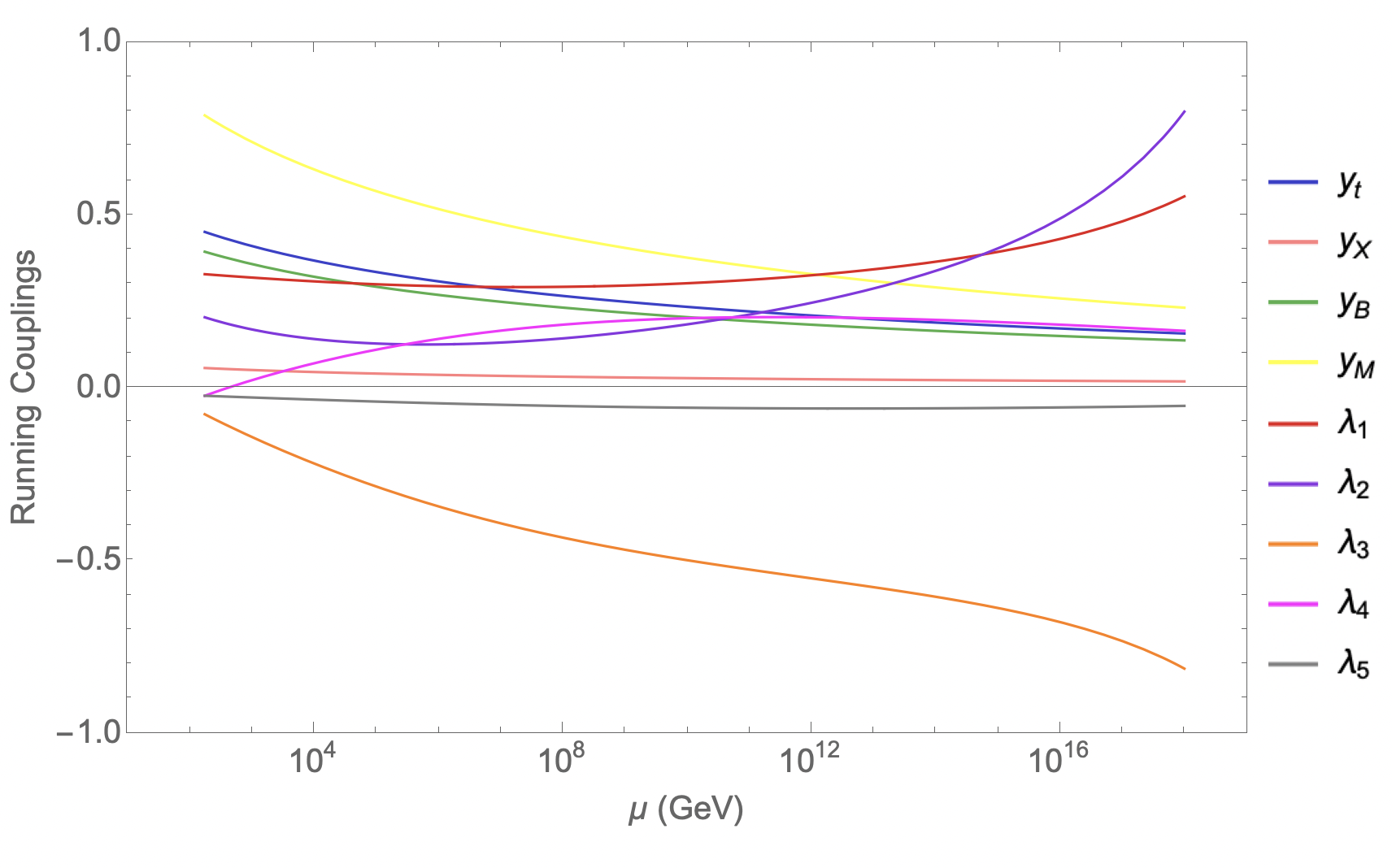}
		\caption{$\mathcal{D}_{X}$ (XT) - Type II}
	\end{subfigure}\\
\begin{subfigure}{.5\textwidth}\hspace{-1.5cm}
		\includegraphics[height=1.8in]{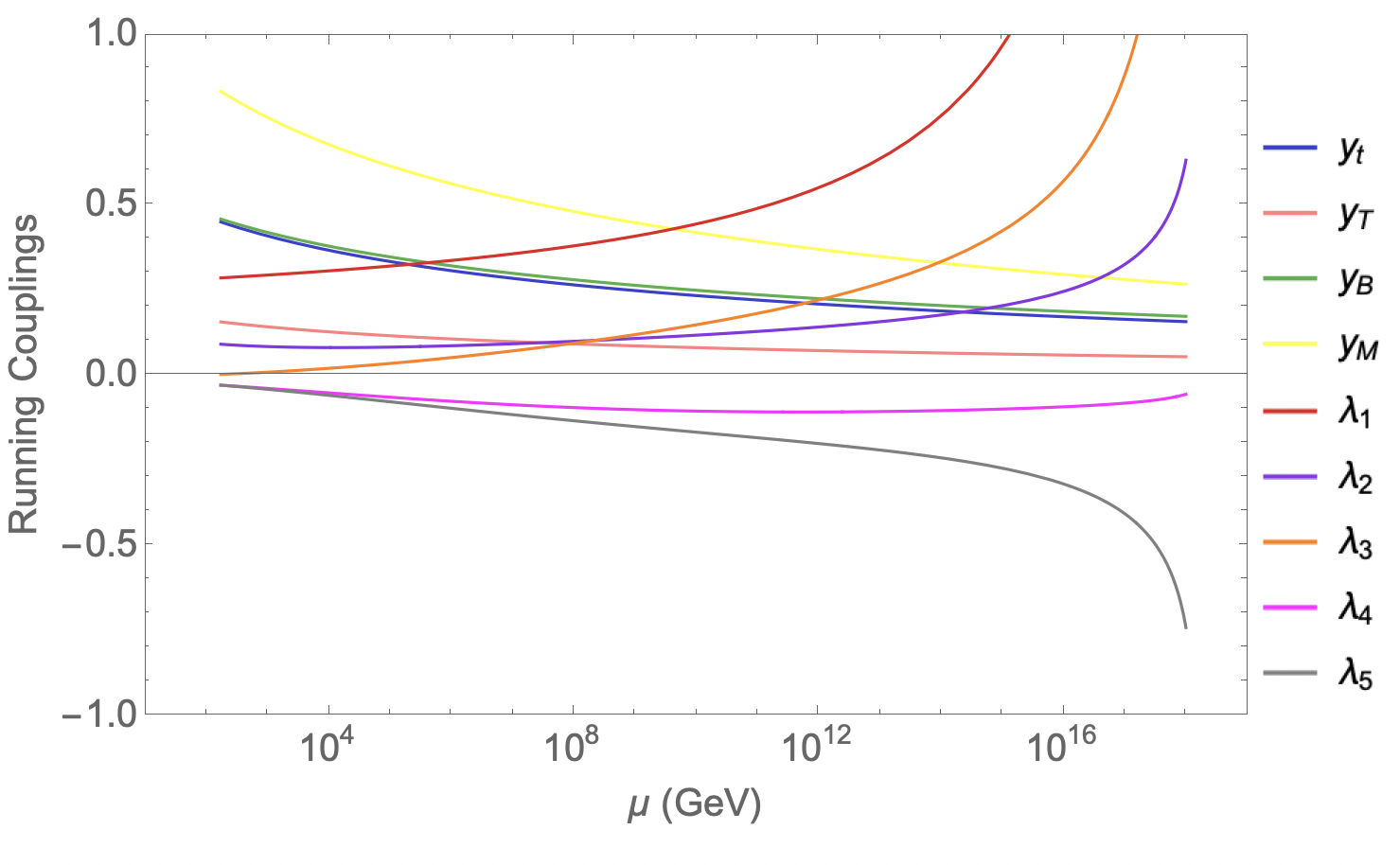}
		\caption{$\mathcal{D}_{2}$ (TB) - Type I}
	\end{subfigure}\hspace{-0.5cm}
	\begin{subfigure}{.5\textwidth}
		\includegraphics[height=1.8in]{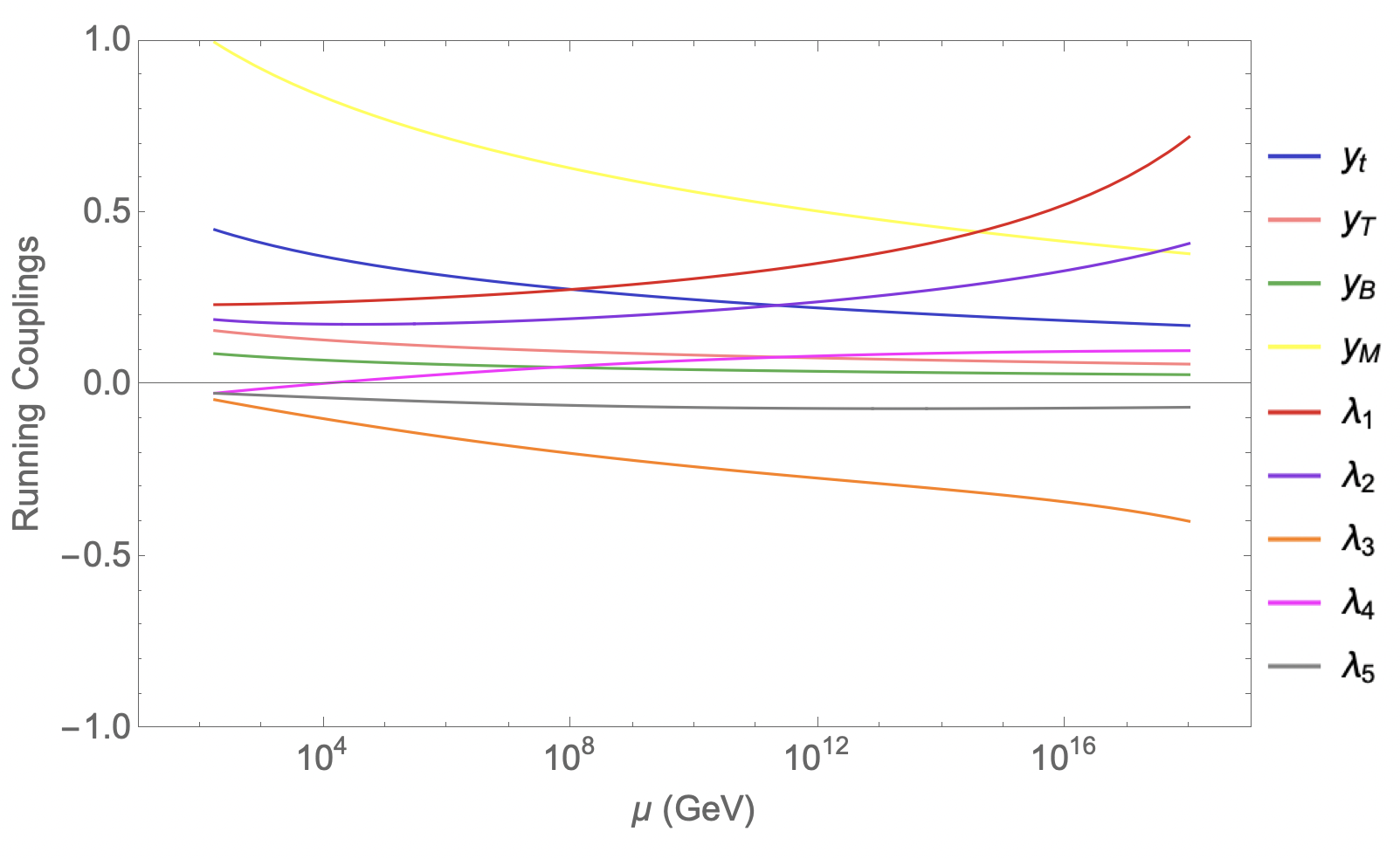}
		\caption{$\mathcal{D}_{2}$ (TB) - Type II}
	\end{subfigure}\\
\begin{subfigure}{.5\textwidth}\hspace{-1.5cm}
		\includegraphics[height=1.8in]{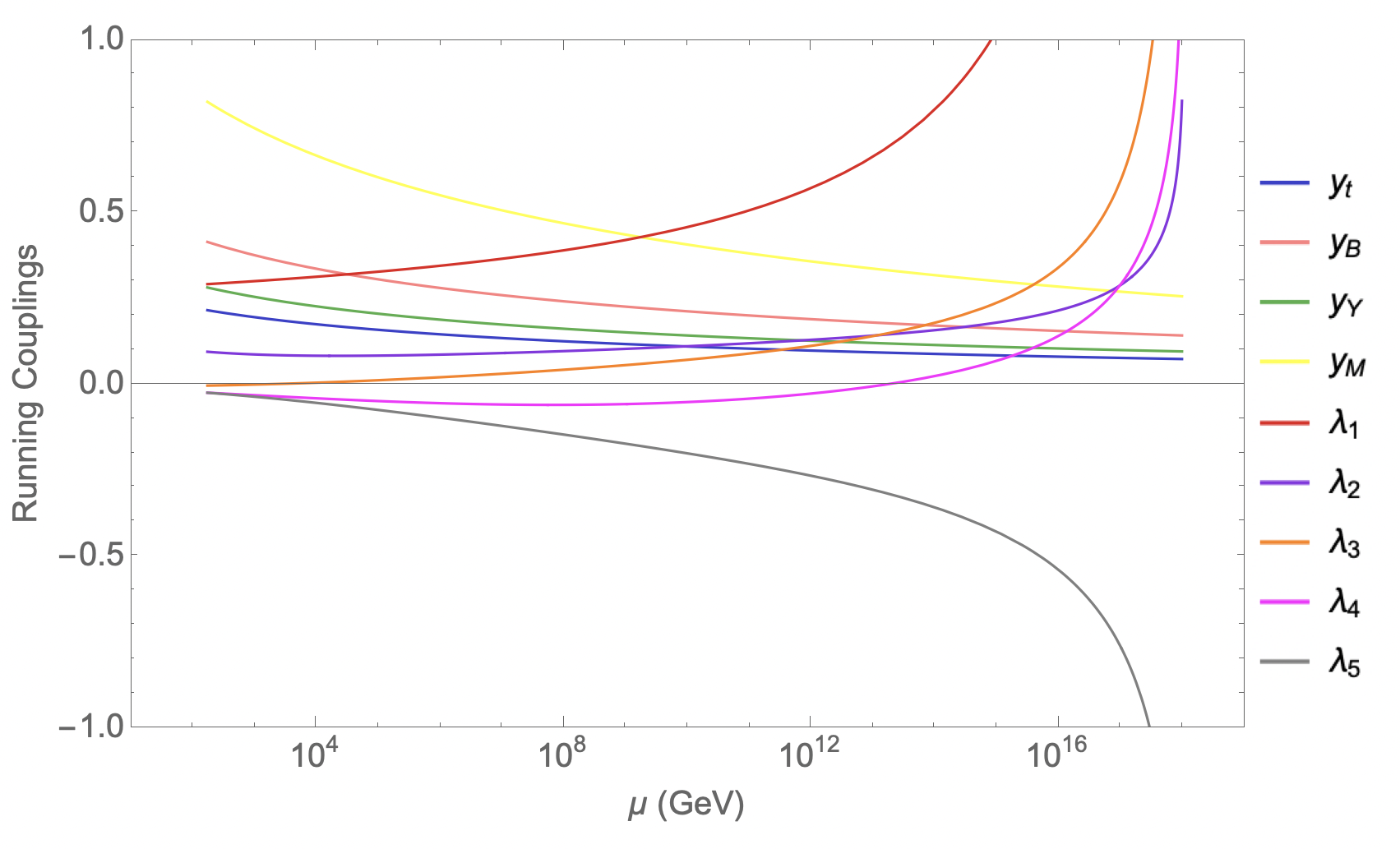}
		\caption{$\mathcal{D}_{Y}$ (BY) - Type I}
	\end{subfigure}\hspace{-0.5cm}
	\begin{subfigure}{.5\textwidth}
		\includegraphics[height=1.8in]{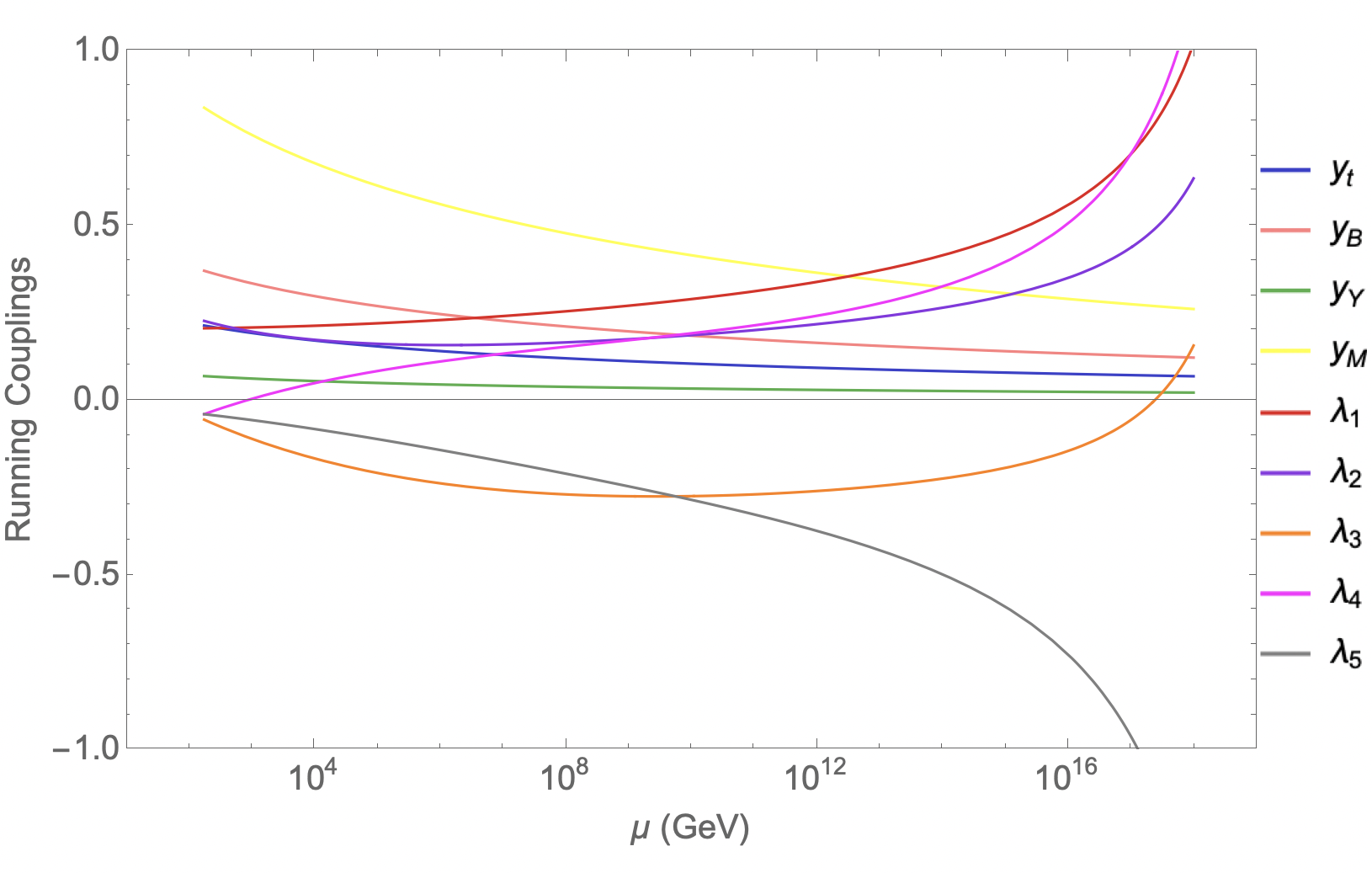}
		\caption{$\mathcal{D}_{Y}$ (BY) - Type II}
	\end{subfigure}	
	\caption{The RGE running of the Yukawa and scalar couplings for models with vector-like fermions. As before, we plot results for Type-I on the left column and for Type -II on the right column. Top panel: doublet vector-like representation, ${\cal D}_X$. Middle panel: doublet vector-like representation, ${\cal D}_2$. Bottom panel: doublet vector-like representation, ${\cal D}_Y$. For doublet models, we have set $m_T=0.85$ TeV, $m_B=1$ TeV, $m_X=1$ TeV, $m_Y=1$ TeV, $M_H=450$ GeV, $M_{H^{\pm}}=M_A=230$ GeV $\mu_0=m_t$, $\tan\beta=6$, and mixing angles  $\sin \alpha=0.1$ and $\sin \theta_L=0.08$.}
  \label{fig:rgedoublets}
\end{figure}

\begin{figure}[htbp]
	\centering
	\begin{subfigure}{.5\textwidth}\hspace{-1.5cm}
		\includegraphics[height=1.8in]{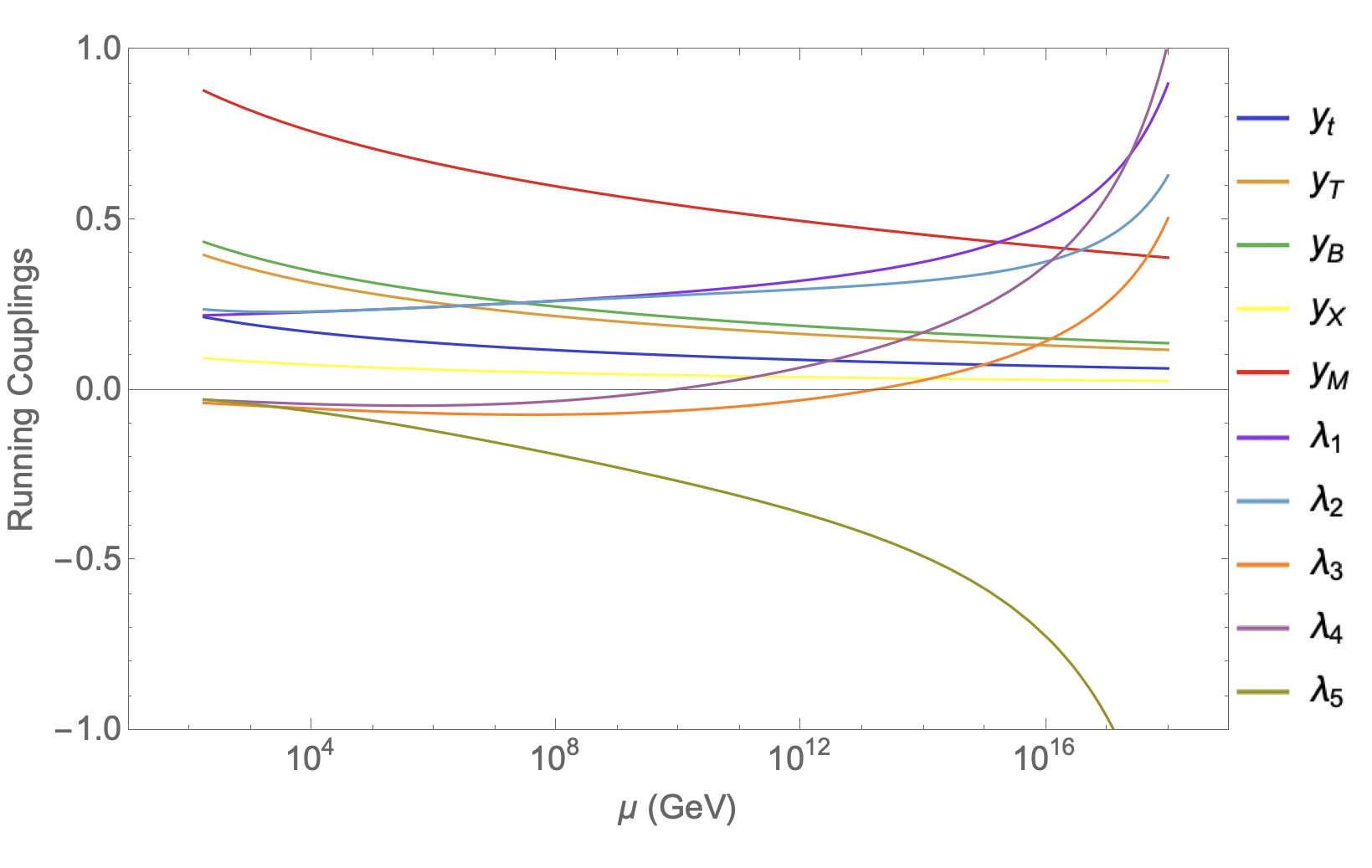}
		\caption{$\mathcal{T}_{X}$ (XTB) - Type I}
	\end{subfigure}\hspace{-0.5cm}
	\begin{subfigure}{.5\textwidth}
		\includegraphics[height=1.8in]{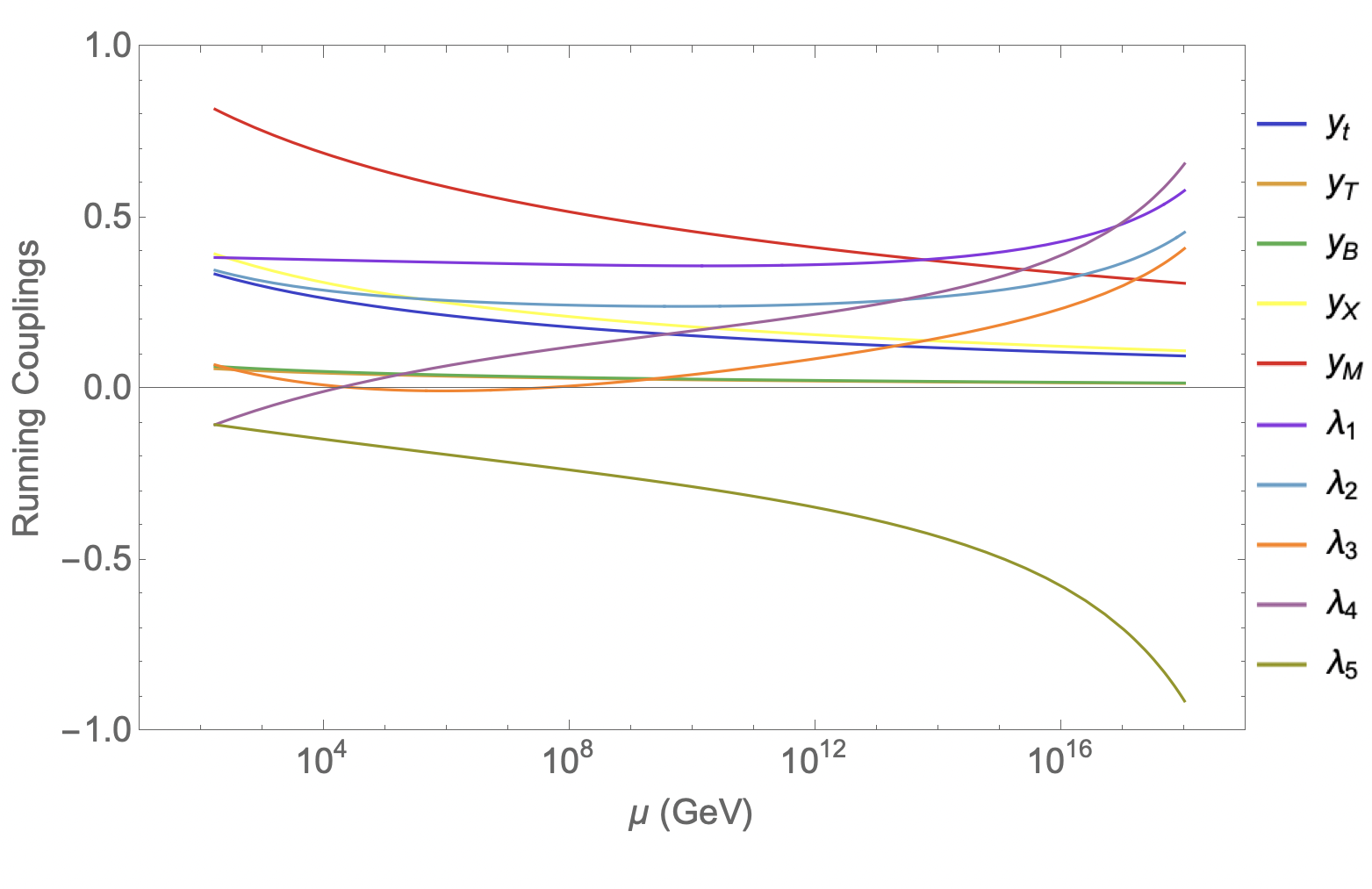}
		\caption{$\mathcal{T}_{X}$ (XTB) - Type II}
	\end{subfigure}\\
\begin{subfigure}{.5\textwidth}\hspace{-1.5cm}
		\includegraphics[height=1.8in]{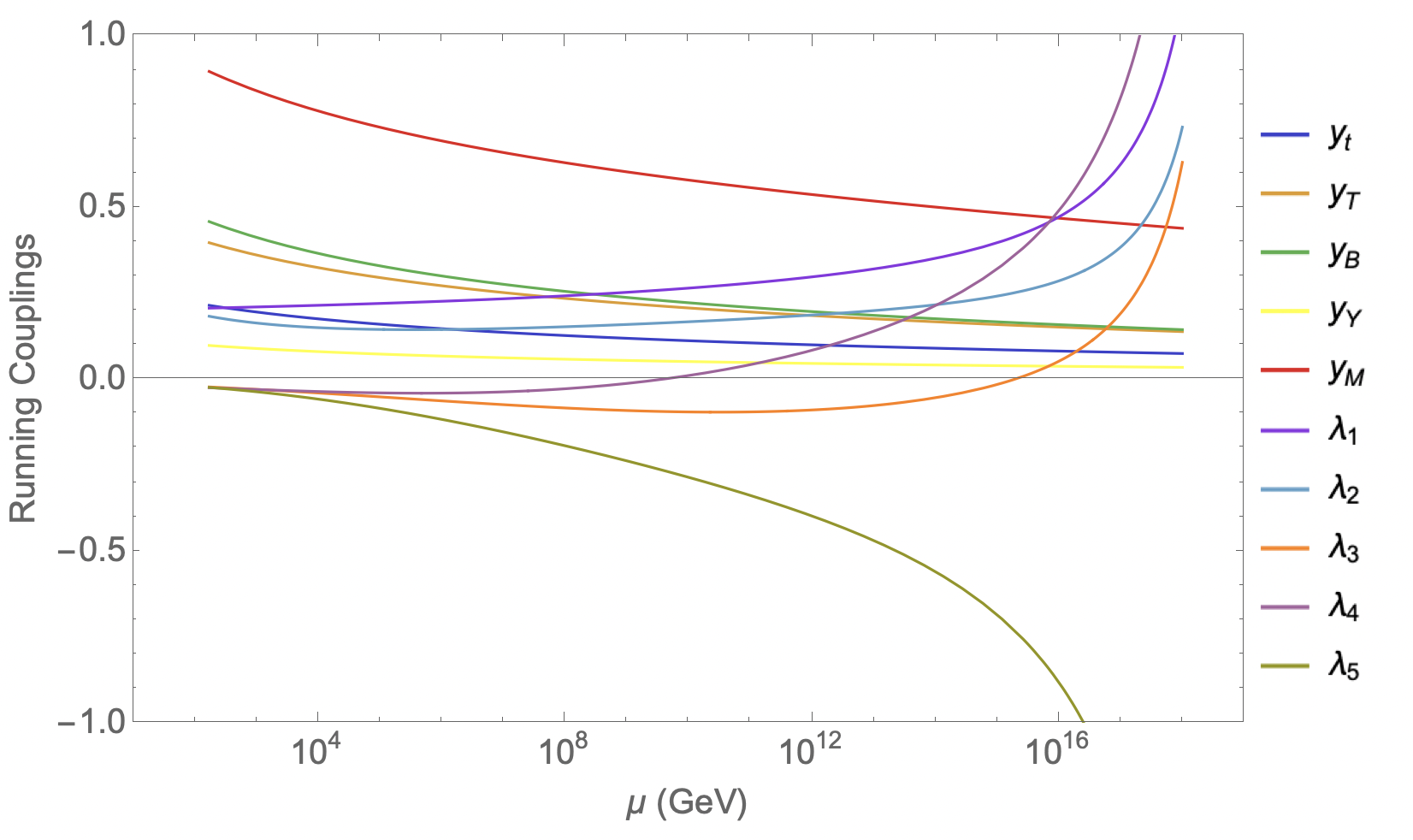}
		\caption{$\mathcal{T}_{Y}$ (TBY) - Type I}
	\end{subfigure}\hspace{-0.5cm}
	\begin{subfigure}{.5\textwidth}
		\includegraphics[height=1.8in]{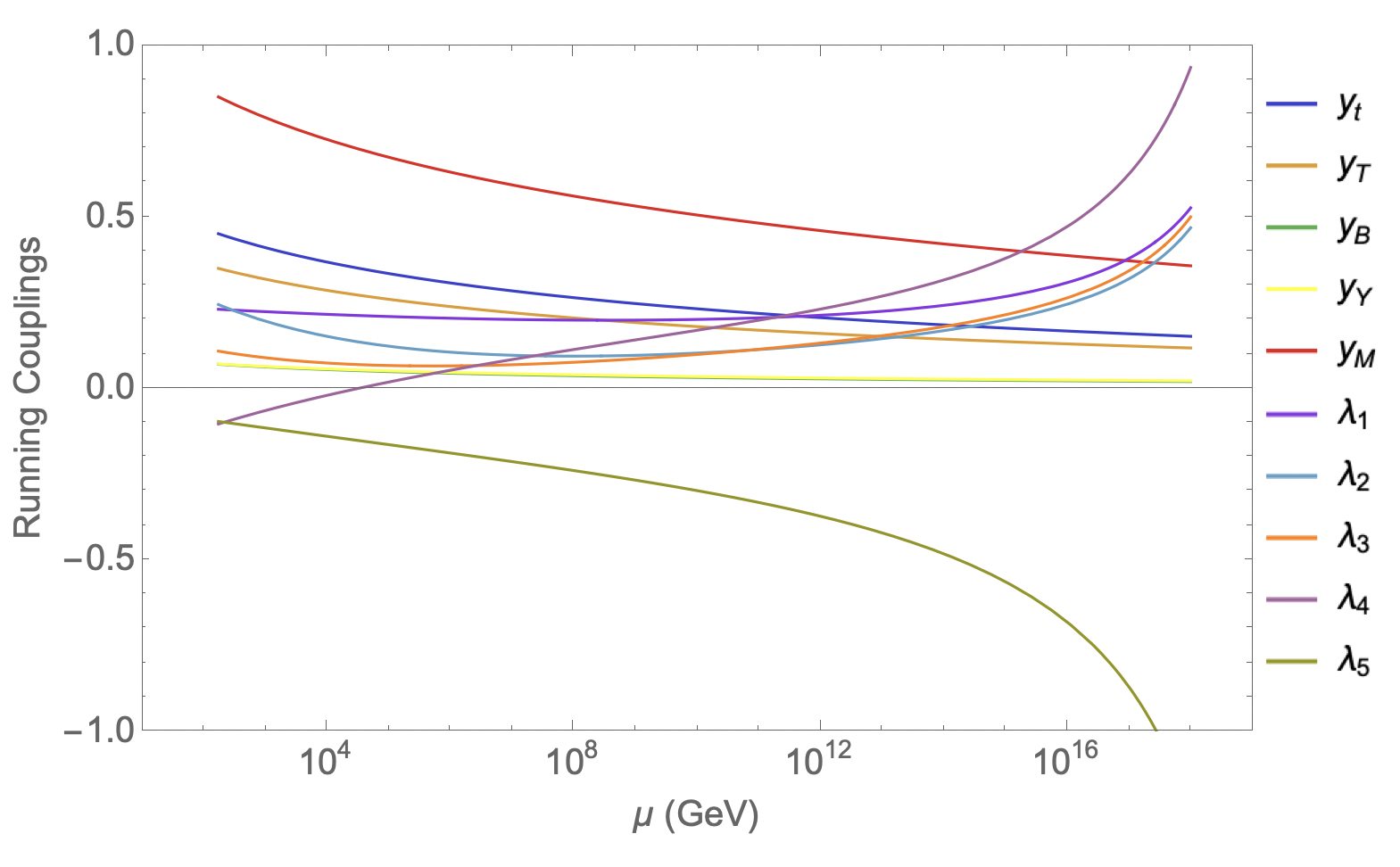}
		\caption{$\mathcal{T}_{Y}$ (TBY) - Type II}
	\end{subfigure}	
  \caption{The RGE running of the Yukawa and scalar couplings for models with vector-like fermions. As before, we plot results for Type-I on the left column and for Type -II on the right column. Top panel: triplet vector-like representation, ${\cal T}_X$. Bottom panel: triplet vector-like representation, ${\cal T}_Y$. For triplet models, we have set $m_T=0.9$ TeV, $m_B=1$ TeV, $m_X=1$ TeV, $m_Y=1$ TeV, $M_H=700$ GeV, $M_{H^{\pm}}=M_A=500$ GeV $\mu_0=m_t$, $\tan\beta=6$, and mixing angles  $\sin \alpha=0.1$ and $\sin \theta_L=0.08$. }
  \label{fig:rgetriplets}
\end{figure}

 In Fig. \ref{fig:rgesinglets}-\ref{fig:rgetriplets} we present the RGE evolution for all vector-like quark representations given in Eqs. \ref{eq:lagtype1}-\ref{eq:lagtype2}, combined with 2HDM couplings respectively, for Type-I and Type-II, in the case where   the lightest CP-even scalar is taken to be the observed 125 GeV Higgs boson. Among various bare 2HDM constraints, the limits on $M_A$ and $M_{H^{\pm}}$ are extremely sensitive  to the VEV ratios $\tan\beta$ and to experimental measurement of $M_W$. Higgs couplings \cite{ALEPH:2013htx,Basler:2017nzu,Kang:2022mdy} and $\Delta M_W$ corrections \cite{Kanemura:2019kjg} led us to take $M_A, M_{H^{\pm}}=[80,320]$ GeV in decoupling limit for 2HDM Type-I. However, we found that stabilizing the electroweak vacuum opens up more space for the upper limit of masses $M_{H^{\pm}},\, M_ A$, consistent with the alignment region ($\cos(\beta-\alpha)=0$) in the light Higgs scenario. The resulting electroweak precision observables (EWPO) in 2HDM Type-II \cite{Accomando:2019jrb} and Higgs couplings in VLQ+Type-II model \cite{Benbrik:2022kpo} are consistent with our input parameters in the scalar sector. Lastly, we choose the recent  unitarity constraints for VLQ mixing \cite{Albergaria:2023vls} at ${\cal O}$(TeV) scale. 
\\

Type-I 2HDM +singlets yield the most stringent mass limits for VLQs required to satisfy the stability bound, as expected from the insufficiency of multiplicity. While the interval $m_T=[770,850]$ GeV satisfies the stability, requirement,  $m_T>{\cal O}$(TeV) leads to negative top Yukawa coupling in sub-Planckian region. Since the RGEs of singlets in Type-I are similar, the difference in the $\lambda_2$ behaviour stems from the mass splitting between the top and bottom sector. The quartic coupling $\lambda_2$ in ${\cal D}_1$ singlets evolves to the non-perturbative region faster than for ${\cal U}_1$ singlets in Fig. \ref{fig:rgesinglets}. As discussed previously, Type-II allows for larger parameter space in the scalar sector of the combined model. In Type-I, we obtained the stability constraints on scalars for $\tan\beta=6$ as $M_h=[250,450]$ GeV and $M_{H^\pm}, \, M_A=[40,250]$ GeV, whereas Type-II dictates $M_h=[250,680]$ GeV and $M_{H^\pm},\, M_A=[80,320]$ GeV for singlets+2HDM. A unique feature of ${\cal D}_1$  is $y^{I}_t > y^{II}_t$. The overall shift stems from the fact that $m_B$ gets its mass from $v_1$, hence generating larger radiative corrections on $y_t$ and thus it shifts its overall running downward. Moreover, radiative corrections for Yukawa couplings in singlets are extremely important, because the $t$-quark cannot couple to any particle in the ${\cal D}_1$ model while the same is true for  the $b$-quark  in the ${\cal U}_1$ model.
\\

 A few comments regarding the behaviour of the couplings are in order. 
 \begin{itemize}
 \item Both $\lambda_1$ and $\lambda_2$ have to remain positive up to $\Lambda_{Pl}$ to satisfy the vacuum stability in Type-II models. 
 \item The maximum mass limit on the additional CP-even scalar $H$ can be extended further if $\tan\beta$ is increased accordingly. However, this in fact leads to a $\lambda_i$ suppression in a narrow perturbation region and flattens all scalar RGE flows for $\tan\beta>12$. Nonetheless, we note that $\tan\beta$ suppression does not break the vacuum stability. This characteristic can  be read from the denominator term through initial conditions in Eq. \ref{eq:intscalarcouplings}.  Ref. \cite{Hiller:2023bdb} discusses the details of ``squeezing'' regions of stability  for VLQ in various models. 
 \item The high energy suppression of $\lambda_2$ in Type-II models occurs because of the presence of the vector-like Yukawa terms $y_M^2$ and $y_f^2y_M^2$ appearing in $\lambda_{3,4,5}$, weighs upon the running coupling constants as $y_M$ approaches to $\Lambda_{Pl}$,  being the largest contribution among all Yukawa couplings. 
 \item Due to the splitting of Yukawa terms between $\Phi_1$ and $\Phi_2$, Type-II+singlet VLQ models are safer for vacuum stability due the fact that $\lambda_2^I$ stays closer to zero  as compared to $\lambda_2^{II}$. 
 \end{itemize}
In Fig. \ref{fig:rgedoublets}(d), the most interesting feature of ${\cal D}_2$ model with Type-II is  $\lambda_1$ remaining within perturbation scale while preserving the stability up to $\Lambda_{Pl}$ because the $B$-quark couples to $\Phi_1$ via $v_1$. So, the $B$-quark affects $\lambda_1$  to keep its evolution in perturbativity range. This is not the case for  ${\cal U}_1$ singlet model, however, it holds for multiplets  with up and down type VLQs. The same fact explains why Type-II models yield better results for the stability for $\tan\beta\sim5$. Requiring larger $v_1$ values also decreases the initial condition on $y_B$. In return, it would have a significant positive effect on $\lambda_2^{II}$ up to $\Lambda_{Pl}$ through the region where $\tan\beta\geq6$ stabilizes the vacua as well. We found that the perturbativity and the stability constraints in Type-I are, $M_H=[250,510]$ GeV, $M_{H^\pm},\, M_A=[40,260]$ GeV, while Type-II extends the parameter space to $M_H=[280,570]$ GeV and $M_{H^\pm},\, M_A=[40,330]$ GeV. 
\\

Note that, the parameter space of VLQ+2HDM that satisfies the vacuum stability constraint extends to $m_{VLQ}< {\cal O}$(TeV) and to larger mixing angle $\sin\theta_{L,R}^{t,b}$ (not shown). However, the recent experimental constraints \cite{ATLAS:2022ozf,ATLAS:2022hnn,ATLAS:2023bfh,ATLAS:2023pja} and constraints from EWPO, Fig. \ref{fig:ewpospaceVLQ}, discard large mixings and the light $m_{VLQ}$ domain. 
\\

Unlike ${\cal D}_2$ model, for doublet representations ${\cal D}_X$ and ${\cal D}_Y$, VEV ratios outside the interval $\tan\beta=[6,12]$ break perturbativity. Among all the doublet models, ${\cal D}_X$ is the most sensitive to the mass of the scalars, yielding a very narrow parameter space for the stability. For Type-I, we obtained $M_H=[375,460]$ GeV, $M_{H^\pm},\, M_A=[80,235]$ GeV, and for Type-II, $M_H=[540,680]$ GeV, $M_{H^\pm},\, M_A=[80,440]$ GeV. It should also be emphasized that $\lambda_2^{II}$ extends to larger perturbation scales in ${\cal D}_X^{II}$ and ${\cal D}_Y^{II}$ as compared to the ${\cal D}_2^{II}$ model because the mass splitting between upper components of these models is more effective than for Type-I analogues, as they separately interact with $\Phi_1$ and $\Phi_2$. Furthermore, unless VLQ mixing regime is modified, the mass limit of $m_Y=[0.92 , 1]$ TeV is quite restricted. Beyond this limit, the ${\cal D}_Y^{II}$ model ends up with negative Yukawa couplings as seen from Fig. \ref{fig:rgedoublets}(f). Because the initial conditions on $y_B$ and $y_Y$, this model becomes more effective due to radiative corrections as $Y$ and $X$ vector-like quarks do not mix with other fermions and are pure mass eigenstates\footnote{A similar coupling characteristic also occurs for $y_X$ in ${\cal D}_X^{II}$ model.}.
In contrast to the narrower stability bounds of ${\cal D}_X^{I,II}$, we extracted the stability constraints of ${\cal D}_Y^{I}$ as $M_H=[280,485]$ GeV and $M_{H^\pm},\, M_A=[40,290]$ GeV, whereas the ${\cal D}_Y^{II}$ model requires $M_H=[570,765]$ GeV and $M_{H^\pm},\, M_A=[80,280]$ GeV. Additionally, in the RGE for $\lambda_2^{II}$, perturbativity breaks down if $M_H>855$ GeV.
\\

Finally, for triplets Fig. \ref{fig:rgetriplets} we went beyond the 2HDM $\Delta M_W$ constraint \cite{Heo:2022dey} on scalar masses, which  does not assume presence of additional Yukawa couplings. It was shown previously that how a cut off scale $\Lambda$ opens a wider spectrum for $M_A$ and $M_{H^{\pm}}$ through the high-energy behaviour of the 2HDM via RGE \cite{Kang:2022mdy}. Since the Yukawa multiplicity appearing at the RGE level entails a larger mass spectrum for the CP-even neutral scalar, we checked the parameter space from different theoretical constraints, such as electroweak precision tests from Section \ref{subsec:2hdmstu} and all $B$-physics constraints \cite{Benbrik:2022kpo}. The latter reference also confirms that the $\tan\beta$ scale from $B$-physics is consistent with our analysis on the vacuum stability. 
\newpage
Except for ${\cal T}_Y^{II}$, the evolution of $\lambda_{1,2}$ is much safer compared to the rest of multiplets discussed earlier due to the fact that masses of scalars are proportional to the initial conditions in Eq. \ref{eq:intscalarcouplings}. The only exception here is the $\lambda_2^{II}$ in Fig. \ref{fig:rgetriplets}(f), which stays much closer to zero than those in all VLQ+2HDM-II representations because $B$ and $Y$ VLQs are assumed heavier than $T$ and $X$, thus generating stronger effects on $\lambda_2^{II}$. However, this critical behaviour can be ameliorated if $M_H\sim{\cal O}$(TeV), which is also allowed to leading order at $\tan\beta=6$ \cite{Xu:2017vpq}. Consequently, we extracted a less stringent parameter space for the scalar sector that satisfies the vacuum stability. For ${\cal T}_X^{I}$,  $M_H=[370,710]$ GeV, $M_{H^\pm},\, M_A=[80,525]$ GeV and for ${\cal T}_Y^{I}$, $M_H=[500,980]$ GeV, $M_{H^\pm},\, M_A=[80,623]$ GeV. On the other hand, the ${\cal T}_Y^{I}$ model allows, $M_H=[310,650]$ GeV, $M_{H^\pm},\, M_A=[80,295]$ GeV, whereas  the ${\cal T}_Y^{II}$ model significantly extends the interval to  $M_H=[520,1010]$ GeV, $M_{H^\pm},\, M_A=[80,675]$ GeV.
\\

For completeness, all the relevant RGE for the Yukawa couplings, the couplings between the bosons and coupling constants are included in the Appendix \ref{sec:appendix}.

\section{Electroweak Precision Constraints}
\label{sec:electroweakprecision}
Signals from new physics are also constrained through electroweak precision observables, which are highly correlated to large logarithms of extra masses when the scale of new model is significantly larger than electroweak scale \cite{Machacek:1983tz,Machacek:1983fi,Machacek:1984zw}. The modifications to electroweak gauge boson loops at loop level are calculated through the oblique parameters, $\mathbb{S}$,$\mathbb{T}$, and $\mathbb{U}$, defined as \cite{Peskin:1991sw} 
\begin{eqnarray}
\label{eq:ewpgeneral}
\mathbb{S}&=&16\pi \Re\left[\bar{\Pi}^{3Q}_{\gamma}(M_{Z}^{2}) -\bar{\Pi}^{33}_{Z}(0)\right] ,\nonumber \\
\mathbb{T}&=&\frac{4\sqrt{2}G_{F}}{\alpha_{e}}\Re\left[\bar{\Pi}^{3Q}(0) -\bar{\Pi}^{11}(0)\right]   ,\nonumber \\
\mathbb{U}&=&16\pi \Re \left[\bar{\Pi}^{33}_{Z}(0) -\bar{\Pi}^{11}_{W}(0)\right] \,.
\end{eqnarray}
The $\mathbb{S}$,\, $\mathbb{T}$ parameters in new physics models, such as VLQ scenarios and 2HDMs, are different from those in the SM due to extra scalars and fermions appearing in gauge boson self energies at the loop level. Additionally, the mixing between the SM fields and the new particles modifies the Higgs and electroweak couplings as well. Consequently, electroweak precision observables are universal.  
The current experimental values \cite{ParticleDataGroup:2020ssz} are obtained by fixing the differences between the new physics and the SM contributions by setting $\Delta \mathbb{U}=0$, yielding $\Delta \mathbb{T}=0.09\pm0.07$, $\Delta \mathbb{S}=0.05\pm0.08$ (and $\rho_{S,T}=0.92\pm0.11$). For the work carried here, we can split the oblique parameters calculation of $\mathbb{S, T}$ and $\mathbb{U}$ parameters via loop contributions  into two independent contributions, one  due to bosons and  the other to fermions circulating in self-energy diagrams. We  extracted gauge boson self energies using {\tt LoopTools} and {\tt FormCalc} \cite{Hahn:1999mt}, and implemented analytical expressions of Passarino-Veltman (PV) functions in {\tt FeynCalc} \cite{Shtabovenko:2020gxv} to obtain oblique parameters.

\subsection{Contributions to the $\mathbb{S}$ and $\mathbb{T}$-parameters from 2HDM}
\label{subsec:2HDMSTU}

Further expanding Eq. \ref{eq:ewpgeneral} explicitly in terms of the scalar loop contributions to the gauge boson two point functions
\label{subsec:2hdmstu}
\begin{eqnarray}
\mathbb{S}_{2HDM}&=&16\pi\Re\left[\frac{\Pi_{2HDM}^{Z\gamma}(M_Z^2)}{s_Wc_Wg_Z^2}+\frac{\Pi_{2HDM}^{\gamma\gamma}(M_Z^2)}{c_W^2g_Z^2}-\frac{\Pi_{2HDM}^{ZZ}(M_Z^2)-\Pi_{2HDM}^{ZZ}(0)}{g_Z^2}-\frac{2s_W\Pi_{2HDM}^{Z\gamma}(0)}{c_Wg_Z^2} \right] , \nonumber \\
\mathbb{T}_{2HDM}&=&\frac{4\sqrt{2}G_{F}}{\alpha_{e}}\Re\left[ \frac{\Pi_{2HDM}^{ZZ}(0)}{g_Z^2}+\frac{2s_W\Pi_{2HDM}^{Z\gamma}(0)}{c_Wg_Z^2}-\frac{\Pi_{2HDM}^{WW}(0)}{c_W^2g_Z^2}\right]
\end{eqnarray}
The coupling factors are $g_Z=g/c_W$ and the photon two-point function in the 2HDM is
\begin{eqnarray}
\Pi_{2HDM}^{\gamma\gamma}(p^2)&=&e^2B_5(p^2,M^2_{H^{\pm}},M^2_{H^{\pm}})-e^2p^2[5B_0(p^2,M_W^2,M_W^2)+12B_3(p^2,M_W^2,M_W^2)+\frac{2}{3}],
\end{eqnarray}
The photon-$Z$ mixing is given by:
\begin{eqnarray}
\label{eq:Zgamma2pfunc}
\Pi_{2HDM}^{Z\gamma}(p^2)&=& \frac{eg_Z}{2}B_5(p^2,M^2_{H^{\pm}},M^2_{H^{\pm}})-eg_Zp^2\left(\frac{11}{2}B_0(p^2,M_W^2,M_W^2)+10B_3(p^2,M_W^2,M_W^2)+\frac{2}{3}\right)\,  \nonumber \\
&-&\frac{s_W}{c_W} \left[e^2B_5(p^2,M^2_{H^{\pm}},M^2_{H^{\pm}})-e^2p^2[5B_0(p^2,M_W^2,M_W^2)+12B_3(p^2,M_W^2,M_W^2)+\frac{2}{3}] \right],
\end{eqnarray}
The $Z$-boson two-point function in the 2HDM is
\begin{eqnarray}
\label{eq:ZZ2pfunc}
\Pi_{2HDM}^{ZZ}(p^2)&=&g_Z^2\left[ \frac{s^2_{\beta-\alpha}}{4}B_5(p^2,M_H^2,M_A^2)+\frac{c^2_{\beta-\alpha}}{4}B_5(p^2,M_h^2,M_A^2)+\frac{1}{4}B_5(p^2,M^2_{H^{\pm}},M^2_{H^{\pm}}) -\frac{2p^2}{3}\right. \nonumber \\
&+&\left. s^2_{\beta-\alpha}[M_Z^2B_0(p^2,M_h^2,M_Z^2)+\frac{1}{4}B_5(p^2,M_h^2,M_Z^2)]+c^2_{\beta-\alpha}[M_Z^2B_0(p^2,M_H^2,M_Z^2)\right. \nonumber \\
&+&\left. \frac{1}{4}B_5(p^2,M_H^2,M_Z^2)] +2M_W^2B_0(p^2,M_W^2,M_W^2)-\frac{23p^2}{4}B_0(p^2,M_W^2,M_W^2)-9p^2B_3(p^2,M_W^2,M_W^2)\right]\, \nonumber \\
&-& \frac{s_W^2}{c_W^2}\left(e^2B_5(p^2,M^2_{H^{\pm}},M^2_{H^{\pm}})-e^2p^2[5B_0(p^2,M_W^2,M_W^2)+12B_3(p^2,M_W^2,M_W^2)+\frac{2}{3}]\right)\, \nonumber \\
&-& \frac{2s_W}{c_W}\left[\frac{eg_Z}{2}B_5(p^2,M^2_{H^{\pm}},M^2_{H^{\pm}})-eg_Zp^2\left(\frac{11}{2}B_0(p^2,M_W^2,M_W^2)+10B_3(p^2,M_W^2,M_W^2)+\frac{2}{3}\right)\right. \nonumber \\
&-&\left. \frac{s_W}{c_W} \left(e^2B_5(p^2,M^2_{H^{\pm}},M^2_{H^{\pm}})-e^2p^2[5B_0(p^2,M_W^2,M_W^2)+12B_3(p^2,M_W^2,M_W^2)+\frac{2}{3}]\right) \right]\,.
\end{eqnarray}
The $W$-boson two-point function in the 2HDM follows as
\begin{eqnarray}
\label{eq:WW2pfunc}
\Pi_{2HDM}^{WW}(p^2)&=&g^2\left[\frac{1}{4}B_5(p^2,M^2_A,M^2_{H^{\pm}})+\frac{s^2_{\beta-\alpha}}{4}B_5(p^2,M_H^2,M^2_{H^{\pm}})+\frac{c^2_{\beta-\alpha}}{4}B_5(p^2,M_h^2,M^2_{H^{\pm}})-\frac{2p^2}{3}\right. \nonumber \\
&+&\left.s^2_{\beta-\alpha}[M_W^2B_0(p^2,M_h^2,M_W^2)+\frac{1}{4}B_5(p^2,M_h^2,M_W^2)]-8p^2c_W^2B_0(p^2,M_Z^2,M_W^2) \right. \nonumber \\
&+&\left.c^2_{\beta-\alpha}[M_W^2B_0(p^2,M_H^2,M_W^2)+\frac{1}{4}B_5(p^2,M_H^2,M_W^2)]+(\frac{1}{4}+2c_W^2)B_5(p^2,M_Z^2,M_W^2)\right. \nonumber \\
&+&\left. M_W^2(1-4s_W^2)B_0(p^2,M_Z^2,M_W^2)+M_Z^2B_0(p^2,M_Z^2,M_W^2)+2s_W^2B_5(p^2,0,M_W^2) \right. \nonumber \\
&+&\left. 2M_W^2B_0(p^2,0,M_W^2)-4p^2B_0(p^2,0,M_W^2) \right]
\end{eqnarray}
The Passarino-Veltman functions and relevant identities are given in Appendix \ref{sec:PVinteg}. Subtracting the SM contributions from $\mathbb{S}$ and $\mathbb{T}$ parameters of the 2HDM yields the new physics contributions to oblique parameters:
\begin{eqnarray}
\label{eq:2HDMobliqueT}
\Delta\mathbb{T}_{2HDM}&=& \frac{1}{4 \pi M_Z^2c_W^2s_W^2}\left[ B_{00}(0,M_A^2,M^2_{H^{\pm}})+B_{00}(0,M_Z^2,M_W^2)-B_{00}(0,M_W^2,M_h^2)-B_{00}(0,M_W^2,M_Z^2)\right. \nonumber \\
&+&\left. (4s_W^4-1)B_{00}(0,M^2_{H^{\pm}},M^2_{H^{\pm}})-2M^2_{H^{\pm}}s_W^4[1+B_{0}(0,M^2_{H^{\pm}},M^2_{H^{\pm}}] +s^2_{\beta-\alpha}[B_{00}(0,M^2_H,M^2_{H^{\pm}})\right. \nonumber \\
&+&\left. B_{00}(0,M^2_h,M^2_W)-B_{00}(0,M^2_H,M^2_A)]+c^2_{\beta-\alpha}[M_W^2B_{0}(0,M^2_h,M^2_W)+B_{00}(0,M^2_h,M^2_Z)\right. \nonumber \\
&-&\left.M_Z^2B_{0}(0,M^2_h,M^2_Z)+B_{00}(0,M^2_h,M^2_{H^{\pm}})-B_{00}(0,M^2_h,M^2_A )+B_{00}(0,M^2_H,M^2_W)\right. \nonumber \\
&-&\left.M_W^2B_{0}(0,M^2_H,M^2_W)+M_Z^2B_{0}(0,M^2_H,M^2_Z)-B_{00}(0,M^2_H,M^2_Z)]\right]
\end{eqnarray}

\begin{eqnarray}
\label{eq:2HDMobliqueS}
\Delta\mathbb{S}_{2HDM}&=& \frac{1}{\pi M_Z^2}\left[2s_W^2c_W^2A_0(M^2_{H^{\pm}})-B_{00}(M_Z^2,M^2_{H^{\pm}},M^2_{H^{\pm}})+(c_W^2-s_W^2)^2B_{00}(0,M^2_{H^{\pm}},M^2_{H^{\pm}}) \right. \nonumber \\
&+&\left.s^2_{\beta-\alpha}[B_{00}(M_Z^2,M^2_H,M^2_A)-B_{00}(0,M^2_H,M^2_A)] +c^2_{\beta-\alpha}[B_{00}(M_Z^2,M^2_H,M^2_A)\right. \nonumber \\
&+&\left.B_{00}(0,M^2_h,M^2_Z)+B_{00}(M_Z^2,M^2_h,M^2_Z)-B_{00}(M_Z^2,M^2_H,M^2_Z)-B_{00}(0,M^2_H,M^2_A)\right. \nonumber \\
&-&\left.B_{00}(0,M^2_H,M^2_Z)+M_Z^2B_{0}(M_Z^2,M^2_h,M^2_Z)+M_Z^2B_{0}(0,M^2_H,M^2_Z)\right. \nonumber \\
&-&\left.M_Z^2B_{0}(0,M^2_h,M^2_Z)-M_Z^2B_{0}(M_Z^2,M^2_H,M^2_Z) ]
\right]
\end{eqnarray}
\begin{figure}[htbp]
	\centering
	\begin{subfigure}{.5\textwidth}\hspace{-1.5cm}
		\includegraphics[height=2.5in]{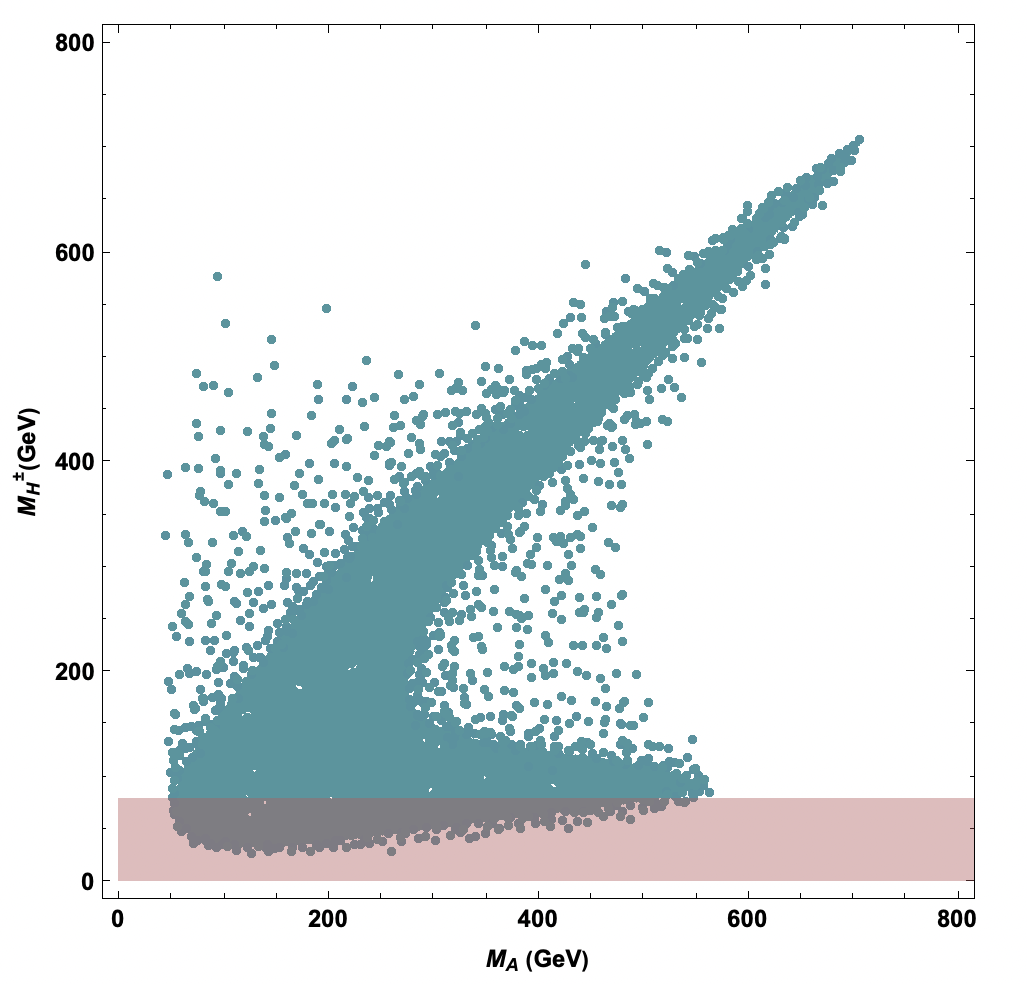}
		\caption{}
	\end{subfigure}\hspace{-1.4cm}
	\begin{subfigure}{.5\textwidth}
		\includegraphics[height=2.5in]{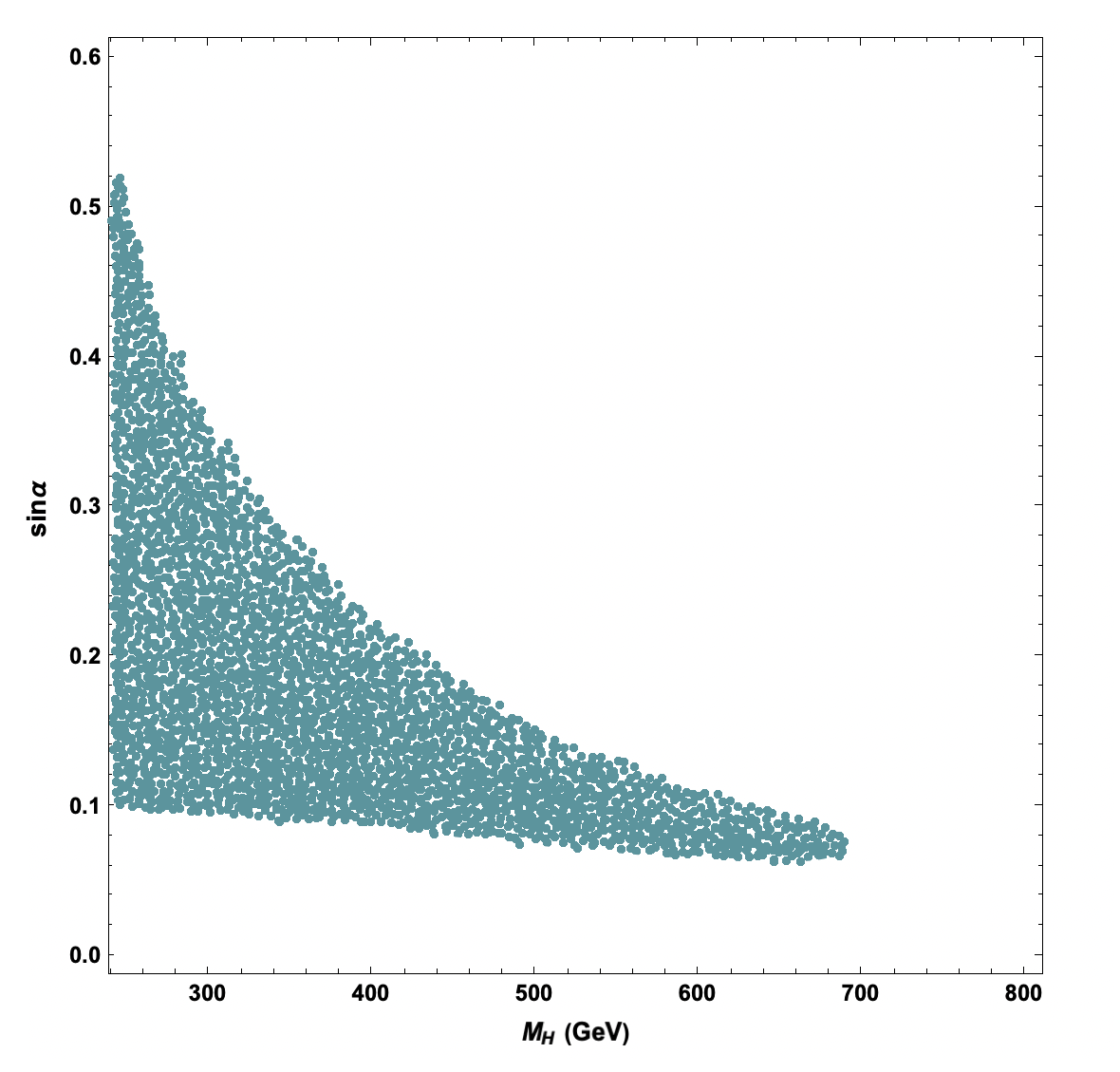}
		\caption{}
	\end{subfigure}
  \caption{The allowed mass regions from EWPO for the pseudoscalar boson mass $M_A$ and charged Higgs mass $M_{H^{\pm}}$  (left panel). The allowed parameter space from EWPO for scalar mass $M_H$ and scalar mixing angle with the SM Higgs $\sin\alpha$  (right panel) in 2HDM. The $\sin \alpha=0$ limit of CP-even scalars  mixing is allowed by EWPO but excluded due to the vacuum stability constraint. We have set $\tan\beta=6$.}
  \label{fig:ewpspacescalars}
\end{figure}
\begin{figure}
	\begin{center}
	\includegraphics[width=3.0in]{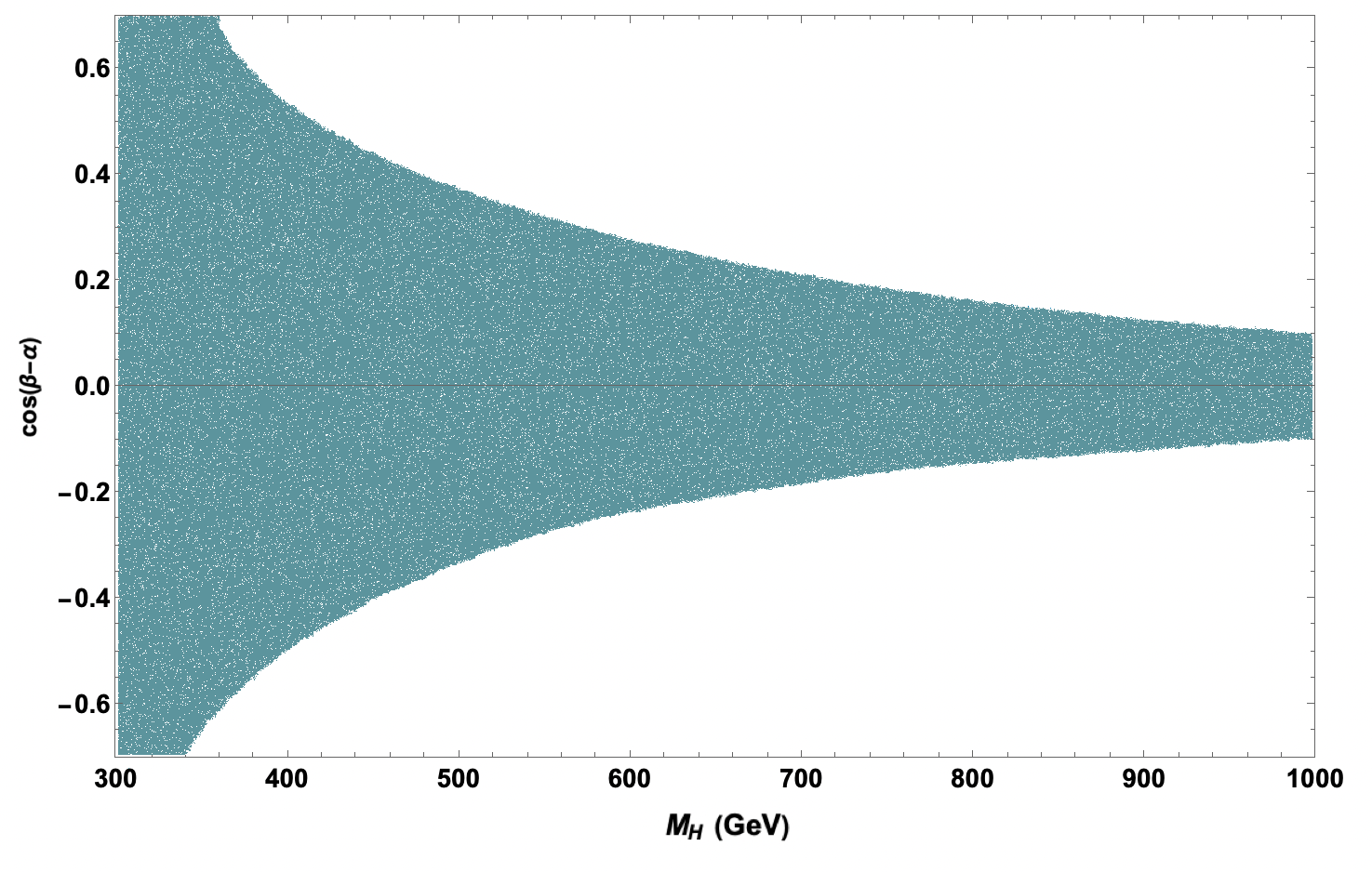}
	\end{center}
	\caption{The allowed mass regions extracted from EWPO for $M_H$ versus $\cos(\beta-\alpha)$ mixing between CP-even scalars in 2HDM.}
	\label{fig:ewpspacebeta}
\end{figure}
In Fig. \ref{fig:ewpspacescalars} (left panel), the correlation between $M_A$ and $M_{H^{\pm}}$ due to EWPO does not constrain the masses in a stringent way, though when $M_A>550$ GeV, the correlation becomes significantly important. The red region has already been discarded by direct searches at LEP \cite{ALEPH:2013htx}. Considering the imposing theoretical bounds only, the findings from EWPO are consistent with the unitary bounds in $M_A-M_{H^{\pm}}$ plane \cite{Horejsi:2005da}. Note that, $\tan\beta$ dependence of the oblique parameters alone is more relaxed, allowing a wide mass spectrum. This is due to fact that the mixing between CP-even scalars can be shifted away from the $\sin \alpha=0$ (decoupling) limit, hence the variation in $\tan\beta$  compensates for the Higgs data requirement of the near-alignment limit, $\cos (\beta-\alpha)\approx 0$.   Consequently, imposing the alignment limit on the mass spectrum of scalars is by choice (to fit the Higgs data) rather than requirement of the theory when RG running $\mu<$1 TeV. This consequence is highlighted particularly for Type-I with various $\tan\beta$ values \cite{Kang:2022mdy}. On the other hand, as seen from  Fig. \ref{fig:ewpspacescalars} (right panel), the limit is stronger in the $M_H-\sin\alpha$ plane for fixed value of $\tan\beta$ in both types of 2HDMs. It is seen that for $\tan\beta=6$, EWPO constrain the masses in a way that the decoupling limit of CP-even scalars occurs in a natural way at a scale $\sim{\cal O}$ (TeV). Although the $\sin \alpha=0$ limit is not forbidden by EWPO, we combined it with the minimum stability requirement on $\sin\alpha$ near the decoupling limit, which in fact is in agreement with the minimum of $M_H\sim250$ GeV for the stability analysis in both Type-I and Type-II 2HDMs. Moreover, the constraints on $M_H$  obtained from EWPO and from the vacuum stability match with the constraint for signal rates of $H\to WW^{\star}\to e\nu\mu\nu$ \cite{ATLAS:2023hyd,Kanemura:2015mxa}. Furthermore, we excluded the $\sin \alpha=0$ region because a nonzero mixing between CP-even scalars ($\sin\alpha\neq0$) is required to preserve the vacuum stability all the time. As keeping $\cos (\beta-\alpha)$ closer to zero is motivated by the alignment limit from the Higgs data \cite{Accomando:2019jrb}, we impose this along with the requirement that the  couplings evolved with the RGEs remain away from the vacuum instability.  Hence, using the mass spectrum allowed from EWPO constraints fit with the stability analysis. It is important that the theoretical constraints align with each other as the limits rising from vacuum stability become stronger at the scale $[10^3,10^{10}]$ GeV \cite{Hiller:2023bdb}, thus  restricting  $M_A, M_{H^{\pm}}>500$ GeV,  limit  also favoured by  collider bounds \cite{Basler:2017nzu}.
\\

In Fig. \ref{fig:ewpspacebeta}, we have considered the energy scale  $\mu= [800-1000]$ GeV in self energy diagrams, and scan over values up to 1 TeV with respect to the oblique parameters. The upper limit of $M_H$ is chosen to be in a good agreement with the limits from vacuum stability on scalars+VLQs at $\tan\beta=[5,6]$. The current experimental data constraining $h$  to having SM-like Higgs behaviour  restrict values of $\cos(\beta-\alpha)$ much closer to the decoupling limit. Consequently, the electroweak vacuum stability requirements and EWPO impose naturally occurring near decoupling limit when $M_A, M_H>600$ GeV. We should also note that, Type-I and Type-II dependent effects are highly manifestable through Higgs channels, for which the signal strengths $\kappa_{hbb},\kappa_{hcc}$ also favour regions slightly beyond the decoupling limit, particularly for $\tan\beta\sim[2 , 12]$ \cite{Accomando:2019jrb}. The contribution to $\mathbb{T}_{2HDM}$  is twofold, depending on the mass parameter space of scalars and on $\sin(\beta-\alpha)$ whereas a negative contribution to  $\mathbb{T}_{2HDM}$ can always be generated by varying $M_{H^{\pm}}$. For general scale of $\sin(\beta-\alpha)$, $M_h$ and $M_H$ splitting has to be small for pushing $\mathbb{T}_{2HDM}$ to be large and negative values. Hence, negative corrections to $\mathbb{T}$ in 2HDM can render overall positive corrections rising from various fermion representations and  further enhancing limitations on additional scalars and mixing among Higgs bosons. 
\newpage
\subsection{VLQ contributions to the $\mathbb{S}$ and $\mathbb{T}$ parameters}
\label{subsec:VLQSTU}
The contributions of VLQs  to $\mathbb{S}$ and $\mathbb{T}$ parameters are different for each representation (singlets, doublets or triplets)  in the current framework. Since the electroweak Lagrangian is constructed with gauge eigenstate fields, any mixing of fermions with extra anomaly-free fields alters the structure of the bare electroweak Lagrangian, as seen from Eq. \ref{eq:ewlag}. As we have already seen in Section \ref{subsec:VLQtheory}, the mixing regime is model dependent.  Ref. \cite{He:2022zjz} highlighted the emergence of disagreement of the oblique parameters for triplets in \cite{Lavoura:1992np}, where the external momenta of gauge bosons are omitted in self-energy diagrams $\Pi_{VV}$. This leads to a discrepancy in the $\mathbb{S}$ parameter, which becomes positive in triplet representations in the large logarithm of $m_T\sim {\cal O}$(TeV), as in Ref. \cite{Chen:2017hak}. Following the corrections carried in Ref. \cite{He:2022zjz}, we obtained better approximations to $\Delta\mathbb{S}_{{\cal T}_X,{\cal T}_Y}$ and $\Delta\mathbb{T}_{{\cal T}_X,{\cal T}_Y}$. Consequently, in to our calculations  $\Delta\mathbb{S}<0$ and $\Delta\mathbb{T}>0$, and we found agreement with the results in \cite{Cao:2022mif}. As we mentioned in Section \ref{subsec:2hdmstu}, the self energies of gauge bosons are extracted so that UV divergences are properly cancelled.  Here we present fthe contributions of VLQs to the oblique parameters in terms of PV functions,  more complete expressions being available in the Appendix \ref{sec:appendix}. 
\\\\
The couplings to $W$-boson and $Z$-boson are been modified by the VLQs through their mixing with SM quarks
\begin{eqnarray}
\label{eq:ewlag}
{\cal L}_{W}&=&\frac{g}{\sqrt{2}}\bar{Q_i}\gamma^{\mu}(C^L_{Q_{i}Q_{j}}P_L+C^R_{Q_{i}Q_{j}}P_R)Q_jW_{\mu}^{+}+h.c.\,,\nonumber \\
{\cal L}_{Z}&=&\frac{g}{2c_W}\bar{Q_i}\gamma^{\mu}(N^L_{Q_{i}Q_{j}}P_L+N^R_{Q_{i}Q_{j}}P_R-2\delta_{ij}\mathbb{Q}s_W^2)Q_jZ^{\mu}\, ,
\end{eqnarray}
where $Q_{i,j}$ are any type of quarks in our convention of electroweak Lagrangian. The condition $|\mathbb{Q}_i-\mathbb{Q}_j|=1$ holds for all form of  $W-Q_{i}-Q_{j}$ interactions. 
\\\\
Further compression of the modified electroweak couplings take the following forms:
\begin{eqnarray}
{\cal L}_{W}&\supset&\gamma^{\mu}(\Omega^L_{WQ_{i}Q_{j}}\mathbb{L}+\Omega^R_{WQ_{i}Q_{j}}\mathbb{R})W_{\mu}^{+}\,,\nonumber \\
{\cal L}_{Z}&\supset&\gamma^{\mu}(\Omega^L_{ZQ_{i}Q_{j}}\mathbb{L}+\Omega^R_{ZQ_{i}Q_{j}}\mathbb{R})Z_{\mu}\,.
\end{eqnarray}
For the cases $i=j$ of $Z-Q_{i}-Q_{j}$ interactions, the last term of Eq. \ref{eq:ewlag} is  absorbed in $\Omega^{L,R}_{ZQ_{i}Q_{j}}$ throughout all VLQ representations.
\begin{eqnarray}
\label{eq:obliqueTvlq}
\mathbb{T}_{VLQ}&=&\frac{1}{\alpha_e}\Re\left[\frac{2s_W}{c_WM_Z^2}\sum_{i}\mathcal{F}_{Z\gamma}(\Omega^L_{ZQ_{i}Q_{i}},\Omega^R_{ZQ_{i}Q_{i}},\mathbb{Q}_i,m_i^2,p^2=0)\right. \nonumber \\
&+&\left.\frac{-2}{M_Z^2}\sum_{i\neq j}\delta(\mathbb{Q}_i-\mathbb{Q}_j)\mathcal{F}_{ZZ}(\Omega^L_{ZQ_{i}Q_{j}},\Omega^R_{ZQ_{i}Q_{j}},m_i^2,m_j^2,p^2=0)\right. \nonumber \\
&+&\left.\frac{1}{M_Z^2}\sum_{i}\mathcal{F}_{ZZ}(\Omega^L_{ZQ_{i}Q_{i}},\Omega^R_{ZQ_{i}Q_{i}},m_i^2,m_i^2,p^2=0)\right. \nonumber \\
&+&\left.\frac{1}{M_W^2}\sum_{i\neq j}\delta(\mathbb{Q}_i-\mathbb{Q}_j)\mathcal{F}_{WW}(\Omega^L_{WQ_{i}Q_{j}},\Omega^R_{WQ_{i}Q_{j}},m_i^2,m_j^2,p^2=0) \right]
\end{eqnarray}
\begin{eqnarray}
\label{eq:obliqueSvlq}
\mathbb{S}_{VLQ}&=&\frac{4s_W^2c_W^2}{\alpha_e}\Re\left[(\frac{c_W^2-s_W^2}{s_Wc_WM_Z^2})\left(\sum_{i}\mathcal{F}_{Z\gamma}(\Omega^L_{ZQ_{i}Q_{i}},\Omega^R_{ZQ_{i}Q_{i}},\mathbb{Q}_i,m_i^2,M_Z^2)+\sum_{i}\mathcal{F}_{Z\gamma}(\Omega^L_{ZQ_{i}Q_{i}},\Omega^R_{ZQ_{i}Q_{i}},\mathbb{Q}_i,m_i^2,0)\right)\right.\nonumber \\
&-&\left.\frac{1}{M_Z^2}\sum_{i}\mathcal{F}_{\gamma\gamma}(\mathbb{Q}_i,\mathbb{Q}_i,m_i^2,m_i^2,M_Z^2)+\frac{2}{M_Z^2}\sum_{i\neq j}\delta(\mathbb{Q}_i-\mathbb{Q}_j)\mathcal{F}_{ZZ}(\Omega^L_{ZQ_{i}Q_{j}},\Omega^R_{ZQ_{i}Q_{j}},m_i^2,m_j^2,M_Z^2)\right. \nonumber \\
&-&\left.\frac{2}{M_Z^2}\sum_{i\neq j}\delta(\mathbb{Q}_i-\mathbb{Q}_j)\mathcal{F}_{ZZ}(\Omega^L_{ZQ_{i}Q_{j}},\Omega^R_{ZQ_{i}Q_{j}},m_i^2,m_j^2,0)+\frac{1}{M_Z^2}\sum_{i}\mathcal{F}_{ZZ}(\Omega^L_{ZQ_{i}Q_{i}},\Omega^R_{ZQ_{i}Q_{i}},m_i^2,m_i^2,M_Z^2)\right.\nonumber \\
&-&\left.\frac{1}{M_Z^2}\sum_{i}\mathcal{F}_{ZZ}(\Omega^L_{ZQ_{i}Q_{i}},\Omega^R_{ZQ_{i}Q_{i}},m_i^2,m_i^2,0) \right]
\end{eqnarray}
where the fermion functions $\mathcal{F}_{VV,Z\gamma}$ contributing to the gauge boson two-point functions are calculated as
\begin{eqnarray}
\mathcal{F}_{Z\gamma}(\Omega_1,\Omega_2,\mathbb{Q},m^2,p^2)&=&\frac{N_c}{8\pi^2}[\mathbb{Q}(\Omega_1+\Omega_2)\left(2B_{00}(p^2,m^2,m^2)-p^2B_1(p^2,m^2,m^2)-A_0(m^2) \right)]\; \nonumber \\
\mathcal{F}_{VV}(\Omega_1,\Omega_2,m_1^2,m_2^2,p^2)&=&\frac{N_c}{8\pi^2}\left[\left((\Omega_1^2+\Omega_2^2)m_1^2-2\Omega_1\Omega_2m_1m_2 \right)B_0(p^2,m_1^2,m_2^2)\right. \nonumber \\
&+&\left.(\Omega_1^2+\Omega_2^2)\left(p^2B_1(p^2,m_1^2,m_2^2)-2B_{00}(p^2,m_1^2,m_2^2)+A_0(m_2^2) \right) \right]
\end{eqnarray}
Complete expressions of the oblique parameters for doublets and triplets are lengthy. Thus, we give the full contributions to $\mathbb{S}$ and $\mathbb{T}$ parameters from singlet VLQ representations ${\cal U}_1$ and ${\cal D}_1$, while approximated expressions for all multiplets are  given in Appendix \ref{sec:STapextypes}. The deviations $\Delta \mathbb{T}, \Delta \mathbb{S}$ of the oblique parameters from their SM values are
\begin{eqnarray}
\label{eq:exactobliqU1}
\Delta \mathbb{T}_{{\cal U}_1}&=& \frac{N_{c}m_{t}^{4}(s_{L}^{t})^2}{16\pi s_{W}^{2}M_{W}^{2}(m_{T}^{2}-m_{t}^{2})}\left[\frac{m_{T}^{2}-m_{t}^{2}}{m_{t}^{2}} (s_{L}^{t})^2(\frac{m_{T}^{2}+m_{t}^{2}}{m_{t}^{2}}) -4\frac{m_{T}^{2}}{m_{t}^{2}}(c_{L}^{t})^2\ln(x_{T}) +2-2x_{T}^{2} \right] \end{eqnarray}
\begin{eqnarray}
\Delta \mathbb{S}_{{\cal U}_1}&=&\frac{N_{c}}{12\pi M_{Z}^{2}}\left[(s_{L}^{t})^2A_{0}(m_t^2)\left(6m_T^2(c_{L}^{t})^2-6m_t^2(c_{L}^{t})^2 -M_Z^2(9(s_{L}^{t})^2-10) \right) -32A_0(m_T^2)M_Z^2s_W^2c_W^2 \right. \nonumber \\ 
&-&\left.18m_t^2(s_{L}^{t})^2(c_{L}^{t})^2B_{0}(0,m_t^2,m_T^2)+6(s_{L}^{t})^2(c_{L}^{t})^2B_{0}(M_Z^2,m_t^2,m_T^2)(M_Z^2(m_t^2+m_T^2)-2M_Z^4+(m_T^2-m_t^2)^2) \right. \nonumber \\ 
&+&\left.(s_{L}^{t})^2A_{0}(m_T^2)\left(M_Z^2(9(s_{L}^{t})^2-10)-6m_T^2(c_{L}^{t})^2+6m_t^2(c_{L}^{t})^2 \right) + m_T^2((s_{L}^{t})^2+32s_W^2c_W^2)-(s_{L}^{t})^2m_t^2 \right. \nonumber \\
&-& \left.3m_t^2(s_{L}^{t})^2B_{0}(0,m_t^2,m_t^2)(3(s_{L}^{t})^2-10)+m_T^2B_{0}(0,m_T^2,m_T^2)(32s_W^2c_W^2-12(s_{L}^{t})^2)\right. \nonumber \\
&+&\left. 2(s_{L}^{t})^2B_{0}(M_Z^2,m_T^2,m_T^2)(m_T^2((s_{L}^{t})^2+8)-M_Z^2(3(s_{L}^{t})^2-4))\right. \nonumber \\
&+&\left.2(s_{L}^{t})^2B_{0}(M_Z^2,m_t^2,m_t^2)(m_t^2(3(s_{L}^{t})^2-16)-M_Z^2(3(s_{L}^{t})^2-4))
 \right]
\end{eqnarray}
\begin{eqnarray}
\label{eq:exactobliqD1}
\Delta \mathbb{T}_{{\cal D}_1}&=& \frac{N_{c}}{8\pi c_{W}^{2}s_{W}^{2}M_{Z}^{2}}\left[ \left( (s_{L}^{b})^2(c_{L}^{b})^2[B_{00}(0,m_b^2,m_B^2)-\frac{A_0(m_B^2)}{2}]-\frac{m_b^2(s_{L}^{b})^2(c_{L}^{b})^2}{2}B_0(0,m_b^2,m_B^2)\right)\right. \nonumber \\  
&+&\left. \left(\frac{1}{16}(1-\frac{4s_W^2}{3}^2)+\frac{4s_W^4}{9})[A_0(m_t^2)-2B_{00}(0,m_t^2,m_t^2)]\right)\right. \nonumber \\
&+&\left.
\left(\frac{m_t^2}{4}(1-\frac{4s_W^2}{3})^2+\frac{m_t^2}{9}(4s_W^4-18)+\frac{2m_t^2s_W^2}{3}(1-\frac{4s_W^2}{3})\right)B_0(0,m_t^2,m_t^2)\right. \nonumber \\
&+&\left. \left((\frac{1}{4}(\frac{2s_W^2}{3}-(c_{L}^{b})^2)^2+\frac{s_W^4}{9})[A_0(m_b^2)-2B_{00}(0,m_b^2,m_b^2)] \right) \right. \nonumber \\
&+&\left. \left( \frac{m_b^2}{4}(\frac{2s_W^2}{3}-(c_{L}^{b})^2)^2+\frac{m_b^2s_W^4}{9}-\frac{m_b^2s_W^2}{3}(\frac{2s_W^2}{3}-(c_{L}^{b})^2)\right)B_0(0,m_b^2,m_b^2)\right. \nonumber \\
&+&\left. \left((\frac{1}{4}(\frac{2s_W^2}{3}-(s_{L}^{b})^2)^2+\frac{s_W^4}{9})[A_0(m_B^2)-2B_{00}(0,m_B^2,m_B^2)] \right) \right. \nonumber \\
&+&\left. \left( \frac{m_B^2}{4}(\frac{2s_W^2}{3}-(s_{L}^{b})^2)^2+\frac{m_B^2s_W^4}{9}-\frac{m_B^2s_W^2}{3}(\frac{2s_W^2}{3}-(s_{L}^{b})^2)\right)B_0(0,m_B^2,m_B^2)\right. \nonumber \\
&+&\left. \frac{2}{3}(\frac{8s_W^4}{3}-s_W^2)[A_0(m_t^2)-2B_{00}(0,m_t^2,m_t^2)]+\frac{s_W^2}{3}(\frac{4s_W^2}{3}-(c_{L}^{b})^2))[A_0(m_b^2)-2B_{00}(0,m_b^2,m_b^2)] \right. \nonumber \\
&+&\left.\frac{s_W^2}{3}(\frac{4s_W^2}{3}-(s_{L}^{b})^2))[A_0(m_B^2)-2B_{00}(0,m_B^2,m_B^2)]+\frac{(c_{L}^{b})^2}{2}[A_0(m_b^2)-2B_{00}(0,m_t^2,m_b^2)]\right. \nonumber \\
&+&\left. \frac{m_t^2(c_{L}^{b})^2}{2}B_0(0,m_t^2,m_b^2)+\frac{(s_{L}^{b})^2}{2}[A_0(m_B^2)-2B_{00}(0,m_t^2,m_b^2)]+\frac{m_t^2(s_{L}^{b})^2}{2}B_0(0,m_t^2,m_B^2)
\right] 
\end{eqnarray}
\begin{eqnarray}
\Delta \mathbb{S}_{{\cal D}_1}&=&\frac{N_c}{12\pi M_Z^2}\left[\frac{2}{3}\ln(\frac{m_t^2}{m_b^2})-2 +(s_{L}^{b})^2(c_{L}^{b})^2\left(6B_{00}(m_Z^2,m_b^2,m_B^2)-6B_{00}(0,m_b^2,m_B^2)+3m_Z^2B_1(M_Z^2,m_b^2,m_B^2) \right)\right. \nonumber \\
&+&\left. m_b^2(s_{L}^{b})^2(c_{L}^{b})^2\left(B_0(M_Z^2,m_b^2,m_B^2)-B_0(0,m_b^2,m_B^2)\right)\right. \nonumber \\
&-&\left. \frac{1}{c_W^2s_W^2}\left(\frac{16}{3}[A_0(m_t^2)-2B_{00}(M_Z^2,m_t^2,m_t^2)+M_Z^2B_1(M_Z^2,m_t^2,m_t^2)]\right) \right. \nonumber \\
&-&\left. \frac{1}{c_W^2s_W^2}\left(\frac{4}{3}[A_0(m_b^2)-2B_{00}(M_Z^2,m_b^2,m_b^2)+M_Z^2B_1(M_Z^2,m_b^2,m_b^2)]\right) \right. \nonumber \\
&-&\left. \frac{1}{c_W^2s_W^2}\left(\frac{4}{3}[A_0(m_B^2)-2B_{00}(M_Z^2,m_B^2,m_B^2)+M_Z^2B_1(M_Z^2,m_B^2,m_B^2)]\right) \right. \nonumber \\
&+& \left.\left(\frac{3}{2}(1-\frac{4s_W^2}{3})^2+\frac{8s_W^4}{3}\right)[M_Z^2B_1(M_Z^2,m_t^2,m_t^2)+2B_{00}(0,m_t^2,m_t^2)-2B_{00}(M_Z^2,m_t^2,m_t^2)]\right. \nonumber \\
&+& \left. \left(\frac{3m_t^2}{2}(1-\frac{4s_W^2}{3})^2+\frac{8m_t^2s_W^4}{3}+4m_t^2s_W^2(1-\frac{4s_W^2}{3})\right)[B_0(M_Z^2,m_t^2,m_t^2)-B_0(0,m_t^2,m_t^2)]\right. \nonumber \\
&+&\left. \left(\frac{3}{2}(\frac{2s_W^2}{3}-(c_{L}^{b})^2)^2+\frac{2s_W^4}{3}\right)[M_Z^2B_1(M_Z^2,m_b^2,m_b^2)+2B_{00}(0,m_b^2,m_b^2)-2B_{00}(M_Z^2,m_b^2,m_b^2)]\right. \nonumber \\
&+&\left. \left(\frac{3m_b^2}{2}(\frac{2s_W^2}{3}-(c_{L}^{b})^2)^2+\frac{2m_b^2s_W^4}{3}-2m_b^2s_W^2(\frac{2s_W^2}{3}-(c_{L}^{b})^2)\right)[B_0(M_Z^2,m_b^2,m_b^2)-B_0(0,m_b^2,m_b^2)]\right. \nonumber \\
&+&\left. \left(\frac{3}{2}(\frac{2s_W^2}{3}-(s_{L}^{b})^2)^2+\frac{2s_W^4}{3}\right)[M_Z^2B_1(M_Z^2,m_B^2,m_B^2)+2B_{00}(0,m_B^2,m_B^2)-2B_{00}(M_Z^2,m_B^2,m_B^2)]\right. \nonumber \\
&+&\left. \left(\frac{3m_B^2}{2}(\frac{2s_W^2}{3}-(s_{L}^{b})^2)^2+\frac{2m_B^2s_W^4}{3}-2m_B^2s_W^2(\frac{2s_W^2}{3}-(s_{L}^{b})^2)\right)[B_0(M_Z^2,m_B^2,m_B^2)-B_0(0,m_B^2,m_B^2)]\right. \nonumber \\
&+&\left. \left(c_W^2-s_W^2\right)\big[ (\frac{16s_W^2}{3}-2)[M_Z^2B_1(M_Z^2,m_t^2,m_t^2)-2B_{00}(M_Z^2,m_t^2,m_t^2)-2B_{00}(0,m_t^2,m_t^2)]\right. \nonumber \\
&+&\left. (\frac{4s_W^2}{3}-(c_{L}^{b})^2)[M_Z^2B_1(M_Z^2,m_b^2,m_b^2)-2B_{00}(M_Z^2,m_b^2,m_b^2)-2B_{00}(0,m_b^2,m_b^2)] \big]\right.\nonumber \\
&+&\left. (\frac{4s_W^2}{3}-(s_{L}^{b})^2)[M_Z^2B_1(M_Z^2,m_B^2,m_B^2)-2B_{00}(M_Z^2,m_B^2,m_B^2)-2B_{00}(0,m_B^2,m_B^2)] \big]
\right]
\end{eqnarray}
Contributions to $\mathbb{S}$ and $\mathbb{T}$ parameters from doublet and triplet VLQ representations follow from Eq. \ref{eq:obliqueTvlq}-\ref{eq:obliqueSvlq} by a straightforward calculation with the relevant electroweak couplings as in the Appendix \ref{sec:EWcouplings}.
\begin{figure}[htbp]
	\centering
	\begin{subfigure}{.5\textwidth}\hspace{-1.5cm}
		\includegraphics[height=2.0in]{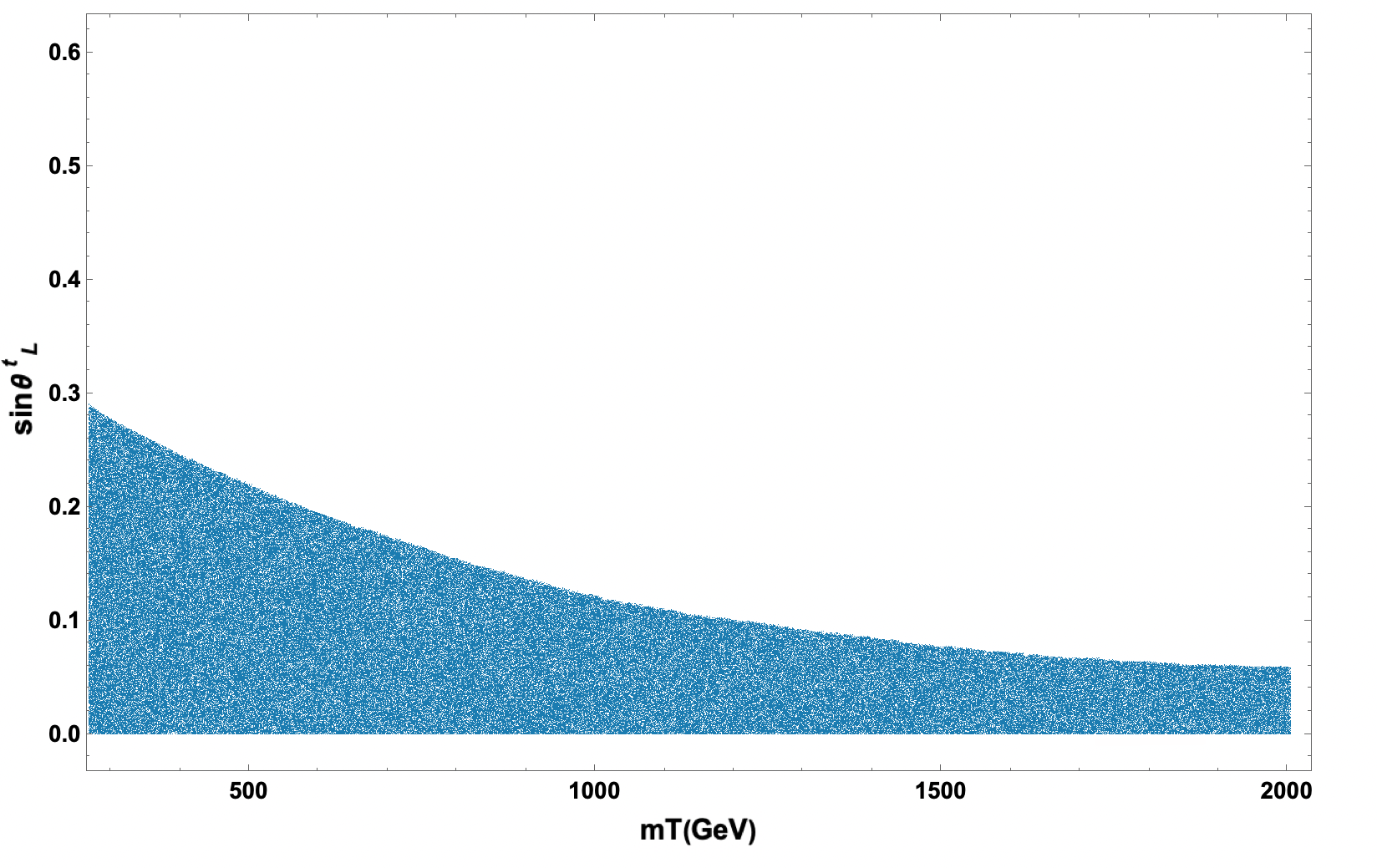}
		\caption{$\mathcal{U}_{1}$ (T)}
	\end{subfigure}\hspace{-1.5cm}
	\begin{subfigure}{.5\textwidth}
		\includegraphics[height=2.0in]{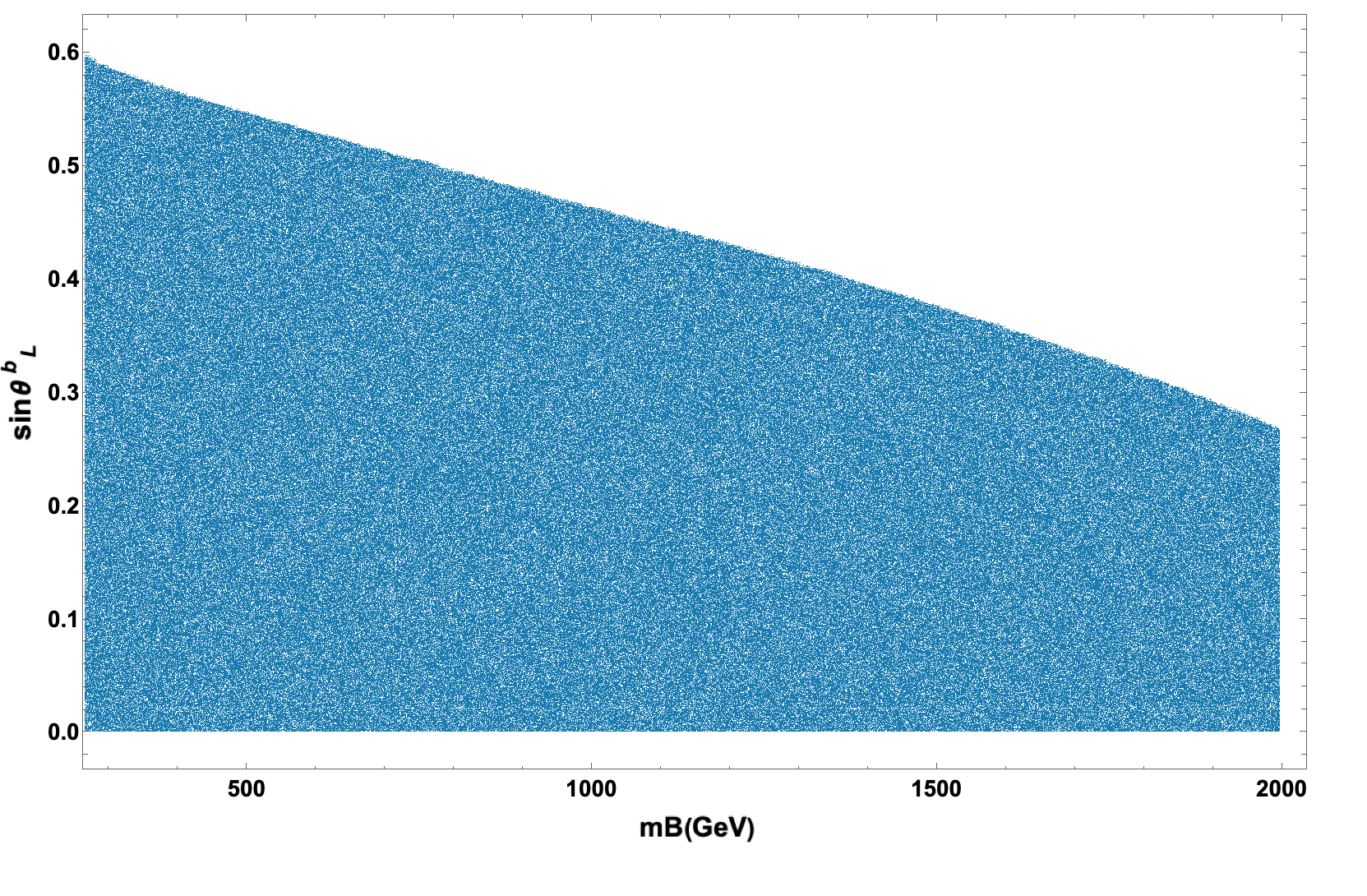}
		\caption{$\mathcal{D}_{1}$ (B)}
	\end{subfigure}\\
	\begin{subfigure}{.33\textwidth}\hspace{-2.8cm}
		\includegraphics[height=2.0in]{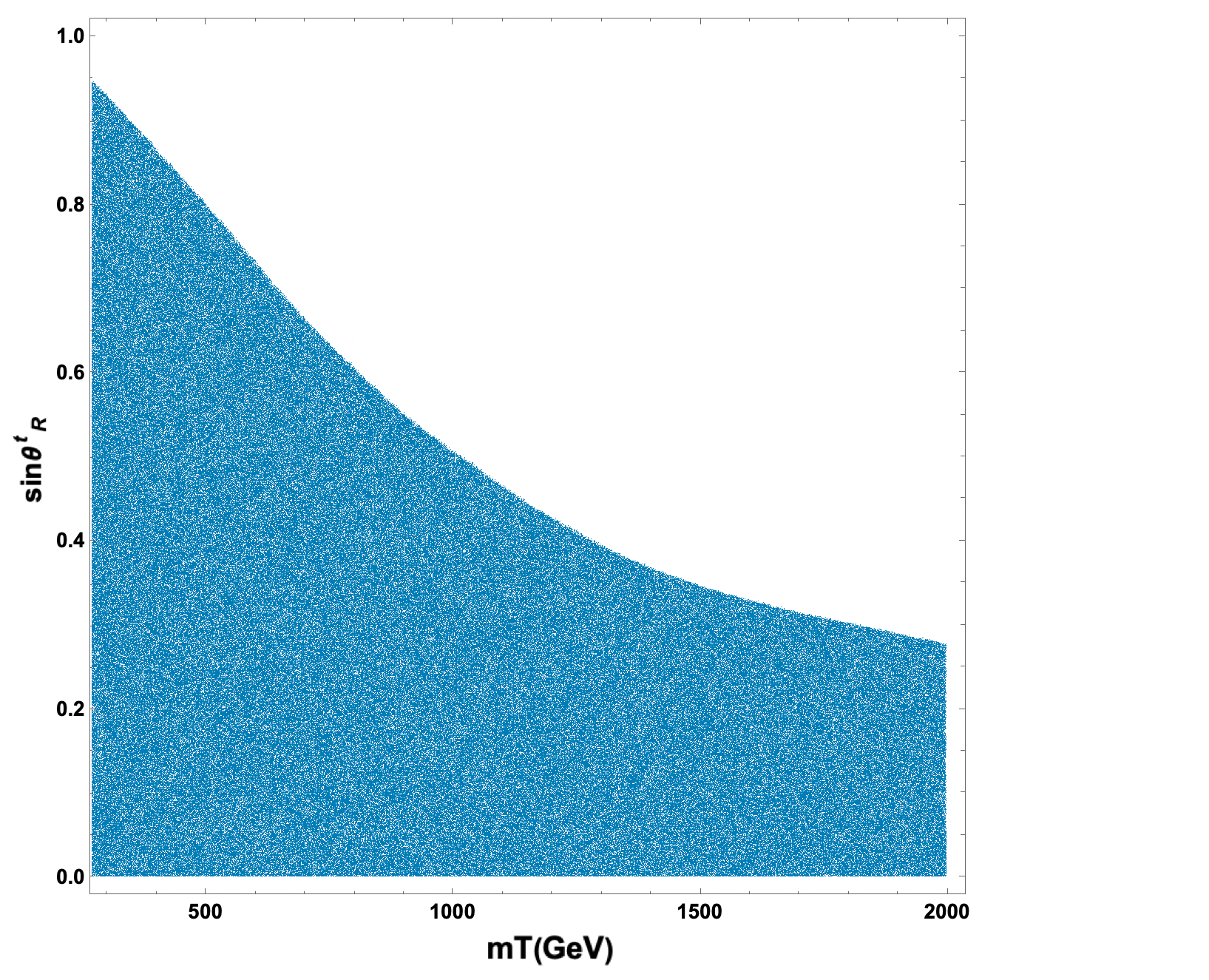}
		\caption{${\cal D}_X$ (XT)}
	\end{subfigure}\hspace{-1.2cm}
	\begin{subfigure}{.33\textwidth}
		\includegraphics[height=2.0in]{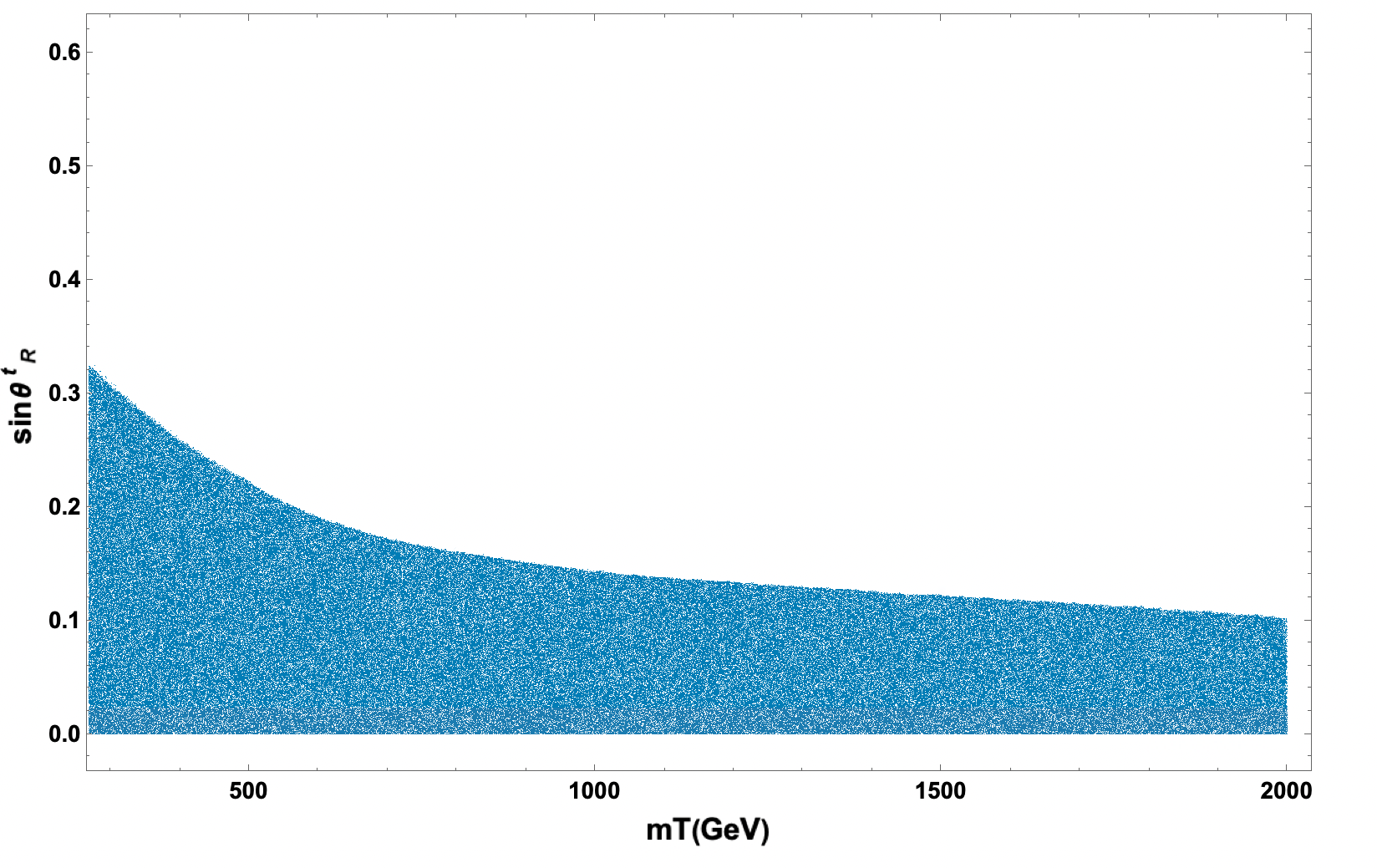}
		\caption{${\cal D}_2$ (TB)}
	\end{subfigure}\\ \hspace{-1.7cm}
	\begin{subfigure}{.33\textwidth}
		\includegraphics[height=2.0in]{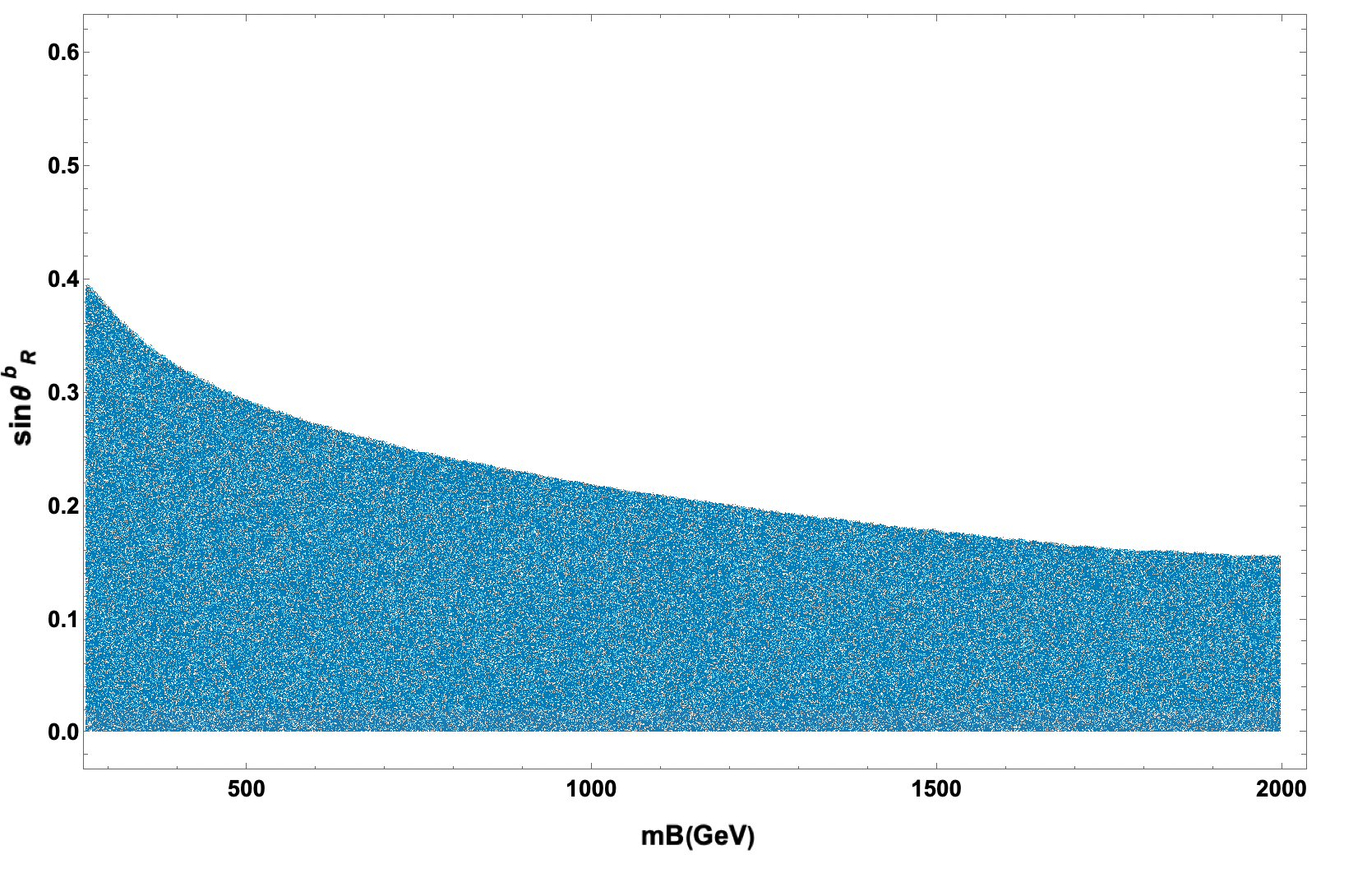}
		\caption{${\cal D}_Y$ (BY)}
	\end{subfigure}\\
\begin{subfigure}{.5\textwidth}\hspace{-1.5cm}
		\includegraphics[height=2.0in]{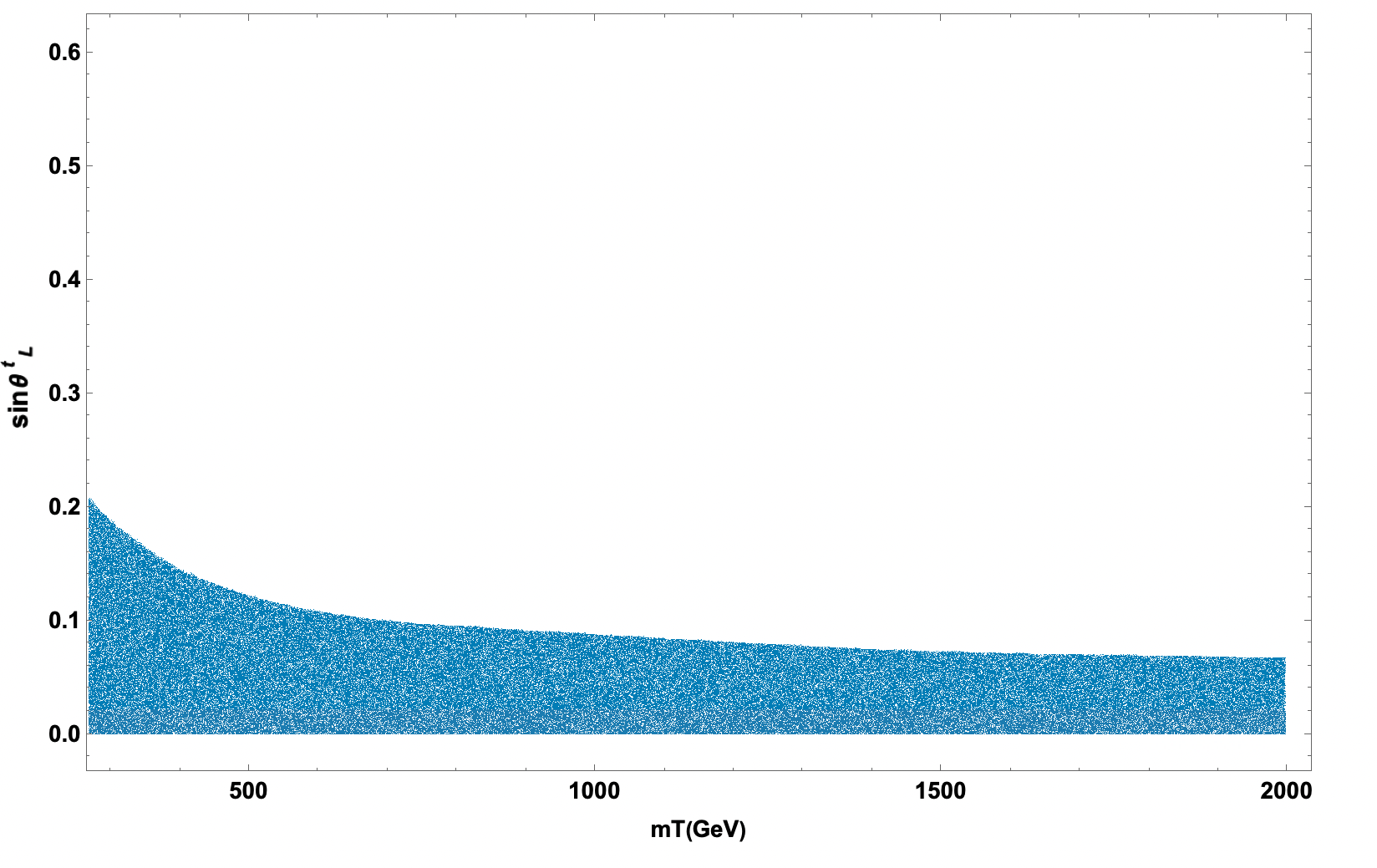}
		\caption{$\mathcal{T}_{X}$ (XTB)}
	\end{subfigure}\hspace{-1.5cm}
	\begin{subfigure}{.5\textwidth}
		\includegraphics[height=2.0in]{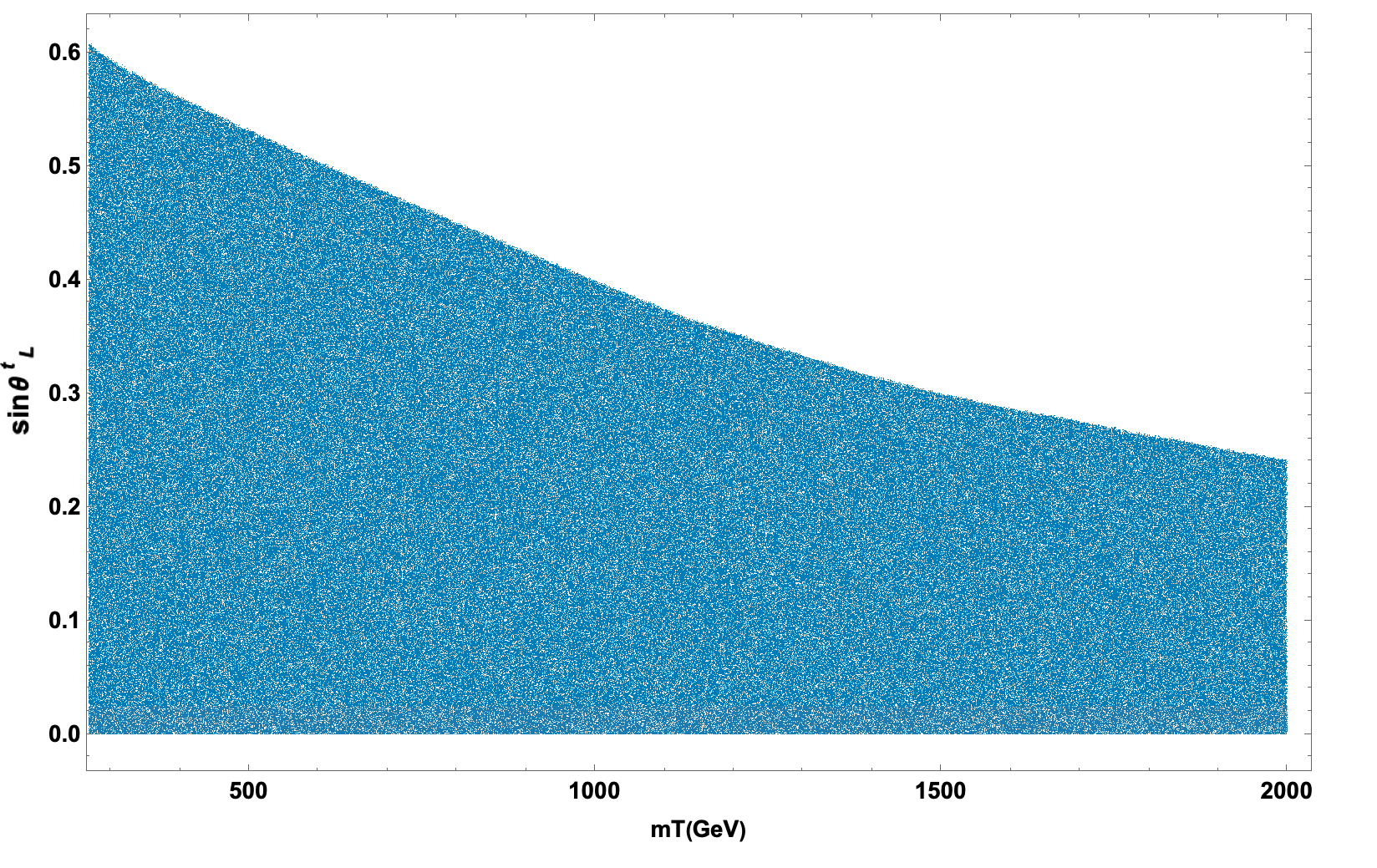}
		\caption{$\mathcal{T}_{Y}$ (TBY)}
	\end{subfigure}
\caption{The allowed parameter space from EWPO: $T$ fermion mass and mixing angle with the top quark for singlet $\mathcal{U}_{1}$ model (top left panel), doublet $\mathcal{D}_{X}$ (middle left panel), doublet $\mathcal{D}_{2}$ (middle centre panel), triplet $\mathcal{T}_{X}$ (bottom left panel) and triplet $\mathcal{T}_{Y}$ (bottom right panel) models. $B$ fermion mass and mixing angle with the bottom quark for singlet $\mathcal{D}_{1}$ model (top right panel) and doublet $\mathcal{D}_{Y}$ model (middle right panel). Loop functions are calculated at energy scale $\mu=m_t$.}
  \label{fig:ewpospaceVLQ}
\end{figure}
\\

In Fig. \ref{fig:ewpospaceVLQ}, we plot the parameter space restricting the mixing between $t-T$ and $b-B$ versus the corresponding VLQ masses satisfying EWPO, in accordance with the expressions given before. The largest deviations arise from the $\mathbb{T}$ parameter due to large logarithm of $(m_T/m_t)^2$, yielding a wide range for  the mass-mixing spectrum compared to the $\mathbb{S}$ parameter for all multiplets. In analogy with the case of  $\sin \alpha =0$  behaviour in the scalar sector, decoupling between the VLQs and the SM quarks becomes much prominent as $m_{VLQ}\to {\cal O}$(TeV) scale. The behaviour of the decoupling zone due to larger values of $m_{VLQ}$ can be seen from Eq. \ref{eq:exactobliqU1}, $\frac{\Delta}{m_{VLQ}}\sim s_L^{t^{2}}$. This consequence can always be viewed as a rule-of-thumb to explain why EWPO constraints are already satisfied in decoupling limit. However, regardless of $m_{VLQ}$, there are no model parameter contributions to $\mathbb{S}$ and $\mathbb{T}$ parameters in zero mixing ($\sin \alpha =0$) domain.
\\

The mixing angle in the singlet   $\mathcal{D}_{1}$ model, Fig. \ref{fig:ewpospaceVLQ} (b), is much more relaxed compared to that in the $\mathcal{U}_{1}$ model due to the fact that up and down type mixings are exclusively dependent on mass splitting between VLQ and the SM quark as seen from Eq. \ref{eq:mixing}. In fact, this holds true for all models if the parameter space allows for $b-B$ mixing. The allowed space for $t-T$ mixing in $\mathcal{U}_{1}$ model matches input values we used to assure the stability in Type-I and Type-II models, whereas scenario $\mathcal{D}_{1}$ lifts the upper bound of $b-B$ mixing to a scale which cannot stabilize the electroweak vacuum around $m_B\geq1$ TeV. Hence, the stability requirements is much severe than the oblique parameters requirements for singlets. The values of $\Delta \mathbb{T}_{VLQ}$ in $\mathcal{U}_{1}$ are always positive and accordingly have more potential to compensate the negative effect of $\Delta\mathbb{T}_{2HDM}$, whereas $\mathcal{D}_{1}$ features negative corrections to $\Delta \mathbb{T}_{VLQ}$. Thus, in terms  of the oblique corrections between both sectors, $\mathcal{U}_{1}$ is capable of imposing more bounds on $M_A, \, M_{H^{\pm}}$ and $\cos(\beta-\alpha)$.
\\

For doublets, the parameter space is larger and similar to $\mathcal{D}_{1}$ except for $\mathcal{D}_{2}$ where $\Delta\mathbb{S}$ contributes  negative values. We should emphasize that, for cases where $\mathbb{S}_{VLQ}$ contributes  negatively  to cancel the positive effect of the $\mathbb{T}$ parameter, the allowed parameter spaces are  effectively enlarged as seen in Fig. \ref{fig:ewpospaceVLQ}(c),(e) for $\mathcal{D}_{X}$ and $\mathcal{D}_{Y}$ models. In contrast, $\Delta\mathbb{S}$ is positive in $\mathcal{D}_{2}$ for $m_T\geq 645$ GeV. We  also observe the behaviour  from Eq. \ref{eq:obliqueSvlq}, where the $\mathcal{D}_{2}$ model does not contribute to (FCNC case) $ZtT$ and $ZbB$ channels, hence $\mathbb{S}$ is relatively larger than those in other doublets. 
Among all doublet models only $\mathcal{D}_{X}$ has a negative $\Delta \mathbb{T}_{VLQ}$ contribution. On the other hand,  $\Delta \mathbb{T}_{VLQ}$ stays close to zero in $\mathcal{D}_{Y}$ model, making it more limited for rendering  $\Delta \mathbb{T}_{2HDM}$ negative, compared to the $\mathcal{D}_{2}$ model, where the correction $\Delta \mathbb{T}_{VLQ}\geq0.08$ yields $m_T>1$ TeV.
\\

Furthermore, at the TeV scale, the EWPO parameter space of the $\mathcal{D}_{2}$ model is in good agreement with the vacuum stability requirements for $ t-T$ mixing, while constraints in $\mathcal{D}_{X}$ and $b-B$ mixing in $\mathcal{D}_{Y}$ allow angles beyond the maximum allowed in the stability analysis. 

The parameter space of the triplet $\mathcal{T}_{X}$ model is quite restricted, and $t-T$ and $b-B$ mixing allowed by the oblique parameters do no cross beyond the vacuum stability requirements. However, for the model $\mathcal{T}_{Y}$, constraints are more relaxed, though $\sin\theta^t_L>0.2$ only exacerbate the constraints on vacuum stability. The relaxation of the mixing in the $\mathcal{T}_{Y}$ scenario  compared to that in $\mathcal{T}_{X}$, can  be described in terms of mixing relations Eqs. \ref{eq:mixingrelations}. Since up and down type mixing angles are not independent for triplets, $s_L^t\simeq\frac{s_L^b}{\sqrt{2}}$\footnote{This relation is valid for the triplet model $\mathcal{T}_{X}$.}, which enhances the $Zbb$ coupling over the one in $\mathcal{T}_{Y}$, and thus leads to more severe corrections in $\mathbb{S}$ \cite{Cao:2022mif}. $\Delta\mathbb{T}_{VLQ}$ is always positive in $\mathcal{T}_{Y}$, while $\mathcal{T}_{X}$ has positive corrections to the $\mathbb{T}$ parameter for $m_T>400$ GeV. As a consequence, the $\mathcal{T}_{Y}$ model is more relaxed as it compensates the negative corrections in 2HDM and it expands the parameter space through combined analysis of the oblique parameters.

\section{Conclusions}
\label{sec:conclusion}

We analyzed the stability of the electroweak vacuum resulting from the interplay between vector-like quarks and the extended bosonic sector of the Two Higgs Doublet Model by 
adopting various representations to scrutinize the potential effects of vector-like quarks on the Higgs sector. In particular, our work zooms in the effects of renormalization group flow that governs the energy scale and flavour dependent behaviour of interactions in the theory. Our investigation remains agnostic to specific parameter choices, while restricting  the mixing of vector-like quarks to solely with the third generation SM quarks. The core of the analysis revolves around the delicate balance of the Higgs potential stability. It has been long assumed that the SM lies in a  metastable state or there is an alternative mechanics  behind the absolute stability of the vacuum. In fact, there is an effective approach to extend the Higgs sector of the SM with additional scalar bosons, as allowed by certain symmetries of the  model. To this end, an auxiliary scalar doublet is introduced here to ameliorate the SM vacuum predicament. Using RGEs, we showed the mere presence  in the 2HDM of the scalar mixing alone does not insure the absolute stability of vacua without constraining the parameter space of 2HDM. 
\\

We then added all anomaly-free representations of vector-like quarks (two singlet, three doublet and two triplet representations). We showed that the inclusion of vector-like quarks, although analogous in their couplings to SM quarks, has complicated consequences. Although fermions contribute negatively to the couplings at RGE level, vector-like quarks effectively modify beta functions through the gauge  and Yukawa portals. Even though the gauge portal effects are weaker than those of the Yukawa couplings, the corrections are multiplicative with respect to number of fermions in the family included. A natural and straightforward attempt could be add more vector-like quarks, considering their effect on gauge couplings modifications.  However, there is a relationship between the number of vector-like quarks and their masses that imposes an upper bound on each, for which the vacuum can be stabilized. If $m_{VLQ}$ is too large and $n_F$ is too small, then RG evolutions fall into the negative perturbativity region before lifting it up. On the other hand, if $n_F$ is too large and $m_{VLQ}$ too small, RG evolutions are too strong and abruptly diverge, thus predictability is lost due to a Landau pole around $\mu<\Lambda_{Pl}$. Considering the strong gauge portal alone, this imposes the upper bounds: $m_{VLQ}\leq10^6$ TeV and $n_F=[2,18]$. Additionally, the hypercharge portal vanishes either by increasing $m_{VLQ}$,  thus leaving insufficient RG evolution for the parameter space restrictions to be operative, or by increasing $n_F$ causing a sub-Planckian theory breakdown. Increasing the hypercharge limits $n_F$ to small values and  to a narrower interval. Thus, allowed hypercharge values are obtained for smallest number of flavours $n_F$, and there is a fine-tuned mutual relation between mass, flavour and hypercharge of vector-like quarks that is capable to generate absolute stability of the vacuum.
\\

We imposed perturbativity and stability constraints for $\lambda_{1,2}$ in both Type-I+VLQ  and Type-II+VLQ 2HDMs, however not for $\lambda_{3,4,5}$ since the latter are not directly related to the stability requirements. Except for scenario $(T,B)$, all VLQ representations require $\tan\beta\geq6$, with small mixing, $\sin\theta_L<0.2$ for mixing in the left-handed sector, for $m_{VLQ}\sim {\cal O}$(TeV). Although larger $\tan\beta$ values satisfy the stability requirements, we observed that they lead to a heavy suppression of the quartic couplings in small perturbation regions. Overall the behaviour of $\lambda_{1,2}$ in Type-II+VLQ, except for the $(T,B,Y)$ representation, is not as sensitive as in Type-I due to the fact $ B$ and $Y$ vector-like quarks are assumed to be heavier than  $X$ and $T$ VLQs. Generally, Type-II involves larger masses from scalar sector of the combined model compared to the Type-I counterparts. However, $T$ and $ B$ singlet models in Type-II are in a more critical situation due to the absence of Yukawa couplings in $\lambda_1$ and $\lambda_2$ respectively, thus the Yukawa portal is momentarily lost. Nonetheless, this critical evolution is ameliorated for larger $M_H$ values. Compared to $T$ vector-like quarks, constraints on the $B$-like fermion masses and mixing angles are much more relaxed. This is simply a consequence of the fact that the mixing between vector-like quarks and the SM quarks is described in terms of the inverse of mass splitting between two quarks. Due to excessive number of negative quartic Yukawa terms appearing at  RGE level, the constraints rising in bare 2HDM have to be enlarged from the above considerations. To this end, we checked both theoretically and experimentally allowed regions of 2HDM and VLQ models. 
\\

We also scanned over EWPO and found the space for $t-T$ and $b-B$ mixings versus the mass of vector-like quarks which includes stability regions, especially in the near decoupling limit. Lower bounds on the masses of pseudo and charged scalars extracted from the oblique parameters fall behind LEP results, accordingly we discarded these. Furthermore, since the scalar and fermion part of the oblique parameters are calculated separately and then combined in our work, we found that the upper bound on the heavier CP-even scalar is insufficient to preserve vacuum stability when combined with triplet VLQs. For this reason, we assumed mass values of the heavier neutral scalar beyond the limits of 2HDM oblique parameters. However, the extension of the upper limit of $M_H$ as $\cos(\beta-\alpha)$ approaches the alignment limit,  also confirms the stability requirement on scalar masses near TeV scale. 
Although  mixing between CP-even states $\cos (\beta-\alpha)\neq 0$  is allowed by the oblique parameters, the stability requires at least near-alignment limit as $\cos (\beta-\alpha)$ remains close to zero. In fact, we observed that RGE running of $\lambda_1$ and $\lambda_2$ deteriorate and the condition for the potential to be bounded from below cannot be satisfied as $\cos (\beta-\alpha)$ strays away from the alignment limit. Accordingly, the lower limit on CP-even mixing angle from the stability and EWPO requirements also match  the experimental Higgs bounds.  

For the VLQ part of the oblique parameters, the allowed parameter space for $t-T$ and $b-B$ mixing is largest for cases where $\Delta \mathbb{S}$ contributes negatively and we have  shown that, for all vector-like multiplets, the EWPO constraints lead to the dalignment limit occurring naturally   as $m_{VLQ} > 1$ TeV. In turn,  constraints at higher TeV range  from the oblique parameters become more consistent with the stability requirements.
Thus the constraints to oblique parameters from vector-like quarks, combined with $\Delta \mathbb{S}$ and $\Delta \mathbb{S}$ from the  2HDM are  VLQ representation-dependent as well as differring for Type-I and Type-II 2HDM and can be used to distinguish among different scenarios. In a specific representation and model-type, these corrections may indicate an allowed deviation from the required cancellations, and this would impose further restrictions on the extra scalar and its mixing with the  SM Higgs boson. 
\begin{acknowledgments}
 This work is funded in part by NSERC  under grant number SAP105354.
\end{acknowledgments}

 \section{Appendix}
 \label{sec:appendix}
In the appendices below, for completeness,  we give the renormalization group equations with respect to Type-I and Type-II models studied in the text, as well as expressions for the contributions of VLQs to the $\mathbb{S}$ and $\mathbb{T}$ parameters, together with the Passrino-Veltman integrals used. We also list the electroweak couplings of the VLQs of different representations.
 \subsection{RGEs for 2HDM + VLQ - Type I}
\label{sec:apprge1}
\subsubsection{Singlet ${\cal U}_1~(T)$, $Y=2/3$}
The relevant RGE for the Yukawa couplings are
\begin{eqnarray}
\label{eq:rgeTsingletfermiont1}
\frac{dy_t^2}{d \ln \mu^2}&=& \frac{y_t^2}{16 \pi^2}\left (\frac{9y_t^2}{2}+\frac{9y_T^2}{2}-\frac{17g_1^2}{12}-\frac{9g_2^2}{4}- 8g_3^2 \right)\, ,\nonumber\\
\frac{dy_T^2}{d \ln \mu^2}&=& \frac{y_T^2}{16 \pi^2}\left (\frac{9y_t^2}{2}+ \frac{9y_T^2}{2}+\frac{3y_M^2}{2}-\frac{17g_1^2}{12}-\frac{9g_2^2}{4}- 8g_3^2 \right)\, ,\nonumber\\
\frac{dy_M^2}{d \ln \mu^2}&=& \frac{y_M^2}{16 \pi^2}\left (y_T^2+\frac{9y_M^2}{2}-\frac{41g_1^2}{20}-8g_3^2\right).
\end{eqnarray}
The Higgs sector RGEs, describing the interactions between the two bosons are:
\begin{eqnarray}
\label{eq:rgeTsingletscalart1}
\frac{d \lambda_1}{d \ln \mu^2}&=& \frac{1}{16 \pi^2} \left[ -4\lambda_1 \left (\frac{3g_{1}^{2}}{4} + \frac{9g_{2}^{2}}{4}\right) +12\lambda_{1}^{2}+4\lambda_{3}^{2}+4\lambda_{3}\lambda_{4}+2\lambda_{4}^{2}+2\lambda_{5}^{2}
 \right. \nonumber \\
 &+& \left. \frac{3g_{1}^{4}}{4}+\frac{9g_{2}^{4}}{4}+\frac{3g_{1}^{2}g_{2}^{2}}{2}\right ] \, , \nonumber \\
\frac{d \lambda_2}{d \ln \mu^2}&=& \frac{1}{16 \pi^2} \left[ 4\lambda_2 \left (6y_{t}^{2}+6y_{T}^{2}-\frac{3g_{1}^{2}}{4} - \frac{9g_{2}^{2}}{4}\right) +12\lambda_{2}^{2}+4\lambda_{3}^{2}+4\lambda_{3}\lambda_{4}+2\lambda_{4}^{2}+2\lambda_{5}^{2}
 \right. \nonumber \\
 &+& \left. \frac{3g_{1}^{4}}{16}+\frac{9g_{2}^{4}}{4}+\frac{3g_{1}^{2}g_{2}^{2}}{2}-12y_{t}^{4}-24y_{T}^{4}-24y_{t}^{2}y_{T}^{2}\right ] \, , \nonumber \\
 \frac{d \lambda_3}{d \ln \mu^2}&=& \frac{1}{16 \pi^2} \left[ 2\lambda_{3} \left (6y_{t}^{2}+6y_{T}^{2}+12y_{M}^{2}-\frac{3g_{1}^{2}}{2} - \frac{9g_{2}^{2}}{2}\right) +4\lambda_{3}^{2}+2\lambda_{4}^{2}+2\lambda_{5}^{2}+ (\lambda_{1}+\lambda_{2})(6\lambda_{3}+2\lambda_{4})
 \right. \nonumber \\
 &+& \left. \frac{3g_{1}^{4}}{4}+\frac{9g_{2}^{4}}{4}-\frac{3g_{1}^{2}g_{2}^{2}}{2}\right ] \, , \nonumber \\
 \frac{d \lambda_4}{d \ln \mu^2}&=& \frac{1}{16 \pi^2} \left[ 2\lambda_{4} \left (6y_{t}^{2}+6y_{T}^{2}+12y_{M}^{2}-\frac{3g_{1}^{2}}{2} - \frac{9g_{2}^{2}}{2}\right) +3g_{1}^{2}g_{2}^{2} +4\lambda_{4}^{2}+8\lambda_{5}^{2}+8\lambda_{3}\lambda_{4}+ 2\lambda_{4}(\lambda_{1}+\lambda_{2})\right ] \, , \nonumber \\
\frac{d \lambda_5}{d \ln \mu^2}&=& \frac{1}{16 \pi^2} \left[ 2\lambda_{5} \left (6y_{t}^{2}+6y_{T}^{2}+12y_{M}^{2}-\frac{3g_{1}^{2}}{2} - \frac{9g_{2}^{2}}{2}\right) +2\lambda_{5}(\lambda_{1}+\lambda_{2}+4\lambda_{3}+6\lambda_{4})\right ] 
 \end{eqnarray}
\subsubsection{Singlet  ${\cal D}_1~(B)$, $Y=-1/3$}
The relevant RGE for the Yukawa couplings are
\begin{eqnarray}
\label{eq:rgeBsingletfermiont1}
\frac{dy_t^2}{d \ln \mu^2}&=& \frac{y_t^2}{16 \pi^2}\left (\frac{9y_t^2}{2}+\frac{3y_B^2}{2}-\frac{17g_1^2}{12}-\frac{9g_2^2}{4}- 8g_3^2 \right)\, ,\nonumber\\
\frac{dy_B^2}{d \ln \mu^2}&=& \frac{y_B^2}{16 \pi^2}\left (\frac{3y_t^2}{2}+ \frac{9y_B^2}{2}+\frac{3y_M^2}{2}-\frac{5g_1^2}{12}-\frac{9g_2^2}{4}- 8g_3^2 \right)\, ,\nonumber\\
\frac{dy_M^2}{d \ln \mu^2}&=& \frac{y_M^2}{16 \pi^2}\left (\frac{9y_M^2}{2}+y_{B}^{2}-\frac{17g_1^2}{20}-8g_3^2\right).
\end{eqnarray}
\newpage
The Higgs sector RGEs, describing the interactions between the two bosons are:
\begin{eqnarray}
\label{eq:rgeBsingletscalart1}
\frac{d \lambda_1}{d \ln \mu^2}&=& \frac{1}{16 \pi^2} \left[ -4\lambda_1 \left (\frac{3g_{1}^{2}}{4} + \frac{9g_{2}^{2}}{4}\right) +12\lambda_{1}^{2}+4\lambda_{3}^{2}+4\lambda_{3}\lambda_{4}+2\lambda_{4}^{2}+2\lambda_{5}^{2}
 \right. \nonumber \\
 &+& \left. \frac{3g_{1}^{4}}{4}+\frac{9g_{2}^{4}}{4}+\frac{3g_{1}^{2}g_{2}^{2}}{2}\right ] \, , \nonumber \\
\frac{d \lambda_2}{d \ln \mu^2}&=& \frac{1}{16 \pi^2} \left[ 4\lambda_2 \left (6y_{t}^{2}+6y_{B}^{2}-\frac{3g_{1}^{2}}{4} - \frac{9g_{2}^{2}}{4}\right) +12\lambda_{2}^{2}+4\lambda_{3}^{2}+4\lambda_{3}\lambda_{4}+2\lambda_{4}^{2}+2\lambda_{5}^{2}
 \right. \nonumber \\
 &+& \left. \frac{3g_{1}^{4}}{16}+\frac{9g_{2}^{4}}{4}+\frac{3g_{1}^{2}g_{2}^{2}}{2}-12y_{t}^{4}-24y_{B}^{4}\right ] \, , \nonumber \\
 \frac{d \lambda_3}{d \ln \mu^2}&=& \frac{1}{16 \pi^2} \left[ 2\lambda_{3} \left (6y_{t}^{2}+6y_{B}^{2}+12y_{M}^{2}-\frac{3g_{1}^{2}}{2} - \frac{9g_{2}^{2}}{2}\right) +4\lambda_{3}^{2}+2\lambda_{4}^{2}+2\lambda_{5}^{2}+ (\lambda_{1}+\lambda_{2})(6\lambda_{3}+2\lambda_{4})
 \right. \nonumber \\
 &+& \left. \frac{3g_{1}^{4}}{4}+\frac{9g_{2}^{4}}{4}-\frac{3g_{1}^{2}g_{2}^{2}}{2}\right ] \, , \nonumber \\
 \frac{d \lambda_4}{d \ln \mu^2}&=& \frac{1}{16 \pi^2} \left[ 2\lambda_{4} \left (6y_{t}^{2}+6y_{B}^{2}+12y_{M}^{2}-\frac{3g_{1}^{2}}{2} - \frac{9g_{2}^{2}}{2}\right) +3g_{1}^{2}g_{2}^{2} +4\lambda_{4}^{2}+8\lambda_{5}^{2}+8\lambda_{3}\lambda_{4}+ 2\lambda_{4}(\lambda_{1}+\lambda_{2})\right ] \, , \nonumber \\
\frac{d \lambda_5}{d \ln \mu^2}&=& \frac{1}{16 \pi^2} \left[ 2\lambda_{5} \left (6y_{t}^{2}+6y_{B}^{2}+12y_{M}^{2}-\frac{3g_{1}^{2}}{2} - \frac{9g_{2}^{2}}{2}\right) +2\lambda_{5}(\lambda_{1}+\lambda_{2}+4\lambda_{3}+6\lambda_{4})\right ] 
\end{eqnarray}
Finally the coupling constants gain additional terms due to the new fermion, for both models ${\cal U}_1,~{\cal D}_1$ with singlet fermions as follows:
\begin{eqnarray}
\frac{dg_1^2}{d \ln \mu^2}&=&\frac{g_1^4}{16 \pi^2}\left(7+\frac{4}{15}\right )\, , \nonumber \\
\frac{dg_2^2}{d \ln \mu^2}&=&\frac{g_2^4}{16 \pi^2}\left( -3 \right ) \, , \nonumber \\
\frac{dg_3^2}{d \ln \mu^2}&=&\frac{g_3^4}{16 \pi^2}\left(-7+\frac{2}{3}\right )\, .
\end{eqnarray}
\subsubsection{Doublet ${\cal D}_2$ $(T,B), \,Y=1/6$}
The relevant RGE for the Yukawa couplings are
\begin{eqnarray}
\label{eq:rgeTBdoubletfermiont1}
\frac{dy_t^2}{d \ln \mu^2}&=& \frac{y_t^2}{16 \pi^2}\left (\frac{9y_t^2}{2}+\frac{9y_T^2}{2}+\frac{3y_B^2}{2}+3y_{M}^{2}-\frac{17g_{1}^{2}}{12}-\frac{9g_2^2}{4}- 8g_3^2 \right)\, ,\nonumber\\
\frac{dy_T^2}{d \ln \mu^2}&=& \frac{y_T^2}{16 \pi^2}\left (\frac{9y_t^2}{2}+ \frac{9y_T^2}{2}+\frac{3y_B^2}{2}+3y_{M}^{2}-\frac{17g_{1}^{2}}{12}-\frac{9g_2^2}{4}- 8g_3^2 \right)\, ,\nonumber\\
\frac{dy_B^2}{d \ln \mu^2}&=& \frac{y_B^2}{16 \pi^2}\left (\frac{9y_t^2}{2}+ \frac{3y_T^2}{2}+\frac{9y_B^2}{2}+3y_{M}^{2}-\frac{5g_{1}^{2}}{12}-\frac{9g_2^2}{4}- 8g_3^2 \right)\, ,\nonumber\\
\frac{dy_M^2}{d \ln \mu^2}&=& \frac{y_M^2}{16 \pi^2}\left (y_T^2+y_B^2+\frac{11y_M^2}{2}-\frac{19g_{1}^{2}}{40}-\frac{9g_{2}^{2}}{2}-8g_3^2\right).
\end{eqnarray}
The Higgs sector RGEs, describing the interactions between the two bosons are:
\begin{eqnarray}
\label{eq:rgeTBdoubletTBscalart1}
\frac{d \lambda_1}{d \ln \mu^2}&=& \frac{1}{16 \pi^2} \left[ -4\lambda_1 \left (\frac{3g_{1}^{2}}{4} + \frac{9g_{2}^{2}}{4}\right) +12\lambda_{1}^{2}+4\lambda_{3}^{2}+4\lambda_{3}\lambda_{4}+2\lambda_{4}^{2}+2\lambda_{5}^{2}
 \right. \nonumber \\
 &+& \left. \frac{3g_{1}^{4}}{4}+\frac{9g_{2}^{4}}{4}+\frac{3g_{1}^{2}g_{2}^{2}}{2}\right ] \, , \nonumber \\
\frac{d \lambda_2}{d \ln \mu^2}&=& \frac{1}{16 \pi^2} \left[ 4\lambda_2 \left (3y_{t}^{2}+6y_{T}^{2}+6y_{B}^{2}-\frac{3g_{1}^{2}}{4} - \frac{9g_{2}^{2}}{4}\right) +12\lambda_{2}^{2}+4\lambda_{3}^{2}+4\lambda_{3}\lambda_{4}+2\lambda_{4}^{2}+2\lambda_{5}^{2}
 \right. \nonumber \\
 &+& \left. \frac{3g_{1}^{4}}{4}+\frac{9g_{2}^{4}}{4}+\frac{3g_{1}^{2}g_{2}^{2}}{2}-12y_{t}^{4}-24y_{T}^{4}-24y_{B}^{4}-24y_{t}^{2}y_{T}^{2}-24y_{T}^{2}y_{B}^{2}\right ] \, , \nonumber \\
 \frac{d \lambda_3}{d \ln \mu^2}&=& \frac{1}{16 \pi^2} \left[ 2\lambda_{3} \left (3y_{t}^{2}+6y_{T}^{2}+6y_{B}^{2}+12y_{M}^{2}-\frac{3g_{1}^{2}}{2} - \frac{9g_{2}^{2}}{2}\right) +4\lambda_{3}^{2}+2\lambda_{4}^{2}+2\lambda_{5}^{2}+ (\lambda_{1}+\lambda_{2})(6\lambda_{3}+2\lambda_{4})
 \right. \nonumber \\
 &+& \left. \frac{3g_{1}^{4}}{4}+\frac{9g_{2}^{4}}{4}-\frac{3g_{1}^{2}g_{2}^{2}}{2}\right ] \, , \nonumber \\
 \frac{d \lambda_4}{d \ln \mu^2}&=& \frac{1}{16 \pi^2} \left[ 2\lambda_{4} \left (3y_{t}^{2}+6y_{T}^{2}+6y_{B}^{2}+12y_{M}^{2}-\frac{3g_{1}^{2}}{2} - \frac{9g_{2}^{2}}{2}\right) +3g_{1}^{2}g_{2}^{2} +4\lambda_{4}^{2}+8\lambda_{5}^{2}+8\lambda_{3}\lambda_{4}+ 2\lambda_{4}(\lambda_{1}+\lambda_{2})\right ] \, , \nonumber \\
\frac{d \lambda_5}{d \ln \mu^2}&=& \frac{1}{16 \pi^2} \left[ 2\lambda_{5} \left (3y_{t}^{2}+6y_{T}^{2}+6y_{B}^{2}+12y_{M}^{2}-\frac{3g_{1}^{2}}{2} - \frac{9g_{2}^{2}}{2}\right) +2\lambda_{5}(\lambda_{1}+\lambda_{2}+4\lambda_{3}+6\lambda_{4})\right ] 
 \end{eqnarray}

\subsubsection{Doublet ${\cal D}_X$ $(X,T), \,Y=7/6$}
The relevant RGE for the Yukawa couplings are
\begin{eqnarray}
\label{eq:rgeXTdoubletfermiont1}
\frac{dy_t^2}{d \ln \mu^2}&=& \frac{y_t^2}{16 \pi^2}\left (\frac{9y_t^2}{2}+\frac{3y_T^2}{2}+\frac{9y_X^2}{2}+3y_{M}^{2}-\frac{17g_1^2}{12}-\frac{9g_2^2}{4}- 8g_3^2 \right)\, ,\nonumber\\
\frac{dy_T^2}{d \ln \mu^2}&=& \frac{y_T^2}{16 \pi^2}\left (\frac{9y_t^2}{2}+\frac{3y_X^2}{2}+ \frac{9y_T^2}{2}+3y_{M}^{2}-\frac{5g_1^2}{12}-\frac{9g_2^2}{4}- 8g_3^2 \right)\, ,\nonumber\\
\frac{dy_X^2}{d \ln \mu^2}&=& \frac{y_X^2}{16 \pi^2}\left (\frac{9y_t^2}{2}+ \frac{9y_X^2}{2}+ \frac{3y_T^2}{2}+3y_M^{2}-\frac{17g_1^2}{12}-\frac{9g_2^2}{4}- 8g_3^2 \right)\, ,\nonumber\\
\frac{dy_M^2}{d \ln \mu^2}&=& \frac{y_M^2}{16 \pi^2}\left (y_T^2+y_X^2+\frac{11y_M^2}{2}-\frac{67g_1^2}{40}-\frac{9g_2^2}{2}-8g_3^2\right).
\end{eqnarray}
The Higgs sector RGEs, describing the interactions between the two bosons are:
\begin{eqnarray}
\label{eq:rgeXTdoubletscalart1}
\frac{d \lambda_1}{d \ln \mu^2}&=& \frac{1}{16 \pi^2} \left[ -4\lambda_1 \left (\frac{3g_{1}^{2}}{4} + \frac{9g_{2}^{2}}{4}\right) +12\lambda_{1}^{2}+4\lambda_{3}^{2}+4\lambda_{3}\lambda_{4}+2\lambda_{4}^{2}+2\lambda_{5}^{2}
 \right. \nonumber \\
 &+& \left. \frac{3g_{1}^{4}}{4}+\frac{9g_{2}^{4}}{4}+\frac{3g_{1}^{2}g_{2}^{2}}{2}\right ] \, , \nonumber \\
\frac{d \lambda_2}{d \ln \mu^2}&=& \frac{1}{16 \pi^2} \left[ 4\lambda_2 \left (3y_{t}^{2}+6y_{T}^{2}+6y_{X}^{2}-\frac{3g_{1}^{2}}{4} - \frac{9g_{2}^{2}}{4}\right) +12\lambda_{2}^{2}+4\lambda_{3}^{2}+4\lambda_{3}\lambda_{4}+2\lambda_{4}^{2}+2\lambda_{5}^{2}
 \right. \nonumber \\
 &+& \left. \frac{3g_{1}^{4}}{4}+\frac{9g_{2}^{4}}{4}+\frac{3g_{1}^{2}g_{2}^{2}}{2}-12y_{t}^{4}-24y_{X}^{4}-24y_{T}^{4}-24y_{t}^{2}y_{T}^{2}-24y_{X}^{2}y_{T}^{2}\right ] \, , \nonumber \\
 \frac{d \lambda_3}{d \ln \mu^2}&=& \frac{1}{16 \pi^2} \left[ 2\lambda_{3} \left (3y_{t}^{2}+6y_{T}^{2}+6y_{X}^{2}+12y_{M}^{2}-\frac{3g_{1}^{2}}{2} - \frac{9g_{2}^{2}}{10}\right) +4\lambda_{3}^{2}+2\lambda_{4}^{2}+2\lambda_{5}^{2}+ (\lambda_{1}+\lambda_{2})(6\lambda_{3}+2\lambda_{4})
 \right. \nonumber \\
 &+& \left. \frac{3g_{1}^{4}}{4}+\frac{9g_{2}^{4}}{4}-\frac{3g_{1}^{2}g_{2}^{2}}{2}\right ] \, , \nonumber \\
 \frac{d \lambda_4}{d \ln \mu^2}&=& \frac{1}{16 \pi^2} \left[ 2\lambda_{4} \left (3y_{t}^{2}+6y_{T}^{2}+6y_{X}^{2}+12y_{M}^{2}-\frac{3g_{1}^{2}}{2} - \frac{9g_{2}^{2}}{10}\right) +3g_{1}^{2}g_{2}^{2} +4\lambda_{4}^{2}+8\lambda_{5}^{2}+8\lambda_{3}\lambda_{4}+ 2\lambda_{4}(\lambda_{1}+\lambda_{2})\right ] \, , \nonumber \\
\frac{d \lambda_5}{d \ln \mu^2}&=& \frac{1}{16 \pi^2} \left[ 2\lambda_{5} \left (3y_{t}^{2}+6y_{T}^{2}+6y_{X}^{2}+12y_{M}^{2}-\frac{3g_{1}^{2}}{2} - \frac{9g_{2}^{2}}{10}\right) +2\lambda_{5}(\lambda_{1}+\lambda_{2}+4\lambda_{3}+6\lambda_{4})\right ]  \end{eqnarray}

\subsubsection{Additional non SM-like quark doublet ${\cal D}_Y$ $(B,Y), \,Y=-5/6$}
The relevant RGE for the Yukawa couplings are
\begin{eqnarray}
\label{eq:rgeBYdoubletfermiont1}
\frac{dy_t^2}{d \ln \mu^2}&=& \frac{y_t^2}{16 \pi^2}\left (\frac{9y_t^2}{2}+\frac{9y_B^2}{2}+\frac{3y_Y^2}{2}+3y_M^{2}-\frac{17g_1^2}{12}-\frac{9g_2^2}{4}- 8g_3^2 \right)\, ,\nonumber\\
\frac{dy_B^2}{d \ln \mu^2}&=& \frac{y_T^2}{16 \pi^2}\left (\frac{9y_t^2}{2}+\frac{9y_B^2}{2}+ \frac{3y_Y^2}{2}+3y_M^{2}-\frac{17g_1^2}{12}-\frac{9g_2^2}{4}- 8g_3^2 \right)\, ,\nonumber\\
\frac{dy_Y^2}{d \ln \mu^2}&=& \frac{y_Y^2}{16 \pi^2}\left (\frac{3y_t^2}{2}+ \frac{9y_Y^2}{2}+ \frac{3y_B^2}{2}+3y_M^{2}-\frac{5g_1^2}{12}-\frac{9g_2^2}{4}- 8g_3^2 \right)\, ,\nonumber\\
\frac{dy_M^2}{d \ln \mu^2}&=& \frac{y_M^2}{16 \pi^2}\left (y_Y^2+y_B^2+\frac{11y_M^2}{2}-\frac{43g_1^2}{40}-\frac{9g_{2}^{2}}{2}-8g_3^2\right).
\end{eqnarray}
The Higgs sector RGEs, describing the interactions between the two bosons are:
\begin{eqnarray}
\label{eq:rgeBYdoubletscalart1}
\frac{d \lambda_1}{d \ln \mu^2}&=& \frac{1}{16 \pi^2} \left[ -4\lambda_1 \left (\frac{3g_{1}^{2}}{4} + \frac{9g_{2}^{2}}{4}\right) +12\lambda_{1}^{2}+4\lambda_{3}^{2}+4\lambda_{3}\lambda_{4}+2\lambda_{4}^{2}+2\lambda_{5}^{2}
 \right. \nonumber \\
 &+& \left. \frac{3g_{1}^{4}}{4}+\frac{9g_{2}^{4}}{4}+\frac{3g_{1}^{2}g_{2}^{2}}{2}\right ] \, , \nonumber \\
\frac{d \lambda_2}{d \ln \mu^2}&=& \frac{1}{16 \pi^2} \left[ 4\lambda_2 \left (3y_{t}^{2}+6y_{B}^{2}+6y_{Y}^{2}-\frac{3g_{1}^{2}}{4} - \frac{9g_{2}^{2}}{4}\right) +12\lambda_{2}^{2}+4\lambda_{3}^{2}+4\lambda_{3}\lambda_{4}+2\lambda_{4}^{2}+2\lambda_{5}^{2}
 \right. \nonumber \\
 &+& \left. \frac{3g_{1}^{4}}{4}+\frac{9g_{2}^{4}}{4}+\frac{3g_{1}^{2}g_{2}^{2}}{2}-12y_{t}^{4}-24y_{B}^{4}-24y_{Y}^{4}-24y_{B}^{2}y_{Y}^{2}\right ] \, , \nonumber \\
 \frac{d \lambda_3}{d \ln \mu^2}&=& \frac{1}{16 \pi^2} \left[ 2\lambda_{3} \left (3y_{t}^{2}+6y_{B}^{2}+6y_{Y}^{2}+12y_{M}^{2}-\frac{3g_{1}^{2}}{2} - \frac{9g_{2}^{2}}{2}\right) +4\lambda_{3}^{2}+2\lambda_{4}^{2}+2\lambda_{5}^{2}+ (\lambda_{1}+\lambda_{2})(6\lambda_{3}+2\lambda_{4})
 \right. \nonumber \\
 &+& \left. \frac{3g_{1}^{4}}{4}+\frac{9g_{2}^{4}}{4}-\frac{3g_{1}^{2}g_{2}^{2}}{2}\right ] \, , \nonumber \\
 \frac{d \lambda_4}{d \ln \mu^2}&=& \frac{1}{16 \pi^2} \left[ 2\lambda_{4} \left (3y_{t}^{2}+6y_{Y}^{2}+6y_{B}^{2}+12y_{M}^{2}-\frac{3g_{1}^{2}}{2} - \frac{9g_{2}^{2}}{2}\right) +3g_{1}^{2}g_{2}^{2} +4\lambda_{4}^{2}+8\lambda_{5}^{2}+8\lambda_{3}\lambda_{4}+ 2\lambda_{4}(\lambda_{1}+\lambda_{2})\right ] \, , \nonumber \\
\frac{d \lambda_5}{d \ln \mu^2}&=& \frac{1}{16 \pi^2} \left[ 2\lambda_{5} \left (3y_{t}^{2}+6y_{Y}^{2}+6y_{B}^{2}+12y_{M}^{2}-\frac{3g_{1}^{2}}{2} - \frac{9g_{2}^{2}}{10}\right) +2\lambda_{5}(\lambda_{1}+\lambda_{2}+4\lambda_{3}+6\lambda_{4})\right ]  \end{eqnarray}
The coupling constants gain additional terms due to the new fermion in all doublet models as follows:
\begin{eqnarray}
\frac{dg_1^2}{d \ln \mu^2}&=&\frac{g_1^4}{16 \pi^2}\left(7+\frac{6}{5}\right )\, , \nonumber \\
\frac{dg_2^2}{d \ln \mu^2}&=&\frac{g_2^4}{16 \pi^2}\left( -3 + 2 \right ) \, , \nonumber \\
\frac{dg_3^2}{d \ln \mu^2}&=&\frac{g_3^4}{16 \pi^2}\left(-7+\frac{4}{3}\right )\, .
\end{eqnarray}

\subsubsection{Triplet ${\cal T}_X$ $(X, T, B), \,Y=2/3$}
The relevant RGE for the Yukawa couplings are
\begin{eqnarray}
\label{eq:rgeXTBtripletfermiont1}
\frac{dy_t^2}{d \ln \mu^2}&=& \frac{y_t^2}{16 \pi^2}\left (\frac{9y_t^2}{2}+\frac{9y_X^2}{2}+\frac{9y_T^2}{2}+\frac{3y_B^2}{2}+\frac{3y_M^2}{2}-\frac{17g_1^2}{12}-\frac{9g_2^2}{4}- 8g_3^2 \right)\, ,\nonumber\\
\frac{dy_T^2}{d \ln \mu^2}&=& \frac{y_T^2}{16 \pi^2}\left (\frac{9y_t^2}{2}+\frac{9y_X^2}{2}+ \frac{9y_T^2}{2}+\frac{3y_B^2}{2}+\frac{3y_M^2}{2}-\frac{17g_1^2}{12}-\frac{9g_2^2}{4}- 8g_3^2 \right)\, ,\nonumber\\
\frac{dy_X^2}{d \ln \mu^2}&=& \frac{y_T^2}{16 \pi^2}\left (\frac{9y_t^2}{2}+ \frac{9y_X^2}{2}+ \frac{9y_T^2}{2}+\frac{3y_B^2}{2}+\frac{3y_M^2}{2}-\frac{41g_1^2}{20}-\frac{9g_2^2}{4}- 8g_3^2 \right)\, ,\nonumber\\
\frac{dy_B^2}{d \ln \mu^2}&=& \frac{y_B^2}{16 \pi^2}\left (\frac{3y_t^2}{2}+\frac{3y_X^2}{2}+ \frac{9y_T^2}{2}+ \frac{9y_B^2}{2}+\frac{3y_M^2}{2}-\frac{5g_1^2}{12}-\frac{9g_2^2}{4}- 8g_3^2 \right)\, ,\nonumber\\
\frac{dy_M^2}{d \ln \mu^2}&=& \frac{y_M^2}{16 \pi^2}\left (y_X^2+y_T^2+y_B^2+\frac{15y_M^2}{2}-\frac{41g_1^2}{20}-\frac{9g_2^2}{4}-8g_3^2\right).
\end{eqnarray}
The Higgs sector RGEs, describing the interactions between the two bosons are:
\begin{eqnarray}
\label{eq:rgeXTBtripletscalart1}
\frac{d \lambda_1}{d \ln \mu^2}&=& \frac{1}{16 \pi^2} \left[ -4\lambda_1 \left (\frac{3g_{1}^{2}}{4} + \frac{9g_{2}^{2}}{4}\right) +12\lambda_{1}^{2}+4\lambda_{3}^{2}+4\lambda_{3}\lambda_{4}+2\lambda_{4}^{2}+2\lambda_{5}^{2}
 \right. \nonumber \\
 &+& \left. \frac{3g_{1}^{4}}{4}+\frac{9g_{2}^{4}}{4}+\frac{3g_{1}^{2}g_{2}^{2}}{2}\right ] \, , \nonumber \\
\frac{d \lambda_2}{d \ln \mu^2}&=& \frac{1}{16 \pi^2} \left[ 4\lambda_2 \left (3y_{t}^{2}+6y_{X}^{2}+6y_{T}^{2}+6y_{B}^{2}-\frac{3g_{1}^{2}}{2} - \frac{9g_{2}^{2}}{2}\right) +12\lambda_{2}^{2}+4\lambda_{3}^{2}+4\lambda_{3}\lambda_{4}+2\lambda_{4}^{2}+2\lambda_{5}^{2}
 \right. \nonumber \\
 &+& \left. \frac{3g_{1}^{4}}{2}+\frac{9g_{2}^{2}}{2}+3g_{1}^{2}g_{2}^{2}-12y_{t}^{4}-24y_{X}^{4}-24y_{T}^{4}-24y_{B}^{4}-24y_{t}^{2}y_{T}^{2}-24y_{T}^{2}y_{B}^{2}-24y_{X}^{2}y_{T}^{2}\right ] \, , \nonumber \\
 \frac{d \lambda_3}{d \ln \mu^2}&=& \frac{1}{16 \pi^2} \left[ 2\lambda_{3} \left (3y_{t}^{2}+6y_{T}^{2}+6y_{X}^{2}+6y_{B}^{2}+12y_{M}^{2}-3g_{1}^{2} - 9g_{2}^{2}\right) +4\lambda_{3}^{2}+2\lambda_{4}^{2}+2\lambda_{5}^{2}+ (\lambda_{1}+\lambda_{2})(6\lambda_{3}+2\lambda_{4})
 \right. \nonumber \\
 &+& \left. \frac{3g_{1}^{4}}{2}+\frac{9g_{2}^{4}}{2}-3g_{1}^{2}g_{2}^{2}\right ] \, , \nonumber \\
 \frac{d \lambda_4}{d \ln \mu^2}&=& \frac{1}{16 \pi^2} \left[ 2\lambda_{4} \left (3y_{t}^{2}+6y_{X}^{2}+6y_{T}^{2}+6y_{B}^{2}+12y_{M}^{2}-3g_{1}^{2} - 9g_{2}^{2}\right) +6g_{1}^{2}g_{2}^{2} +4\lambda_{4}^{2}+8\lambda_{5}^{2}+8\lambda_{3}\lambda_{4}+ 2\lambda_{4}(\lambda_{1}+\lambda_{2})\right ] \, , \nonumber \\
\frac{d \lambda_5}{d \ln \mu^2}&=& \frac{1}{16 \pi^2} \left[ 2\lambda_{5} \left (3y_{t}^{2}+6y_{T}^{2}+6y_{X}^{2}+6y_{B}^{2}+12y_{M}^{2}-3g_{1}^{2} - 9g_{2}^{2}\right) +2\lambda_{5}(\lambda_{1}+\lambda_{2}+4\lambda_{3}+6\lambda_{4})\right ] 
 \end{eqnarray}

\subsubsection{Triplet ${\cal T}_Y$ $( T, B, Y), \, Y=-1/3$}
The relevant RGE for the Yukawa couplings are
\begin{eqnarray}
\label{eq:rgeTBYtripletfermiont1}
\frac{dy_t^2}{d \ln \mu^2}&=& \frac{y_t^2}{16 \pi^2}\left (\frac{9y_t^2}{2}+\frac{9y_T^2}{2}+\frac{3y_B^2}{2}+\frac{3y_Y^2}{2}+3y_{M}^{2}-\frac{17g_1^2}{12}-\frac{9g_2^2}{4}- 8g_3^2 \right)\, ,\nonumber\\
\frac{dy_T^2}{d \ln \mu^2}&=& \frac{y_T^2}{16 \pi^2}\left (\frac{9y_t^2}{2}+\frac{9y_T^2}{2}+\frac{3y_B^2}{2}+ \frac{3y_Y^2}{2}+3y_{M}^{2}-\frac{17g_1^2}{12}-\frac{9g_2^2}{4}- 8g_3^2 \right)\, ,\nonumber\\
\frac{dy_B^2}{d \ln \mu^2}&=& \frac{y_B^2}{16 \pi^2}\left (\frac{3y_t^2}{2}+\frac{3y_T^2}{2}+\frac{9y_B^2}{2}+ \frac{9y_Y^2}{2}+\frac{3y_M^2}{2}-\frac{5g_1^2}{12}-\frac{9g_2^2}{4}- 8g_3^2 \right)\, ,\nonumber\\
\frac{dy_Y^2}{d \ln \mu^2}&=& \frac{y_Y^2}{16 \pi^2}\left (\frac{9y_Y^2}{2}+\frac{9y_B^2}{2}+ \frac{3y_t^2}{2}+ \frac{3y_T^2}{2}+3y_{M}^{2}-\frac{17g_1^2}{20}-\frac{9g_2^2}{4}- 8g_3^2 \right)\, ,\nonumber\\
\frac{dy_M^2}{d \ln \mu^2}&=& \frac{y_M^2}{16 \pi^2}\left (y_T^2+y_B^2+y_Y^2+\frac{15y_M^2}{2}-\frac{17g_1^2}{20}-\frac{9g_2^2}{4}-8g_3^2\right).
\end{eqnarray}
The Higgs sector RGEs, describing the interactions between the two bosons are:
\begin{eqnarray}
\label{eq:rgeTBYtripletscalart1}
\frac{d \lambda_1}{d \ln \mu^2}&=& \frac{1}{16 \pi^2} \left[ -4\lambda_1 \left (\frac{3g_{1}^{2}}{4} + \frac{9g_{2}^{2}}{4}\right) +12\lambda_{1}^{2}+4\lambda_{3}^{2}+4\lambda_{3}\lambda_{4}+2\lambda_{4}^{2}+2\lambda_{5}^{2}
 \right. \nonumber \\
 &+& \left. \frac{3g_{1}^{4}}{4}+\frac{9g_{2}^{4}}{4}+\frac{3g_{1}^{2}g_{2}^{2}}{2}\right ] \, , \nonumber \\
\frac{d \lambda_2}{d \ln \mu^2}&=& \frac{1}{16 \pi^2} \left[ 4\lambda_2 \left (3y_{t}^{2}+6y_{T}^{2}+6y_{B}^{2}+6y_{Y}^{2}-\frac{3g_{1}^{2}}{2} - \frac{9g_{2}^{2}}{2}\right) +12\lambda_{2}^{2}+4\lambda_{3}^{2}+4\lambda_{3}\lambda_{4}+2\lambda_{4}^{2}+2\lambda_{5}^{2}
 \right. \nonumber \\
 &+& \left. \frac{3g_{1}^{4}}{2}+\frac{9g_{2}^{4}}{2}+3g_{1}^{2}g_{2}^{2}-12y_{t}^{4}-24y_{T}^{4}-24y_{B}^{4}-24y_{Y}^{4}-24y_{t}^{2}y_{T}^{2}-24y_{T}^{2}y_{B}^{2}-24y_{B}^{2}y_{Y}^{2}\right ] \, , \nonumber \\
 \frac{d \lambda_3}{d \ln \mu^2}&=& \frac{1}{16 \pi^2} \left[ 2\lambda_{3} \left (3y_{t}^{2}+6y_{T}^{2}+6y_{B}^{2}+6y_{Y}^{2}+12y_{M}^{2}-3g_{1}^{2} - 9g_{2}^{2}\right) +4\lambda_{3}^{2}+2\lambda_{4}^{2}+2\lambda_{5}^{2}+ (\lambda_{1}+\lambda_{2})(6\lambda_{3}+2\lambda_{4})
 \right. \nonumber \\
 &+& \left. \frac{3g_{1}^{4}}{2}+\frac{9g_{2}^{2}}{4}-3g_{1}^{2}g_{2}^{2}\right ] \, , \nonumber \\
 \frac{d \lambda_4}{d \ln \mu^2}&=& \frac{1}{16 \pi^2} \left[ 2\lambda_{4} \left (3y_{t}^{2}+6y_{T}^{2}+6y_{B}^{2}+6y_{Y}^{2}+12y_{M}^{2}-3g_{1}^{2} - 9g_{2}^{2}\right) +6g_{1}^{2}g_{2}^{2} +4\lambda_{4}^{2}+8\lambda_{5}^{2}+8\lambda_{3}\lambda_{4}+ 2\lambda_{4}(\lambda_{1}+\lambda_{2})\right ] \, , \nonumber \\
\frac{d \lambda_5}{d \ln \mu^2}&=& \frac{1}{16 \pi^2} \left[ 2\lambda_{5} \left (3y_{t}^{2}+6y_{T}^{2}+6y_{B}^{2}+6y_{Y}^{2}+12y_{M}^{2}-3g_{1}^{2}- 9g_{2}^{2}\right) +2\lambda_{5}(\lambda_{1}+\lambda_{2}+4\lambda_{3}+6\lambda_{4})\right ]  
\end{eqnarray}
The coupling constants gain additional terms due to the new fermions in both triplet models  as follows:
\begin{eqnarray}
\frac{dg_1^2}{d \ln \mu^2}&=&\frac{g_1^4}{16 \pi^2}\left(7+\frac{4}{5}\right )\, , \nonumber \\
\frac{dg_2^2}{d \ln \mu^2}&=&\frac{g_2^4}{16 \pi^2}\left( -3 +4\right ) \, , \nonumber \\
\frac{dg_3^2}{d \ln \mu^2}&=&\frac{g_3^4}{16 \pi^2}\left(-7+2\right )\, .
\end{eqnarray}

 \subsection{RGEs for 2HDM + VLQ - Type II}
 \label{sec:apprge2}

\subsubsection{Singlet ${\cal U}_1~(T)$, $Y=2/3$}
The relevant RGE for the Yukawa couplings are
\begin{eqnarray}
\label{eq:rgeTsingletfermiont1}
\frac{dy_t^2}{d \ln \mu^2}&=& \frac{y_t^2}{16 \pi^2}\left (\frac{9y_t^2}{2}+\frac{9y_T^2}{2}-\frac{17g_1^2}{12}-\frac{9g_2^2}{4}- 8g_3^2 \right)\, ,\nonumber\\
\frac{dy_T^2}{d \ln \mu^2}&=& \frac{y_T^2}{16 \pi^2}\left (\frac{9y_t^2}{2}+ \frac{9y_T^2}{2}+\frac{3y_M^2}{2}-\frac{17g_1^2}{12}-\frac{9g_2^2}{4}- 8g_3^2 \right)\, ,\nonumber\\
\frac{dy_M^2}{d \ln \mu^2}&=& \frac{y_M^2}{16 \pi^2}\left (y_T^2+\frac{9y_M^2}{2}-\frac{41g_1^2}{20}-8g_3^2\right).
\end{eqnarray}
The Higgs sector RGEs, describing the interactions between the two bosons are:
\begin{eqnarray}
\label{eq:rgeTsingletscalart1}
\frac{d \lambda_1}{d \ln \mu^2}&=& \frac{1}{16 \pi^2} \left[ -4\lambda_1 \left (\frac{3g_{1}^{2}}{4} + \frac{9g_{2}^{2}}{4}\right) +12\lambda_{1}^{2}+4\lambda_{3}^{2}+4\lambda_{3}\lambda_{4}+2\lambda_{4}^{2}+2\lambda_{5}^{2}
 \right. \nonumber \\
 &+& \left. \frac{3g_{1}^{4}}{16}+\frac{9g_{2}^{4}}{4}+\frac{3g_{1}^{2}g_{2}^{2}}{2}\right ] \, , \nonumber \\
\frac{d \lambda_2}{d \ln \mu^2}&=& \frac{1}{16 \pi^2} \left[ 4\lambda_2 \left (3y_{t}^{2}+6y_{T}^{2}-\frac{3g_{1}^{2}}{4} - \frac{9g_{2}^{2}}{4}\right) +12\lambda_{2}^{2}+4\lambda_{3}^{2}+4\lambda_{3}\lambda_{4}+2\lambda_{4}^{2}+2\lambda_{5}^{2}
 \right. \nonumber \\
 &+& \left. \frac{3g_{1}^{4}}{16}+\frac{9g_{2}^{4}}{4}+\frac{3g_{1}^{2}g_{2}^{2}}{2}-12y_{t}^{4}-24y_{T}^{4}-24y_{M}^{4}-24y_{t}^{2}y_{T}^{2}\right ] \, , \nonumber \\
 \frac{d \lambda_3}{d \ln \mu^2}&=& \frac{1}{16 \pi^2} \left[ 2\lambda_{3} \left (3y_{t}^{2}+6y_{T}^{2}+6y_{M}^{2}-\frac{3g_{1}^{2}}{2} - \frac{9g_{2}^{2}}{2}\right) +4\lambda_{3}^{2}+2\lambda_{4}^{2}+2\lambda_{5}^{2}+ (\lambda_{1}+\lambda_{2})(6\lambda_{3}+2\lambda_{4})
 \right. \nonumber \\
 &+& \left. \frac{3g_{1}^{4}}{4}+\frac{9g_{2}^{4}}{4}-\frac{3g_{1}^{2}g_{2}^{2}}{2}-24y_{T}^{2}y_{M}^{2}\right ] \, , \nonumber \\
 \frac{d \lambda_4}{d \ln \mu^2}&=& \frac{1}{16 \pi^2} \left[ 2\lambda_{4} \left (3y_{t}^{2}+6y_{T}^{2}+6y_{M}^{2}-\frac{3g_{1}^{2}}{2} - \frac{9g_{2}^{2}}{2}\right) +3g_{1}^{2}g_{2}^{2} +4\lambda_{4}^{2}+8\lambda_{5}^{2}+8\lambda_{3}\lambda_{4}+ 2\lambda_{4}(\lambda_{1}+\lambda_{2})+24y_{T}^{2}y_{M}^{2} \right ] \, , \nonumber \\
\frac{d \lambda_5}{d \ln \mu^2}&=& \frac{1}{16 \pi^2} \left[ 2\lambda_{5} \left (3y_{t}^{2}+6y_{T}^{2}+6y_{M}^{2}-\frac{3g_{1}^{2}}{2} - \frac{9g_{2}^{2}}{2}\right) +2\lambda_{5}(\lambda_{1}+\lambda_{2}+4\lambda_{3}+6\lambda_{4})\right ] 
 \end{eqnarray}
\subsubsection{Singlet  ${\cal D}_1~(B)$, $Y=-1/3$}
The relevant RGE for the Yukawa couplings are
\begin{eqnarray}
\label{eq:rgeBsingletfermiont1}
\frac{dy_t^2}{d \ln \mu^2}&=& \frac{y_t^2}{16 \pi^2}\left (\frac{9y_t^2}{2}+\frac{y_B^2}{2}-\frac{17g_1^2}{12}-\frac{9g_2^2}{4}- 8g_3^2 \right)\, ,\nonumber\\
\frac{dy_B^2}{d \ln \mu^2}&=& \frac{y_B^2}{16 \pi^2}\left (\frac{y_t^2}{2}+ \frac{9y_B^2}{2}+\frac{3y_M^2}{2}-\frac{5g_1^2}{12}-\frac{9g_2^2}{4}- 8g_3^2 \right)\, ,\nonumber\\
\frac{dy_M^2}{d \ln \mu^2}&=& \frac{y_M^2}{16 \pi^2}\left (\frac{9y_M^2}{2}+y_{B}^{2}-\frac{17g_1^2}{20}-8g_3^2\right).
\end{eqnarray}
\newpage
The Higgs sector RGEs, describing the interactions between the two bosons are:
\begin{eqnarray}
\label{eq:rgeBsingletscalart1}
\frac{d \lambda_1}{d \ln \mu^2}&=& \frac{1}{16 \pi^2} \left[ -4\lambda_1 \left (\frac{3g_{1}^{2}}{4} + \frac{9g_{2}^{2}}{4}-6y_{B}^{2}\right) +12\lambda_{1}^{2}+4\lambda_{3}^{2}+4\lambda_{3}\lambda_{4}+2\lambda_{4}^{2}+2\lambda_{5}^{2}
 \right. \nonumber \\
 &+& \left. \frac{3g_{1}^{4}}{16}+\frac{9g_{2}^{4}}{4}+\frac{3g_{1}^{2}g_{2}^{2}}{2}-24y_{B}^{4}-24y_{M}^{4} \right ] \, , \nonumber \\
\frac{d \lambda_2}{d \ln \mu^2}&=& \frac{1}{16 \pi^2} \left[ 4\lambda_2 \left (3y_{t}^{2}-\frac{3g_{1}^{2}}{4} - \frac{9g_{2}^{2}}{4}\right) +12\lambda_{2}^{2}+4\lambda_{3}^{2}+4\lambda_{3}\lambda_{4}+2\lambda_{4}^{2}+2\lambda_{5}^{2}
 \right. \nonumber \\
 &+& \left. \frac{3g_{1}^{4}}{16}+\frac{9g_{2}^{4}}{4}+\frac{3g_{1}^{2}g_{2}^{2}}{2}-12y_{t}^{4}\right ] \, , \nonumber \\
 \frac{d \lambda_3}{d \ln \mu^2}&=& \frac{1}{16 \pi^2} \left[ 2\lambda_{3} \left (3y_{t}^{2}+6y_{B}^{2}+6y_{M}^{2}-\frac{3g_{1}^{2}}{2} - \frac{9g_{2}^{2}}{2}\right) +4\lambda_{3}^{2}+2\lambda_{4}^{2}+2\lambda_{5}^{2}+ (\lambda_{1}+\lambda_{2})(6\lambda_{3}+2\lambda_{4})
 \right. \nonumber \\
 &+& \left. \frac{3g_{1}^{4}}{4}+\frac{9g_{2}^{4}}{4}-\frac{3g_{1}^{2}g_{2}^{2}}{2}-24y_{B}^{2}y_{M}^{2} \right ] \, , \nonumber \\
 \frac{d \lambda_4}{d \ln \mu^2}&=& \frac{1}{16 \pi^2} \left[ 2\lambda_{4} \left (3y_{t}^{2}+6y_{B}^{2}+6y_{M}^{2}-\frac{3g_{1}^{2}}{2} - \frac{9g_{2}^{2}}{2}\right) +3g_{1}^{2}g_{2}^{2} +4\lambda_{4}^{2}+8\lambda_{5}^{2}+8\lambda_{3}\lambda_{4}+ 2\lambda_{4}(\lambda_{1}+\lambda_{2})+24y_{B}^{2}y_{M}^{2} \right ] \, , \nonumber \\
\frac{d \lambda_5}{d \ln \mu^2}&=& \frac{1}{16 \pi^2} \left[ 2\lambda_{5} \left (3y_{t}^{2}+6y_{B}^{2}+6y_{M}^{2}-\frac{3g_{1}^{2}}{2} - \frac{9g_{2}^{2}}{2}\right) +2\lambda_{5}(\lambda_{1}+\lambda_{2}+4\lambda_{3}+6\lambda_{4})\right ] 
\end{eqnarray}
Finally the coupling constants gain additional terms due to the new fermion, for both models ${\cal U}_1,~{\cal D}_1$ with singlet fermions as follows:
\begin{eqnarray}
\frac{dg_1^2}{d \ln \mu^2}&=&\frac{g_1^4}{16 \pi^2}\left(7+\frac{4}{15}\right )\, , \nonumber \\
\frac{dg_2^2}{d \ln \mu^2}&=&\frac{g_2^4}{16 \pi^2}\left( -3 \right ) \, , \nonumber \\
\frac{dg_3^2}{d \ln \mu^2}&=&\frac{g_3^4}{16 \pi^2}\left(-7+\frac{2}{3}\right )\, .
\end{eqnarray}
\subsubsection{Doublet ${\cal D}_2$ $(T,B), \,Y=1/6$}
The relevant RGE for the Yukawa couplings are
\begin{eqnarray}
\label{eq:rgeTBdoubletfermiont1}
\frac{dy_t^2}{d \ln \mu^2}&=& \frac{y_t^2}{16 \pi^2}\left (\frac{9y_t^2}{2}+\frac{9y_T^2}{2}+\frac{y_B^2}{2}+3y_{M}^{2}-\frac{17g_{1}^{2}}{12}-\frac{9g_2^2}{4}- 8g_3^2 \right)\, ,\nonumber\\
\frac{dy_T^2}{d \ln \mu^2}&=& \frac{y_T^2}{16 \pi^2}\left (\frac{9y_t^2}{2}+ \frac{9y_T^2}{2}+\frac{y_B^2}{2}+3y_{M}^{2}-\frac{17g_{1}^{2}}{12}-\frac{9g_2^2}{4}- 8g_3^2 \right)\, ,\nonumber\\
\frac{dy_B^2}{d \ln \mu^2}&=& \frac{y_B^2}{16 \pi^2}\left (\frac{3y_t^2}{2}+ \frac{y_T^2}{2}+\frac{9y_B^2}{2}+3y_{M}^{2}-\frac{5g_{1}^{2}}{12}-\frac{9g_2^2}{4}- 8g_3^2 \right)\, ,\nonumber\\
\frac{dy_M^2}{d \ln \mu^2}&=& \frac{y_M^2}{16 \pi^2}\left (y_T^2+y_B^2+\frac{11y_M^2}{2}-\frac{19g_{1}^{2}}{40}-\frac{9g_{2}^{2}}{2}-8g_3^2\right).
\end{eqnarray}
The Higgs sector RGEs, describing the interactions between the two bosons are:
\begin{eqnarray}
\label{eq:rgeTBdoubletTBscalart1}
\frac{d \lambda_1}{d \ln \mu^2}&=& \frac{1}{16 \pi^2} \left[4\lambda_1 \left (6y_{B}^{2}-\frac{3g_{1}^{2}}{4} - \frac{9g_{2}^{2}}{4}\right) +12\lambda_{1}^{2}+4\lambda_{3}^{2}+4\lambda_{3}\lambda_{4}+2\lambda_{4}^{2}+2\lambda_{5}^{2}
 \right. \nonumber \\
 &+& \left. \frac{3g_{1}^{4}}{4}+\frac{9g_{2}^{4}}{4}+\frac{3g_{1}^{2}g_{2}^{2}}{2}-24y_{B}^{4}-24y_{B}^{2}y_{M}^{2}\right ] \, , \nonumber \\
\frac{d \lambda_2}{d \ln \mu^2}&=& \frac{1}{16 \pi^2} \left[ 4\lambda_2 \left (3y_{t}^{2}+6y_{T}^{2}-\frac{3g_{1}^{2}}{4} - \frac{9g_{2}^{2}}{4}\right) +12\lambda_{2}^{2}+4\lambda_{3}^{2}+4\lambda_{3}\lambda_{4}+2\lambda_{4}^{2}+2\lambda_{5}^{2}
 \right. \nonumber \\
 &+& \left. \frac{3g_{1}^{4}}{4}+\frac{9g_{2}^{4}}{4}+\frac{3g_{1}^{2}g_{2}^{2}}{2}-12y_{t}^{4}-24y_{T}^{4}-24y_{t}^{2}y_{T}^{2}-24y_{T}^{2}y_{M}^{2}\right ] \, , \nonumber \\
 \frac{d \lambda_3}{d \ln \mu^2}&=& \frac{1}{16 \pi^2} \left[ 2\lambda_{3} \left (3y_{t}^{2}+6y_{T}^{2}+6y_{B}^{2}+6y_{M}^{2}-\frac{3g_{1}^{2}}{2} - \frac{9g_{2}^{2}}{2}\right) +4\lambda_{3}^{2}+2\lambda_{4}^{2}+2\lambda_{5}^{2}+ (\lambda_{1}+\lambda_{2})(6\lambda_{3}+2\lambda_{4})
 \right. \nonumber \\
 &+& \left. \frac{3g_{1}^{4}}{4}+\frac{9g_{2}^{4}}{4}-\frac{3g_{1}^{2}g_{2}^{2}}{2}-24y_{T}^{2}y_{M}^{2}-24y_{B}^{2}y_{M}^{2}\right ] \, , \nonumber \\
 \frac{d \lambda_4}{d \ln \mu^2}&=& \frac{1}{16 \pi^2} \left[ 2\lambda_{4} \left (3y_{t}^{2}+6y_{T}^{2}+6y_{B}^{2}+6y_{M}^{2}-\frac{3g_{1}^{2}}{2} - \frac{9g_{2}^{2}}{2}\right) +3g_{1}^{2}g_{2}^{2} +4\lambda_{4}^{2}+8\lambda_{5}^{2}+8\lambda_{3}\lambda_{4}+ 2\lambda_{4}(\lambda_{1}+\lambda_{2})
 \right. \nonumber \\
&+& \left. 24y_{T}^{2}y_{M}^{2}+24y_{B}^{2}y_{M}^{2} \right ] \, , \nonumber \\
\frac{d \lambda_5}{d \ln \mu^2}&=& \frac{1}{16 \pi^2} \left[ 2\lambda_{5} \left (3y_{t}^{2}+6y_{T}^{2}+6y_{B}^{2}+12y_{M}^{2}-\frac{3g_{1}^{2}}{2} - \frac{9g_{2}^{2}}{2}\right) +2\lambda_{5}(\lambda_{1}+\lambda_{2}+4\lambda_{3}+6\lambda_{4})\right ] 
 \end{eqnarray}

\subsubsection{Doublet ${\cal D}_X$ $(X,T), \,Y=7/6$}
The relevant RGE for the Yukawa couplings are
\begin{eqnarray}
\label{eq:rgeXTdoubletfermiont1}
\frac{dy_t^2}{d \ln \mu^2}&=& \frac{y_t^2}{16 \pi^2}\left (\frac{9y_t^2}{2}+\frac{y_T^2}{2}+\frac{9y_X^2}{2}+3y_{M}^{2}-\frac{17g_1^2}{12}-\frac{9g_2^2}{4}- 8g_3^2 \right)\, ,\nonumber\\
\frac{dy_T^2}{d \ln \mu^2}&=& \frac{y_T^2}{16 \pi^2}\left (\frac{3y_t^2}{2}+\frac{y_X^2}{2}+ \frac{9y_T^2}{2}+3y_{M}^{2}-\frac{5g_1^2}{12}-\frac{9g_2^2}{4}- 8g_3^2 \right)\, ,\nonumber\\
\frac{dy_X^2}{d \ln \mu^2}&=& \frac{y_X^2}{16 \pi^2}\left (\frac{9y_t^2}{2}+ \frac{9y_X^2}{2}+ \frac{y_T^2}{2}+3y_M^{2}-\frac{17g_1^2}{12}-\frac{9g_2^2}{4}- 8g_3^2 \right)\, ,\nonumber\\
\frac{dy_M^2}{d \ln \mu^2}&=& \frac{y_M^2}{16 \pi^2}\left (y_T^2+y_X^2+\frac{11y_M^2}{2}-\frac{67g_1^2}{40}-\frac{9g_2^2}{2}-8g_3^2\right).
\end{eqnarray}
The Higgs sector RGEs, describing the interactions between the two bosons are:
\begin{eqnarray}
\label{eq:rgeXTdoubletscalart1}
\frac{d \lambda_1}{d \ln \mu^2}&=& \frac{1}{16 \pi^2} \left[4\lambda_1 \left (6y_{T}^{2}-\frac{3g_{1}^{2}}{4} - \frac{9g_{2}^{2}}{2}\right) +12\lambda_{1}^{2}+4\lambda_{3}^{2}+4\lambda_{3}\lambda_{4}+2\lambda_{4}^{2}+2\lambda_{5}^{2}
 \right. \nonumber \\
 &+& \left. \frac{3g_{1}^{4}}{4}+\frac{9g_{2}^{4}}{4}+\frac{3g_{1}^{2}g_{2}^{2}}{2}-24y_{T}^{4}-24y_{T}^{2}y_{M}^{2}\right ] \, , \nonumber \\
\frac{d \lambda_2}{d \ln \mu^2}&=& \frac{1}{16 \pi^2} \left[ 4\lambda_2 \left (3y_{t}^{2}+6y_{X}^{2}-\frac{3g_{1}^{2}}{4} - \frac{9g_{2}^{2}}{4}\right) +12\lambda_{2}^{2}+4\lambda_{3}^{2}+4\lambda_{3}\lambda_{4}+2\lambda_{4}^{2}+2\lambda_{5}^{2}
 \right. \nonumber \\
 &+& \left. \frac{3g_{1}^{4}}{4}+\frac{9g_{2}^{4}}{4}+\frac{3g_{1}^{2}g_{2}^{2}}{2}-12y_{t}^{4}-24y_{X}^{4}-24y_{X}^{2}y_{M}^{2}\right ] \, , \nonumber \\
 \frac{d \lambda_3}{d \ln \mu^2}&=& \frac{1}{16 \pi^2} \left[ 2\lambda_{3} \left (3y_{t}^{2}+6y_{T}^{2}+6y_{X}^{2}+6y_{M}^{2}-\frac{3g_{1}^{2}}{2} - \frac{9g_{2}^{2}}{10}\right) +4\lambda_{3}^{2}+2\lambda_{4}^{2}+2\lambda_{5}^{2}+ (\lambda_{1}+\lambda_{2})(6\lambda_{3}+2\lambda_{4})
 \right. \nonumber \\
 &+& \left. \frac{3g_{1}^{4}}{4}+\frac{9g_{2}^{4}}{4}-\frac{3g_{1}^{2}g_{2}^{2}}{2}-24y_{X}^{2}y_{M}^{2}-24y_{T}^{2}y_{M}^{2}\right ] \, , \nonumber \\
 \frac{d \lambda_4}{d \ln \mu^2}&=& \frac{1}{16 \pi^2} \left[ 2\lambda_{4} \left (3y_{t}^{2}+6y_{T}^{2}+6y_{X}^{2}+6y_{M}^{2}-\frac{3g_{1}^{2}}{2} - \frac{9g_{2}^{2}}{10}\right) +3g_{1}^{2}g_{2}^{2} +4\lambda_{4}^{2}+8\lambda_{5}^{2}+8\lambda_{3}\lambda_{4}+ 2\lambda_{4}(\lambda_{1}+\lambda_{2})
 \right. \nonumber \\
&+& \left. 24y_{X}^{2}y_{M}^{2}+24y_{T}^{2}y_{M}^{2}\right ] \, , \nonumber \\
\frac{d \lambda_5}{d \ln \mu^2}&=& \frac{1}{16 \pi^2} \left[ 2\lambda_{5} \left (3y_{t}^{2}+6y_{T}^{2}+6y_{X}^{2}+6y_{M}^{2}-\frac{3g_{1}^{2}}{2} - \frac{9g_{2}^{2}}{10}\right) +2\lambda_{5}(\lambda_{1}+\lambda_{2}+4\lambda_{3}+6\lambda_{4})\right ]  \end{eqnarray}

\subsubsection{Additional non SM-like quark doublet ${\cal D}_Y$ $(B,Y), \,Y=-5/6$}
The relevant RGE for the Yukawa couplings are
\begin{eqnarray}
\label{eq:rgeBYdoubletfermiont1}
\frac{dy_t^2}{d \ln \mu^2}&=& \frac{y_t^2}{16 \pi^2}\left (\frac{9y_t^2}{2}+\frac{3y_B^2}{2}+\frac{y_Y^2}{2}+3y_M^{2}-\frac{17g_1^2}{12}-\frac{9g_2^2}{4}- 8g_3^2 \right)\, ,\nonumber\\
\frac{dy_B^2}{d \ln \mu^2}&=& \frac{y_T^2}{16 \pi^2}\left (\frac{3y_t^2}{2}+\frac{9y_B^2}{2}+ \frac{y_Y^2}{2}+3y_M^{2}-\frac{17g_1^2}{12}-\frac{9g_2^2}{4}- 8g_3^2 \right)\, ,\nonumber\\
\frac{dy_Y^2}{d \ln \mu^2}&=& \frac{y_Y^2}{16 \pi^2}\left (\frac{3y_t^2}{2}+ \frac{9y_Y^2}{2}+ \frac{y_B^2}{2}+3y_M^{2}-\frac{5g_1^2}{12}-\frac{9g_2^2}{4}- 8g_3^2 \right)\, ,\nonumber\\
\frac{dy_M^2}{d \ln \mu^2}&=& \frac{y_M^2}{16 \pi^2}\left (y_Y^2+y_B^2+\frac{11y_M^2}{2}-\frac{43g_1^2}{40}-\frac{9g_{2}^{2}}{2}-8g_3^2\right).
\end{eqnarray}
The Higgs sector RGEs, describing the interactions between the two bosons are:
\begin{eqnarray}
\label{eq:rgeBYdoubletscalart1}
\frac{d \lambda_1}{d \ln \mu^2}&=& \frac{1}{16 \pi^2} \left[ 4\lambda_1 \left (6y_{Y}^{2}-\frac{3g_{1}^{2}}{4} - \frac{9g_{2}^{2}}{4}\right) +12\lambda_{1}^{2}+4\lambda_{3}^{2}+4\lambda_{3}\lambda_{4}+2\lambda_{4}^{2}+2\lambda_{5}^{2}
 \right. \nonumber \\
 &+& \left. \frac{3g_{1}^{4}}{4}+\frac{9g_{2}^{4}}{4}+\frac{3g_{1}^{2}g_{2}^{2}}{2}-24y_{Y}^{4}-24y_{Y}^{2}y_{M}^{2}\right ] \, , \nonumber \\
\frac{d \lambda_2}{d \ln \mu^2}&=& \frac{1}{16 \pi^2} \left[ 4\lambda_2 \left (3y_{t}^{2}+6y_{B}^{2}-\frac{3g_{1}^{2}}{4} - \frac{9g_{2}^{2}}{4}\right) +12\lambda_{2}^{2}+4\lambda_{3}^{2}+4\lambda_{3}\lambda_{4}+2\lambda_{4}^{2}+2\lambda_{5}^{2}
 \right. \nonumber \\
 &+& \left. \frac{3g_{1}^{4}}{4}+\frac{9g_{2}^{4}}{4}+\frac{3g_{1}^{2}g_{2}^{2}}{2}-12y_{t}^{4}-24y_{B}^{4}-24y_{B}^{2}y_{M}^{2}\right ] \, , \nonumber \\
 \frac{d \lambda_3}{d \ln \mu^2}&=& \frac{1}{16 \pi^2} \left[ 2\lambda_{3} \left (3y_{t}^{2}+6y_{B}^{2}+6y_{Y}^{2}+6y_{M}^{2}-\frac{3g_{1}^{2}}{2} - \frac{9g_{2}^{2}}{2}\right) +4\lambda_{3}^{2}+2\lambda_{4}^{2}+2\lambda_{5}^{2}+ (\lambda_{1}+\lambda_{2})(6\lambda_{3}+2\lambda_{4})
 \right. \nonumber \\
 &+& \left. \frac{3g_{1}^{4}}{4}+\frac{9g_{2}^{4}}{4}-\frac{3g_{1}^{2}g_{2}^{2}}{2}-24y_{B}^{2}y_{M}^{2}-24y_{Y}^{2}y_{M}^{2}\right ] \, , \nonumber \\
 \frac{d \lambda_4}{d \ln \mu^2}&=& \frac{1}{16 \pi^2} \left[ 2\lambda_{4} \left (3y_{t}^{2}+6y_{Y}^{2}+6y_{B}^{2}+6y_{M}^{2}-\frac{3g_{1}^{2}}{2} - \frac{9g_{2}^{2}}{2}\right) +3g_{1}^{2}g_{2}^{2} +4\lambda_{4}^{2}+8\lambda_{5}^{2}+8\lambda_{3}\lambda_{4}+ 2\lambda_{4}(\lambda_{1}+\lambda_{2}) \right. \nonumber \\
 &+& \left. 24y_{B}^{2}y_{M}^{2}+24y_{Y}^{2}y_{M}^{2}\right ] \, , \nonumber \\
\frac{d \lambda_5}{d \ln \mu^2}&=& \frac{1}{16 \pi^2} \left[ 2\lambda_{5} \left (3y_{t}^{2}+6y_{Y}^{2}+6y_{B}^{2}+6y_{M}^{2}-\frac{3g_{1}^{2}}{2} - \frac{9g_{2}^{2}}{2}\right) +2\lambda_{5}(\lambda_{1}+\lambda_{2}+4\lambda_{3}+6\lambda_{4})\right ]  \end{eqnarray}
The coupling constants gain additional terms due to the new fermion in all doublet models as follows:
\begin{eqnarray}
\frac{dg_1^2}{d \ln \mu^2}&=&\frac{g_1^4}{16 \pi^2}\left(7+\frac{6}{5}\right )\, , \nonumber \\
\frac{dg_2^2}{d \ln \mu^2}&=&\frac{g_2^4}{16 \pi^2}\left( -3 + 2 \right ) \, , \nonumber \\
\frac{dg_3^2}{d \ln \mu^2}&=&\frac{g_3^4}{16 \pi^2}\left(-7+\frac{4}{3}\right )\, .
\end{eqnarray}

\subsubsection{Triplet ${\cal T}_X$ $(X, T, B), \,Y=2/3$}
The relevant RGE for the Yukawa couplings are
\begin{eqnarray}
\label{eq:rgeXTBtripletfermiont1}
\frac{dy_t^2}{d \ln \mu^2}&=& \frac{y_t^2}{16 \pi^2}\left (\frac{9y_t^2}{2}+\frac{9y_X^2}{2}+\frac{9y_T^2}{2}+\frac{y_B^2}{2}+\frac{3y_M^2}{2}-\frac{17g_1^2}{12}-\frac{9g_2^2}{4}- 8g_3^2 \right)\, ,\nonumber\\
\frac{dy_T^2}{d \ln \mu^2}&=& \frac{y_T^2}{16 \pi^2}\left (\frac{9y_t^2}{2}+\frac{9y_X^2}{2}+ \frac{9y_T^2}{2}+\frac{y_B^2}{2}+\frac{3y_M^2}{2}-\frac{17g_1^2}{12}-\frac{9g_2^2}{4}- 8g_3^2 \right)\, ,\nonumber\\
\frac{dy_X^2}{d \ln \mu^2}&=& \frac{y_T^2}{16 \pi^2}\left (\frac{9y_t^2}{2}+ \frac{9y_X^2}{2}+ \frac{9y_T^2}{2}+\frac{y_B^2}{2}+\frac{3y_M^2}{2}-\frac{41g_1^2}{20}-\frac{9g_2^2}{4}- 8g_3^2 \right)\, ,\nonumber\\
\frac{dy_B^2}{d \ln \mu^2}&=& \frac{y_B^2}{16 \pi^2}\left (\frac{3y_t^2}{2}+\frac{y_X^2}{2}+ \frac{3y_T^2}{2}+ \frac{9y_B^2}{2}+\frac{3y_M^2}{2}-\frac{5g_1^2}{12}-\frac{9g_2^2}{4}- 8g_3^2 \right)\, ,\nonumber\\
\frac{dy_M^2}{d \ln \mu^2}&=& \frac{y_M^2}{16 \pi^2}\left (y_X^2+y_T^2+y_B^2+\frac{15y_M^2}{2}-\frac{41g_1^2}{20}-\frac{9g_2^2}{4}-8g_3^2\right).
\end{eqnarray}
The Higgs sector RGEs, describing the interactions between the two bosons are:
\begin{eqnarray}
\label{eq:rgeXTBtripletscalart1}
\frac{d \lambda_1}{d \ln \mu^2}&=& \frac{1}{16 \pi^2} \left[ 4\lambda_1 \left (6y_{B}^{2}+6y_{T}^{2}-\frac{3g_{1}^{2}}{2} - \frac{9g_{2}^{2}}{2}\right) +12\lambda_{1}^{2}+4\lambda_{3}^{2}+4\lambda_{3}\lambda_{4}+2\lambda_{4}^{2}+2\lambda_{5}^{2}
 \right. \nonumber \\
 &+& \left. \frac{3g_{1}^{4}}{2}+\frac{9g_{2}^{4}}{2}+3g_{1}^{2}g_{2}^{2}-24y_{B}^{4}-24y_{T}^{4}-24y_{B}^{2}y_{M}^{2}-24y_{T}^{2}y_{M}^{2}\right ] \, , \nonumber \\
\frac{d \lambda_2}{d \ln \mu^2}&=& \frac{1}{16 \pi^2} \left[ 4\lambda_2 \left (3y_{t}^{2}+6y_{X}^{2}+6y_{T}^{2}-\frac{3g_{1}^{2}}{2} - \frac{9g_{2}^{2}}{2}\right) +12\lambda_{2}^{2}+4\lambda_{3}^{2}+4\lambda_{3}\lambda_{4}+2\lambda_{4}^{2}+2\lambda_{5}^{2}
 \right. \nonumber \\
 &+& \left. \frac{3g_{1}^{4}}{2}+\frac{9g_{2}^{2}}{2}+3g_{1}^{2}g_{2}^{2}-12y_{t}^{4}-24y_{X}^{4}-24y_{T}^{4}-24y_{t}^{2}y_{T}^{2}-24y_{X}^{2}y_{M}^{2}-24y_{T}^{2}y_{M}^{2}\right ] \, , \nonumber \\
 \frac{d \lambda_3}{d \ln \mu^2}&=& \frac{1}{16 \pi^2} \left[ 2\lambda_{3} \left (3y_{t}^{2}+6y_{T}^{2}+6y_{X}^{2}+6y_{B}^{2}+6y_{M}^{2}-3g_{1}^{2} - 9g_{2}^{2}\right) +4\lambda_{3}^{2}+2\lambda_{4}^{2}+2\lambda_{5}^{2}+ (\lambda_{1}+\lambda_{2})(6\lambda_{3}+2\lambda_{4})
 \right. \nonumber \\
 &+& \left. \frac{3g_{1}^{4}}{2}+\frac{9g_{2}^{4}}{2}-3g_{1}^{2}g_{2}^{2}-24y_{X}^{2}y_{M}^{2}-24y_{T}^{2}y_{M}^{2}-24y_{B}^{2}y_{M}^{2}\right ] \, , \nonumber \\
 \frac{d \lambda_4}{d \ln \mu^2}&=& \frac{1}{16 \pi^2} \left[ 2\lambda_{4} \left (3y_{t}^{2}+6y_{X}^{2}+6y_{T}^{2}+6y_{B}^{2}+6y_{M}^{2}-3g_{1}^{2} - 9g_{2}^{2}\right) +6g_{1}^{2}g_{2}^{2} +4\lambda_{4}^{2}+8\lambda_{5}^{2}+8\lambda_{3}\lambda_{4}+ 2\lambda_{4}(\lambda_{1}+\lambda_{2}) \right. \nonumber \\
&+& \left. 24y_{X}^{2}y_{M}^{2}+24y_{T}^{2}y_{M}^{2}+24y_{B}^{2}y_{M}^{2} \right ] \, , \nonumber \\
\frac{d \lambda_5}{d \ln \mu^2}&=& \frac{1}{16 \pi^2} \left[ 2\lambda_{5} \left (3y_{t}^{2}+6y_{T}^{2}+6y_{X}^{2}+6y_{B}^{2}+6y_{M}^{2}-3g_{1}^{2} - 9g_{2}^{2}\right) +2\lambda_{5}(\lambda_{1}+\lambda_{2}+4\lambda_{3}+6\lambda_{4})\right ] 
 \end{eqnarray}

\subsubsection{Triplet ${\cal T}_Y$ $( T, B, Y), \, Y=-1/3$}
The relevant RGE for the Yukawa couplings are
\begin{eqnarray}
\label{eq:rgeTBYtripletfermiont1}
\frac{dy_t^2}{d \ln \mu^2}&=& \frac{y_t^2}{16 \pi^2}\left (\frac{9y_t^2}{2}+\frac{9y_T^2}{2}+\frac{3y_B^2}{2}+\frac{y_Y^2}{2}+3y_{M}^{2}-\frac{17g_1^2}{12}-\frac{9g_2^2}{4}- 8g_3^2 \right)\, ,\nonumber\\
\frac{dy_T^2}{d \ln \mu^2}&=& \frac{y_T^2}{16 \pi^2}\left (\frac{9y_t^2}{2}+\frac{9y_T^2}{2}+\frac{9y_B^2}{2}+ \frac{y_Y^2}{2}+3y_{M}^{2}-\frac{17g_1^2}{12}-\frac{9g_2^2}{4}- 8g_3^2 \right)\, ,\nonumber\\
\frac{dy_B^2}{d \ln \mu^2}&=& \frac{y_B^2}{16 \pi^2}\left (\frac{3y_t^2}{2}+\frac{y_T^2}{2}+\frac{9y_B^2}{2}+ \frac{9y_Y^2}{2}+\frac{3y_M^2}{2}-\frac{5g_1^2}{12}-\frac{9g_2^2}{4}- 8g_3^2 \right)\, ,\nonumber\\
\frac{dy_Y^2}{d \ln \mu^2}&=& \frac{y_Y^2}{16 \pi^2}\left (\frac{9y_Y^2}{2}+\frac{9y_B^2}{2}+ \frac{3y_t^2}{2}+ \frac{y_T^2}{2}+3y_{M}^{2}-\frac{17g_1^2}{20}-\frac{9g_2^2}{4}- 8g_3^2 \right)\, ,\nonumber\\
\frac{dy_M^2}{d \ln \mu^2}&=& \frac{y_M^2}{16 \pi^2}\left (y_T^2+y_B^2+y_Y^2+\frac{15y_M^2}{2}-\frac{17g_1^2}{20}-\frac{9g_2^2}{4}-8g_3^2\right).
\end{eqnarray}
The Higgs sector RGEs, describing the interactions between the two bosons are:
\begin{eqnarray}
\label{eq:rgeTBYtripletscalart1}
\frac{d \lambda_1}{d \ln \mu^2}&=& \frac{1}{16 \pi^2} \left[ 4\lambda_1 \left (6y_{B}^{2}+6y_{Y}^{2}-\frac{3g_{1}^{2}}{2} - \frac{9g_{2}^{2}}{2}\right) +12\lambda_{1}^{2}+4\lambda_{3}^{2}+4\lambda_{3}\lambda_{4}+2\lambda_{4}^{2}+2\lambda_{5}^{2}
 \right. \nonumber \\
 &+& \left. \frac{3g_{1}^{4}}{2}+\frac{9g_{2}^{4}}{2}+3g_{1}^{2}g_{2}^{2}-24y_{B}^{4}-24y_{Y}^{4}-24y_{B}^{2}y_{M}^{2}-24y_{Y}^{2}y_{M}^{2}\right ] \, , \nonumber \\
\frac{d \lambda_2}{d \ln \mu^2}&=& \frac{1}{16 \pi^2} \left[ 4\lambda_2 \left (3y_{t}^{2}+6y_{T}^{2}+6y_{B}^{2}-\frac{3g_{1}^{2}}{2} -\frac{9g_{2}^{2}}{2}\right) +12\lambda_{2}^{2}+4\lambda_{3}^{2}+4\lambda_{3}\lambda_{4}+2\lambda_{4}^{2}+2\lambda_{5}^{2}
 \right. \nonumber \\
 &+& \left. \frac{3g_{1}^{4}}{2}+\frac{9g_{2}^{4}}{2}+3g_{1}^{2}g_{2}^{2}-12y_{t}^{4}-24y_{T}^{4}-24y_{B}^{4}-24y_{t}^{2}y_{T}^{2}-24y_{T}^{2}y_{M}^{2}-24y_{B}^{2}y_{M}^{2}\right ] \, , \nonumber \\
 \frac{d \lambda_3}{d \ln \mu^2}&=& \frac{1}{16 \pi^2} \left[ 2\lambda_{3} \left (3y_{t}^{2}+6y_{T}^{2}+6y_{B}^{2}+6y_{Y}^{2}+6y_{M}^{2}-3g_{1}^{2} - 9g_{2}^{2}\right) +4\lambda_{3}^{2}+2\lambda_{4}^{2}+2\lambda_{5}^{2}+ (\lambda_{1}+\lambda_{2})(6\lambda_{3}+2\lambda_{4})
 \right. \nonumber \\
 &+& \left. \frac{3g_{1}^{4}}{2}+\frac{9g_{2}^{2}}{2}-3g_{1}^{2}g_{2}^{2}-24y_{T}^{2}y_{M}^{2}-24y_{B}^{2}y_{M}^{2}-24y_{Y}^{2}y_{M}^{2}\right ] \, , \nonumber \\
 \frac{d \lambda_4}{d \ln \mu^2}&=& \frac{1}{16 \pi^2} \left[ 2\lambda_{4} \left (3y_{t}^{2}+6y_{T}^{2}+6y_{B}^{2}+6y_{Y}^{2}+6y_{M}^{2}-3g_{1}^{2} - 9g_{2}^{2}\right) +6g_{1}^{2}g_{2}^{2} +4\lambda_{4}^{2}+8\lambda_{5}^{2}+8\lambda_{3}\lambda_{4}+ 2\lambda_{4}(\lambda_{1}+\lambda_{2})  \right. \nonumber \\
 &+& \left.24y_{T}^{2}y_{M}^{2}+24y_{B}^{2}y_{M}^{2}+24y_{Y}^{2}y_{M}^{2}  \right ] \, , \nonumber \\
\frac{d \lambda_5}{d \ln \mu^2}&=& \frac{1}{16 \pi^2} \left[ 2\lambda_{5} \left (3y_{t}^{2}+6y_{T}^{2}+6y_{B}^{2}+6y_{Y}^{2}+6y_{M}^{2}-3g_{1}^{2}- 9g_{2}^{2}\right) +2\lambda_{5}(\lambda_{1}+\lambda_{2}+4\lambda_{3}+6\lambda_{4})\right ]  
\end{eqnarray}
The coupling constants gain additional terms due to the new fermions in both triplet models  as follows:
\begin{eqnarray}
\frac{dg_1^2}{d \ln \mu^2}&=&\frac{g_1^4}{16 \pi^2}\left(7+\frac{4}{5}\right )\, , \nonumber \\
\frac{dg_2^2}{d \ln \mu^2}&=&\frac{g_2^4}{16 \pi^2}\left( -3 +4\right ) \, , \nonumber \\
\frac{dg_3^2}{d \ln \mu^2}&=&\frac{g_3^4}{16 \pi^2}\left(-7+2\right )\, .
\end{eqnarray}


\subsection{Approximate contributions of VLQs to the $\mathbb{S}$ and $\mathbb{T}$ parameters in VLQ models}
\label{sec:STapextypes}
\subsubsection{Singlet ${\cal U}_1~(T)$, $Y=2/3$}
\label{eq:STsingletUt1}
\begin{eqnarray}
\Delta \mathbb{T}&\simeq&\frac{m_{t}^{2}N_{c}(s_{L}^{t})^{2}}{16\pi c_{W}^{2}s_{W}^{2}M_{Z}^{2}}\left[(x_{T}^{2}(s_{L}^{t})^{2} - (c_{L}^{t})^{2}-1+4(c_{L}^{t})^{2}\frac{m_{T}^{2}}{m_{T}^{2}-m_{t}^{2}}\ln(x_{T}) \right], \nonumber \\
\Delta \mathbb{S}&\simeq&\frac{N_{c}(s_{L}^{t})^{2}}{18\pi}\left(\frac{(c_{L}^{t})^{2}}{(x_{T}-1)^{3}} \left[2\ln(x_{T})(3-9x_{T}^{2}-9x_{T}^{4}+3x_{T}^{6}) +5-27x_{T}^{2}-27x_{T}^{4}-5x_{T}^{6}\right]  \right. \nonumber \\
 &-& \left. 2\ln(x_{T}) \right)
\end{eqnarray}

\subsubsection{Singlet ${\cal D}_1~(B)$, $Y=-1/3$}
\label{eq:STsingletDt1}
\begin{eqnarray}
\Delta \mathbb{T}&\simeq&\frac{m_{t}^{2}N_{c}x_{B}}{16\pi c_{W}^{2}s_{W}^{2}M_{Z}^{2}(x_{B}-1)}\left[(s_{L}^{b})^{4}(x_{B}-1) -2(s_{L}^{b})^{2}\ln(x_{B})  \right], \nonumber \\
\Delta \mathbb{S}&\simeq&\frac{N_{c}(s_{L}^{b})^{2}}{18\pi}\left[2\ln(\frac{m_{b}}{m_{B}})(3(s_{L}^{b})^{2}-4)-5(c_{L}^{b})^{2} \right]
\label{eq:TSD1}
\end{eqnarray}

\subsubsection{Doublet ${\cal D}_2$ $(T,B), \,Y=1/6$}
\label{eq:STdoubletTBt1}
\begin{eqnarray}
\Delta \mathbb{T}&\simeq&\frac{m_{t}^{2}N_{c}(s_{R}^{t})^{2}}{8\pi c_{W}^{2}s_{W}^{2}M_{Z}^{2}}\left[2\ln(x_{T})-2 \right], \nonumber \\
\Delta \mathbb{S}&\simeq&\frac{N_{c}}{18\pi}\left[ 8s_R^{t^{2}}\ln(x_T)-7s_R^{t^{2}}-2\ln(x_b)-3 \right]
\end{eqnarray}

\subsubsection{Doublet ${\cal D}_X$ $(X,T), \,Y=7/6$}
\label{eq:STdoubletXTt1}
\begin{eqnarray}
\Delta \mathbb{T}&\simeq&\frac{m_{t}^{2}N_{c}(s_{R}^{t})^{2}}{16\pi c_{W}^{2}s_{W}^{2}M_{Z}^{2}}\left[-8\ln(x_T)+6+\frac{4}{3}s_R^{t^{2}}x_T^2-s_R^{t^{-2}} \right], \nonumber \\
\Delta \mathbb{S}&\simeq&\frac{N_{c}}{18\pi}\left[-8s_R^{t^{2}}\ln(x_T)+15s_R^{t^{2}}-2\ln(x_b)-3\right]
\label{eq:TSDX}
\end{eqnarray}

\subsubsection{Doublet ${\cal D}_Y$ $(B,Y), \,Y=-5/6$}
\label{eq:STdoubletBYt1}
\begin{eqnarray}
\Delta \mathbb{T}&\simeq&\frac{m_{t}^{2}N_{c}x_{B}}{128\pi c_{W}^{2}s_{W}^{2}M_{Z}^{2}}\left[-16c_{R}^{b}\left(-3+c_{R}^{2b}\cot_{R}^{b}\ln(c_{R}^{b}) \right) + s_{R}^{b}\left(-13-20c_{R}^{2b}+4c_{R}^{2b} \right) \right], \nonumber \\
\Delta S&\simeq&\frac{N_{c}}{18\pi}\left[-4s_R^{b^{2}}\ln(\frac{m_B}{m_b})+11s_R^{b^{2}}-2\ln(x_b)-3\right]
\end{eqnarray}

\subsubsection{Triplet ${\cal T}_X$ $(X, T, B), \,Y=2/3$}
\label{eq:STtripletXTBt1}
\begin{eqnarray}
\Delta  \mathbb{T}&\simeq&\frac{m_{t}^{2}N_{c}(s_{L}^{t})^{2}}{16\pi c_{W}^{2}s_{W}^{2}M_{Z}^{2}}\left[12\ln(x_{T})-10+\frac{19}{3}(s_{L}^{t})^2x_T^2-(s_{L}^{t})^{-2} \right], \nonumber \\
\Delta S&\simeq&\frac{N_{c}}{18\pi}\left\{(s_{L}^{t})^{2}\left[4\ln(x_T)-16\ln(\frac{m_B}{m_b})+13\right] +32\ln(\frac{m_B}{m_T})-2\ln(x_b)-3 \right\}
\end{eqnarray}

\subsubsection{Triplet ${\cal T}_Y$ $( T, B, Y), \, Y=-1/3$}
\label{eq:STtripletTBYt1}
\begin{eqnarray}
\Delta  \mathbb{T}&\simeq&\frac{m_{t}^{2}N_{c}(s_{L}^{t})^{2}}{16\pi c_{W}^{2}s_{W}^{2}M_{Z}^{2}}\left[-\ln(x_{T})+6+\frac{19}{12}(s_{L}^{t})^2x_T^2-(s_{L}^{t})^{-2} \right], \nonumber \\
\Delta S&\simeq&\frac{N_{c}}{18\pi}\left\{(s_{L}^{t})^{2}\left[4\ln(\frac{m_B}{m_b})-4\ln(x_T)+\frac{13}{2}\right]-2\ln(x_b)-3 \right\}
\end{eqnarray}
where $x_{i}=\frac{m_F}{m_t}$ for all representations.

\subsection{Electroweak couplings of VLQ and the SM quarks}
\label{sec:EWcouplings}
\subsubsection{Singlet ${\cal U}_1~(T)$, $Y=2/3$}
\label{eq:EWCsingletUt2}
\begin{align}
\Omega^L_{Wtb}&=\frac{ec_L^t}{\sqrt{2}s_W}           &  \Omega^L_{Ztt} &=\frac{e}{2s_Wc_W}(c_L^{t^2}-\frac{4}{3}s_W^2)          \nonumber   \\
\Omega^R_{Wtb}&=0         &  \Omega^R_{Ztt}&=-\frac{2es_W}{3c_W}  \nonumber  \\
\Omega^L_{WTb}&=\frac{es_L^t}{\sqrt{2}s_W}  &  \Omega^L_{ZTT}&=\frac{e}{2s_Wc_W}(s_L^{t^2}-\frac{4}{3}s_W^2)    \nonumber \\
\Omega^R_{WTb}&=0 &      \Omega^R_{ZTT}&=-\frac{2es_W}{3c_W}  \nonumber  \\
&& \Omega^L_{Zbb} &=\frac{e}{2s_Wc_W}(\frac{2}{3}s_W^2-1) \nonumber \\
&& \Omega^R_{Zbb} &=\frac{es_W}{3c_W} \nonumber \\
&& \Omega^L_{ZtT} &=\frac{es_L^tc_L^t}{2s_Wc_W} \nonumber \\
&& \Omega^R_{ZtT} &=0
\end{align}

\subsubsection{Singlet ${\cal D}_1~(B)$, $Y=-1/3$}
\label{eq:EWCsingletDt2}
\begin{align}
\Omega^L_{Wtb}&=\frac{ec_L^b}{\sqrt{2}s_W}           &  \Omega^L_{Ztt} &=\frac{e}{2s_Wc_W}(1-\frac{4}{3}s_W^2)          \nonumber   \\
\Omega^R_{Wtb}&=0         &  \Omega^R_{Ztt}&=-\frac{2es_W}{3c_W}  \nonumber  \\
\Omega^L_{WtB}&=\frac{es_L^b}{\sqrt{2}s_W}  &  \Omega^L_{Zbb}&=\frac{e}{2s_Wc_W}(\frac{2}{3}s_W^2-c_L^{b^2})    \nonumber \\
\Omega^R_{WtB}&=0 &      \Omega^R_{Zbb}&=\frac{es_W}{3c_W}  \nonumber  \\
&& \Omega^L_{ZBB} &=\frac{e}{2s_Wc_W}(\frac{2}{3}s_W^2-s_L^{b^2}) \nonumber \\
&& \Omega^R_{ZBB} &=\frac{es_W}{3c_W} \nonumber \\
&& \Omega^L_{ZbB} &=-\frac{es_L^bc_L^b}{2s_Wc_W} \nonumber \\
&& \Omega^R_{ZbB} &=0
\end{align}

\subsubsection{Doublet ${\cal D}_2$ $(T,B), \,Y=1/6$}
\label{eq:EWCdoubletTBt2}
\begin{align}
\Omega^L_{Wtb}&=\frac{e}{\sqrt{2}s_W}(s_L^ts_L^b+c_L^tc_L^b)           &  \Omega^L_{Ztt} &=\frac{e}{2s_Wc_W}(1-\frac{4}{3}s_W^2)          \nonumber   \\
\Omega^R_{Wtb}&=\frac{es_R^ts_R^b}{\sqrt{2}s_W}        &  \Omega^R_{Ztt}&=\frac{e}{2s_Wc_W}(s_R^{t^2}-\frac{4}{3}s_W^2)  \nonumber  \\
\Omega^L_{WTb}&=\frac{e}{\sqrt{2}s_W}(s_L^tc_L^b-s_L^bc_L^t)  &  \Omega^L_{Zbb}&=\frac{e}{2s_Wc_W}(\frac{2}{3}s_W^2-1)    \nonumber \\
\Omega^R_{WTb}&=-\frac{es_R^bc_R^t}{\sqrt{2}s_W} &      \Omega^R_{Zbb}&=\frac{e}{2s_Wc_W}(\frac{2}{3}s_W^2-s_R^{b^2})   \nonumber  \\
\Omega^L_{WtB}&=\frac{e}{\sqrt{2}s_W}(s_L^bc_L^t-s_L^tc_L^b)& \Omega^L_{ZtT} &=0\nonumber \\
\Omega^R_{WtB}&=-\frac{es_R^tc_R^b}{\sqrt{2}s_W}& \Omega^R_{ZtT} &=-\frac{es_R^tc_R^t}{2s_Wc_W} \nonumber \\
\Omega^L_{WTB}&=\frac{e}{\sqrt{2}s_W}(s_L^ts_L^b+c_L^tc_L^b) & \Omega^L_{ZbB} &=0 \nonumber \\
\Omega^R_{WTB}&=\frac{ec_R^tc_R^b}{\sqrt{2}s_W} & \Omega^R_{ZbB} &=\frac{es_R^bc_R^b}{2s_Wc_W} \nonumber \\
&& \Omega^L_{ZTT} &=\frac{e}{2s_Wc_W}(1-\frac{4}{3}s_W^2)          \nonumber   \\
&& \Omega^R_{ZTT}&=\frac{e}{2s_Wc_W}(c_R^{t^2}-\frac{4}{3}s_W^2)  \nonumber  \\
&& \Omega^L_{ZBB}&=\frac{e}{2s_Wc_W}(\frac{2}{3}s_W^2-1)    \nonumber \\
&&\Omega^R_{ZBB}&=\frac{e}{2s_Wc_W}(\frac{2}{3}s_W^2-c_R^{b^2})
\end{align}

\subsubsection{Doublet ${\cal D}_X$ $(X,T), \,Y=7/6$}
\label{eq:EWCdoubletXTt2}
\begin{align}
\Omega^L_{Wtb}&=\frac{ec_L^t}{\sqrt{2}s_W}           &  \Omega^L_{Ztt} &=\frac{e}{2s_Wc_W}(1-\frac{4}{3}s_W^2-2s_L^{t^2})          \nonumber   \\
\Omega^R_{Wtb}&=0        &  \Omega^R_{Ztt}&=-\frac{e}{2s_Wc_W}(\frac{4}{3}s_W^2+s_R^{t^2})  \nonumber  \\
\Omega^L_{WTb}&=\frac{es_L^t}{\sqrt{2}s_W}  &  \Omega^L_{Zbb}&=\frac{e}{2s_Wc_W}(\frac{2}{3}s_W^2-1)    \nonumber \\
\Omega^R_{WTb}&=0 &      \Omega^R_{Zbb}&=\frac{es_W}{3c_W}  \nonumber  \\
\Omega^L_{WtX}&=-\frac{es_L^t}{\sqrt{2}s_W}& \Omega^L_{ZtT} &=\frac{es_L^tc_L^t}{s_Wc_W}\nonumber \\
\Omega^R_{WtX}&=-\frac{es_R^t}{\sqrt{2}s_W}& \Omega^R_{ZtT} &=\frac{es_R^tc_R^t}{2s_Wc_W} \nonumber \\
\Omega^L_{WTX}&=\frac{ec_L^t}{\sqrt{2}s_W} & \Omega^L_{ZTT} &=\frac{e}{2s_Wc_W}(1-\frac{4}{3}s_W^2-c_L^{t^2}) \nonumber \\
\Omega^R_{WTX}&=\frac{ec_R^t}{\sqrt{2}s_W} & \Omega^R_{ZTT}&=-\frac{e}{2s_Wc_W}(c_R^{t^2}+\frac{4}{3}s_W^2)  \nonumber \\
&&\Omega^L_{ZXX}&=\frac{e}{2s_Wc_W}(1-\frac{10}{3}s_W^2)    \nonumber \\
&&\Omega^R_{ZXX}&=\frac{e}{2s_Wc_W}(1-\frac{10}{3}s_W^2)
\end{align}

\subsubsection{Doublet ${\cal D}_Y$ $(B,Y), \,Y=-5/6$}
\label{eq:EWCdoubletBYt2}
\begin{align}
\Omega^L_{Wtb}&=\frac{ec_L^b}{\sqrt{2}s_W}           &  \Omega^L_{Ztt} &=\frac{e}{2s_Wc_W}(1-\frac{4}{3}s_W^2)          \nonumber   \\
\Omega^R_{Wtb}&=0        &  \Omega^R_{Ztt}&=-\frac{2es_W}{3c_W}  \nonumber  \\
\Omega^L_{WBt}&=\frac{es_L^b}{\sqrt{2}s_W}  &  \Omega^L_{Zbb}&=\frac{e}{2s_Wc_W}(\frac{2}{3}s_W^2+2s_L^{b^2}-1)    \nonumber \\
\Omega^R_{WBt}&=0 &      \Omega^R_{Zbb}&=\frac{e}{2s_Wc_W}(s_R^{b^2}+\frac{2}{3}s_W^2)   \nonumber  \\
\Omega^L_{WbY}&=-\frac{es_L^b}{\sqrt{2}s_W}& \Omega^L_{ZbB} &=-\frac{es_L^bc_L^b}{s_Wc_W}\nonumber \\
\Omega^R_{WbY}&=-\frac{es_R^b}{\sqrt{2}s_W}& \Omega^R_{ZbB} &=-\frac{es_R^bc_R^b}{2s_Wc_W} \nonumber \\
\Omega^L_{WBY}&=\frac{ec_L^b}{\sqrt{2}s_W} & \Omega^L_{ZBB} &=\frac{e}{2s_Wc_W}(\frac{2}{3}s_W^2+2c_L^{b^2}-1) \nonumber \\
\Omega^R_{WBY}&=\frac{ec_R^b}{\sqrt{2}s_W} & \Omega^R_{ZBB}&=\frac{e}{2s_Wc_W}(c_R^{b^2}+\frac{2}{3}s_W^2)  \nonumber \\
&&\Omega^L_{ZYY}&=-\frac{e}{2s_Wc_W}(1-\frac{8}{3}s_W^2)    \nonumber \\
&&\Omega^R_{ZYY}&=-\frac{e}{2s_Wc_W}(1-\frac{8}{3}s_W^2)
\end{align}

\subsubsection{Triplet ${\cal T}_X$ $(X, T, B), \,Y=2/3$}
\label{eq:EWCtripletXTBt2}
\begin{align}
\Omega^L_{Wtb}&=\frac{e}{\sqrt{2}s_W}(\sqrt{2}s_L^ts_L^b+c_L^tc_L^b)        &  \Omega^L_{Ztt} &=\frac{e}{2s_Wc_W}(c_L^{t^2}-\frac{4}{3}s_W^2)          \nonumber   \\
\Omega^R_{Wtb}&=\frac{es_R^ts_R^b}{s_W}        &  \Omega^R_{Ztt}&=-\frac{2es_W}{3c_W}  \nonumber  \\
\Omega^L_{WTb}&=\frac{e}{\sqrt{2}s_W}(s_L^tc_L^b-\sqrt{2}s_L^bc_L^t)        &  \Omega^L_{Zbb} &=\frac{e}{2s_Wc_W}(\frac{2}{3}s_W^2-s_L^{b^2}-1)  \nonumber \\
\Omega^R_{WTb}&=-\frac{es_R^bc_R^t}{s_W}        &  \Omega^R_{Zbb}&=\frac{e}{2s_Wc_W}(\frac{2}{3}s_W^2-2s_R^{b^2})    \nonumber  \\
\Omega^L_{WtB}&=\frac{e}{\sqrt{2}s_W}(s_L^bc_L^t-\sqrt{2}s_L^tc_L^b) &  \Omega^L_{ZtT} &=\frac{es_L^tc_L^t}{2s_Wc_W}\nonumber \\
\Omega^R_{WtB}&=-\frac{es_R^tc_R^b}{s_W} &  \Omega^R_{ZtT} &=0\nonumber \\
\Omega^L_{WtX}&=-\frac{es_L^t}{s_W} & \Omega^L_{ZbB} &=\frac{es_L^bc_L^b}{2s_Wc_W}\nonumber \\
\Omega^R_{WtX}&=-\frac{es_R^t}{s_W} & \Omega^R_{ZbB} &=\frac{es_R^bc_R^b}{s_Wc_W}\nonumber \\
\Omega^L_{WTB}&=\frac{e}{\sqrt{2}s_W}(s_L^ts_L^b+\sqrt{2}c_L^tc_L^b) & \Omega^L_{ZTT} &=\frac{e}{2s_Wc_W}(s_L^{t^2}-\frac{4}{3}s_W^2) \nonumber \\
\Omega^R_{WTB}&=\frac{ec_R^tc_R^b}{s_W} &  \Omega^R_{ZTT}&=-\frac{2es_W}{3c_W}  \nonumber  \\
\Omega^L_{WTX}&=\frac{ec_L^t}{s_W} &  \Omega^L_{ZBB} &=\frac{e}{2s_Wc_W}(\frac{2}{3}s_W^2-c_L^{b^2}-1) \nonumber \\ 
\Omega^R_{WTX}&=\frac{ec_R^t}{s_W} &  \Omega^R_{ZBB} &=\frac{e}{2s_Wc_W}(\frac{2}{3}s_W^2-2c_R^{b^2}) \nonumber \\ 
&&\Omega^L_{ZXX}&=\frac{e}{2s_Wc_W}(2-\frac{10}{3}s_W^2)    \nonumber \\
&&\Omega^R_{ZXX}&=\frac{e}{2s_Wc_W}(2-\frac{10}{3}s_W^2)
\end{align}
\newpage
\subsubsection{Triplet ${\cal T}_Y$ $( T, B, Y), \, Y=-1/3$}
\label{eq:EWCtripletTBYt2}
\begin{align}
\Omega^L_{Wtb}&=\frac{e}{\sqrt{2}s_W}(\sqrt{2}s_L^ts_L^b+c_L^tc_L^b)        &  \Omega^L_{Ztt} &=\frac{e}{2s_Wc_W}(-\frac{4}{3}s_W^2+s_L^{t^2}+1)          \nonumber   \\
\Omega^R_{Wtb}&=\frac{es_R^ts_R^b}{s_W}        &  \Omega^R_{Ztt}&=\frac{e}{2s_Wc_W}(2s_R^{t^2}-\frac{4}{3}s_W^2)     \nonumber  \\
\Omega^L_{WTb}&=\frac{e}{\sqrt{2}s_W}(s_L^tc_L^b-\sqrt{2}s_L^bc_L^t)        &  \Omega^L_{Zbb} &=\frac{e}{2s_Wc_W}(\frac{2}{3}s_W^2-c_L^{b^2})  \nonumber \\
\Omega^R_{WTb}&=-\frac{es_R^bc_R^t}{s_W}        &  \Omega^R_{Zbb}&=\frac{es_W}{3c_W}    \nonumber  \\
\Omega^L_{WtB}&=\frac{e}{\sqrt{2}s_W}(s_L^bc_L^t-\sqrt{2}s_L^tc_L^b) &  \Omega^L_{ZtT} &=-\frac{es_L^tc_L^t}{2s_Wc_W}\nonumber \\
\Omega^R_{WtB}&=-\frac{es_R^tc_R^b}{s_W} &  \Omega^R_{ZtT} &=-\frac{es_R^tc_R^t}{s_Wc_W}\nonumber \\
\Omega^L_{WbY}&=-\frac{es_L^b}{s_W} & \Omega^L_{ZbB} &=-\frac{es_L^bc_L^b}{2s_Wc_W}\nonumber \\
\Omega^R_{WbY}&=-\frac{es_R^b}{s_W} & \Omega^R_{ZbB} &=0\nonumber \\
\Omega^L_{WTB}&=\frac{e}{\sqrt{2}s_W}(s_L^ts_L^b+\sqrt{2}c_L^tc_L^b) & \Omega^L_{ZTT} &=\frac{e}{2s_Wc_W}(-\frac{4}{3}s_W^2+c_L^{t^2}+1) \nonumber \\
\Omega^R_{WTB}&=\frac{ec_R^tc_R^b}{s_W} &  \Omega^R_{ZTT}&=\frac{e}{2s_Wc_W}(2c_R^{t^2}-\frac{4}{3}s_W^2) \nonumber  \\
\Omega^L_{WBY}&=\frac{ec_L^b}{s_W} &  \Omega^L_{ZBB} &=\frac{e}{2s_Wc_W}(\frac{2}{3}s_W^2-s_L^{b^2}) \nonumber \\ 
\Omega^R_{WBY}&=\frac{ec_R^b}{s_W} &  \Omega^R_{ZBB} &=\frac{es_W}{3c_W} \nonumber \\ 
&&\Omega^L_{ZYY}&=-\frac{e}{2s_Wc_W}(2-\frac{8}{3}s_W^2)    \nonumber \\
&&\Omega^R_{ZYY}&=-\frac{e}{2s_Wc_W}(2-\frac{8}{3}s_W^2)
\end{align}

\subsection{Passarino-Veltman Integrals}
\label{sec:PVinteg}
Although a more detailed discussion about Passarino-Veltman (PV) reduction appears elsewhere \cite{Hagiwara:1994pw}, we give a generic one-loop tensor integral as the following
\begin{equation}
\label{eq:genprop}
T^{\nu_i}_{\rho}=\frac{(2\pi\mu)^{4-D}}{i\pi^2}\int d^Dp\frac{p^{\nu_i}...p^{\nu_{n}}}{\mathbb{D}_1...\mathbb{D}_{\rho}}
\end{equation}
where the propagators are described by 
\begin{eqnarray}
\mathbb{D}_1&=&p^2-m_1^2+i\epsilon, \nonumber \\
\mathbb{D}_2&=&(p+q_1)^2-m_2^2+i\epsilon, \nonumber \\
\mathbb{D}_3&=&(p+q_1+q_2)^2-m_3^2+i\epsilon, 
\end{eqnarray}
After factoring out the $i/(16\pi^2)$, scalar, vector and tensor functions are defined from the generic one-loop tensor integral Eq. \ref{eq:genprop}:
\begin{eqnarray}
A(m_1)&=&\mu^{4-D}\int\frac{d^Dp}{(2\pi)^D\mathbb{D}_1},\nonumber \\
\left[B_0,B^{\mu},B^{\mu\nu}\right](q_1^2,m_1^2,m_2^2)&=&\mu^{4-D}\int\frac{d^Dp}{(2\pi)^D}\frac{\left[1,p^{\mu},p^{\mu}p^{\nu}\right]}{\mathbb{D}_1\mathbb{D}_2}, \nonumber \\
\left[C_0,C^{\mu},C^{\mu\nu}\right](q_1^2,q_2^2,(q_1+q_2)^2,m_1^2,m_2^2,m_3^2)&=&\mu^{4-D}\int\frac{d^Dp}{(2\pi)^D}\frac{\left[1,p^{\mu},p^{\mu}p^{\nu}\right]}{\mathbb{D}_1\mathbb{D}_2\mathbb{D}_3}
\end{eqnarray}
Scalar and tensor integrals are not independent. In fact, tensor forms can be decomposed in terms of scalar functions:
\begin{align}
B^{\mu}&=q_1^{\mu}B_1,  &C^{\mu}&=q_1^{\mu}C_{1}+q_2^{\mu}C_{2}, \nonumber \\
B^{\mu\nu}&=q_1^{\mu}q_1^{\nu}B_{11}+g^{\mu\nu}B_{00}, & C^{\mu\nu}&=\sum_{i=1}^{2}q_i^{\mu}q_j^{\nu}C_{ij}+g^{\mu\nu}C_{00}\nonumber \\
&& C^{\mu\nu\delta}&=\sum_{i,j,k=1}^{2}q_i^{\mu}q_j^{\nu}q_k^{\delta}C_{ijk}+\sum_{i=1}^{2}(q_i^{\mu}g^{\nu\delta}+q_i^{\nu}g^{\delta\mu}+q_i^{\delta}g^{\mu\nu})C_{00i}
\end{align} 
Scalar integrals or vacuum integrals play a main role for all intents and purposes throughout this work. Furthermore, there are only four type of independent scalar(vacuum) integrals. The rest of the vacuum integrals carried out throughout this work are combination of the following definitions:
\begin{eqnarray}
A_0(m_1^2)&=&\frac{(2\pi\mu)^{\epsilon}}{i\pi^2}\int d^Dp\frac{1}{p^2-m_1^2},\nonumber \\
B_0(q_1^2,m_1^2,m_2^2)&=&\frac{(2\pi\mu)^{\epsilon}}{i\pi^2}\int d^Dp\frac{1}{\left[p^2-m_1^2\right]}\frac{1}{\left[(p+q_1)^2-m_2^2\right]},\nonumber \\
C_0(q_1^2,q_2^2,q_{12}^2,m_1^2,m_2^2,m_3^2)&=&\frac{(2\pi\mu)^{\epsilon}}{i\pi^2}\int d^Dp\frac{1}{\left[p^2-m_1^2\right]}\frac{1}{\left[(p+q_1)^2-m_2^2\right]}\frac{1}{\left[(p+q_{12})^2-m_3^2\right]},\nonumber \\
D_0(q_1^2,q_2^2,q_3^2,q_{12}^2,q_{23}^2,m_1^2,m_2^2,m_3^2,m_4^2)&=&\frac{(2\pi\mu)^{\epsilon}}{i\pi^2}\int d^Dp\frac{1}{\left[p^2-m_1^2\right]}\frac{1}{\left[(p+q_1)^2-m_2^2\right]}\frac{1}{\left[(p+q_{12})^2-m_3^2\right]}\frac{1}{\left[(p+q_{123})^2-m_4^2\right]}\nonumber \\
\end{eqnarray}
where $\epsilon=4-D$. Explicit analytical expressions of widely used PV functions are defined as
\begin{eqnarray}
A_0(m^2)&=&m^2\left(\Delta_{\epsilon}+1-\ln\frac{m^2}{\mu^2}\right),\\
B_0(0,m_1^2,m_2^2)&=&\frac{A_0(m_1^2)-A_0(m_2^2)}{m_1^2-m_2^2},\\
B_0(0,m_1^2,m_1^2)&=&\frac{A_0(m_1^2)}{m_1^2}-1,\\
B_0(m_1^2,0,m_1^2)&=&\frac{A_0(m_1^2)}{m_1^2}+1,\\
B_1(0,m_1^2,m_2^2)&=&\frac{2y^2\ln y_2-4y_2\ln y_2-y_2^2+4y_2-3}{4(y_2-1)^2}+\frac{1}{2}\ln \frac{m_1^2}{\mu^2}-\frac{\Delta_{\epsilon}}{2},\\
B_{00}(m_1^2,m_2^2,m_3^2)&=&\frac{(m_1-m_2-m_3)(m_1+m_2-m_3)(m_1-m_2+m_3)(m_1+m_2+m_3)B_0(m_1^2,m_2^2,m_3^2)}{4(1-D)m_1^2}\, \nonumber\\
&+&\frac{A_0(m_2^2)(m_1^2+m_2^2-m_3^2)}{4(1-D)m_1^2}-\frac{A_0(m_3^2)(m_1^2-m_2^2+m_3^2)}{4(1-D)m_1^2},\\
B_{00}(0,m_2^2,m_3^2)&=&-\frac{A_0(m_3^2)}{2(1-D)}-\frac{m_2^2B_0(0,m_2^2,m_3^2)}{1-D}-\frac{(m_2^2-m_3^2)B_1(0,m_2^2,m_3^2)}{2(1-D)},\\
B_{00}(0,m^2,m^2)&=&-\frac{A_0(m^2)}{2(1-D)}-\frac{m^2B_0(0,m^2,m^2)}{1-D},\\
B_{00}(m_1^2,m_2^2,m_2^2)&=&\frac{(m_1^2-4m_2^2)B_0(m_1^2,m_2^2,m_2^2)}{4(1-D)}-\frac{A_0(m_2^2)}{2(1-D)},\\
C_0(m_1^2,m_2^2,m_3^2)&=&\frac{1}{m_1^2}\frac{y_2\ln y_2-y_3\ln y_3-y_2y_3\ln y_2+y_2y_3\ln y_3}{(y_2-1)(y_3-1)(y_2-y_3)},\\
C_0(m_1^2,m_2^2,m_2^2)&=&\frac{1}{m_1^2}\frac{\ln y_2-y_2+1}{(y_2-1)^2},\\
C_0(m^2,m^2,m^2)&=&-\frac{1}{2m^2},
\end{eqnarray}
where the divergent part in MS scheme is given by 
\begin{equation}
\Delta_{\epsilon}=\frac{1}{\epsilon}-\gamma_E+\ln 4\pi+\ln \mu^2
\end{equation}
and the mass ratio parameter 
\begin{equation*}
y_i=\frac{m_i^2}{m_1^2}
\end{equation*}
Finally, the complementary relations to the definitions above can be summarized with the following 4 scalar functions:
\begin{eqnarray}
B_2(p^2,m_1^2,m_2^2)&=&B_{21}(p^2,m_1^2,m_2^2),\\
B_3(p^2,m_1^2,m_2^2)&=&-B_1(p^2,m_1^2,m_2^2)-B_{21}(p^2,m_1^2,m_2^2),\\
B_4(p^2,m_1^2,m_2^2)&=&-m_1^2B_1(p^2,m_2^2,m_1^2)-m_2^2B_1(p^2,m_1^2,m_2^2),\\
B_5(p^2,m_1^2,m_2^2)&=&A_0(m_1^2)+A_0(m_2^2)-4B_{22}(p^2,m_1^2,m_2^2).
\end{eqnarray}

\newpage

\bibliography{2HDMVLQ}

\begin{thebibliography}{100}

\bibitem{ATLAS:2012yve}
G.~Aad et~al.,
\newblock Phys. Lett. B {\bf 716}, 1 (2012).

\bibitem{CMS:2012qbp}
S.~Chatrchyan et~al.,
\newblock Phys. Lett. B {\bf 716}, 30 (2012).

\bibitem{Degrassi:2012ry}
G.~Degrassi et~al.,
\newblock JHEP {\bf 08}, 098 (2012).

\bibitem{Buttazzo:2013uya}
D.~Buttazzo et~al.,
\newblock JHEP {\bf 12}, 089 (2013).

\bibitem{Agashe:2004rs}
K.~Agashe, R.~Contino, and A.~Pomarol,
\newblock Nucl. Phys. B {\bf 719}, 165 (2005).

\bibitem{Ferretti:2013kya}
G.~Ferretti and D.~Karateev,
\newblock JHEP {\bf 03}, 077 (2014).

\bibitem{Kaplan:1991dc}
D.~B. Kaplan,
\newblock Nucl. Phys. B {\bf 365}, 259 (1991).

\bibitem{Ferretti:2014qta}
G.~Ferretti,
\newblock JHEP {\bf 06}, 142 (2014).

\bibitem{Chang:1999nh}
S.~Chang, J.~Hisano, H.~Nakano, N.~Okada, and M.~Yamaguchi,
\newblock Phys. Rev. D {\bf 62}, 084025 (2000).

\bibitem{Gherghetta:2000qt}
T.~Gherghetta and A.~Pomarol,
\newblock Nucl. Phys. B {\bf 586}, 141 (2000).

\bibitem{Contino:2003ve}
R.~Contino, Y.~Nomura, and A.~Pomarol,
\newblock Nucl. Phys. B {\bf 671}, 148 (2003).

\bibitem{Gopalakrishna:2011ef}
S.~Gopalakrishna, T.~Mandal, S.~Mitra, and R.~Tibrewala,
\newblock Phys. Rev. D {\bf 84}, 055001 (2011).

\bibitem{Gopalakrishna:2013hua}
S.~Gopalakrishna, T.~Mandal, S.~Mitra, and G.~Moreau,
\newblock JHEP {\bf 08}, 079 (2014).

\bibitem{Couture:2017mbd}
G.~Couture, M.~Frank, C.~Hamzaoui, and M.~Toharia,
\newblock Phys. Rev. D {\bf 95}, 095038 (2017).

\bibitem{Arkani-Hamed:2002iiv}
N.~Arkani-Hamed et~al.,
\newblock JHEP {\bf 08}, 021 (2002).

\bibitem{Perelstein:2003wd}
M.~Perelstein, M.~E. Peskin, and A.~Pierce,
\newblock Phys. Rev. D {\bf 69}, 075002 (2004).

\bibitem{Schmaltz:2002wx}
M.~Schmaltz,
\newblock Nucl. Phys. B Proc. Suppl. {\bf 117}, 40 (2003).

\bibitem{Martin:2010dc}
S.~P. Martin,
\newblock Phys. Rev. D {\bf 82}, 055019 (2010).

\bibitem{Martin:2009bg}
S.~P. Martin,
\newblock Phys. Rev. D {\bf 81}, 035004 (2010).

\bibitem{Babu:2008ge}
K.~S. Babu, I.~Gogoladze, M.~U. Rehman, and Q.~Shafi,
\newblock Phys. Rev. D {\bf 78}, 055017 (2008).

\bibitem{Kang:2007ib}
J.~Kang, P.~Langacker, and B.~D. Nelson,
\newblock Phys. Rev. D {\bf 77}, 035003 (2008).

\bibitem{FlavourLatticeAveragingGroup:2019iem}
S.~Aoki et~al.,
\newblock Eur. Phys. J. C {\bf 80}, 113 (2020).

\bibitem{Bryman:2021teu}
D.~Bryman, V.~Cirigliano, A.~Crivellin, and G.~Inguglia,
\newblock Ann. Rev. Nucl. Part. Sci. {\bf 72}, 69 (2022).

\bibitem{Kirk:2023oez}
M.~Kirk,
\newblock {Cabibbo angle anomalies and a global fit to vector-like quarks},
\newblock in {\em {21st Conference on Flavor Physics and CP Violation}}, 2023.

\bibitem{Albergaria:2023vls}
F.~Albergaria and G.~C. Branco,
\newblock 2307.13073  (2023).

\bibitem{CarcamoHernandez:2021yev}
A.~E. C\'arcamo~Hern\'andez, S.~F. King, and H.~Lee,
\newblock Phys. Rev. D {\bf 105}, 015021 (2022).

\bibitem{Altarelli:1994rb}
G.~Altarelli and G.~Isidori,
\newblock Phys. Lett. B {\bf 337}, 141 (1994).

\bibitem{Ghorbani:2021rgs}
P.~Ghorbani,
\newblock Nucl. Phys. B {\bf 971}, 115533 (2021).

\bibitem{Arsenault:2022xty}
A.~Arsenault, K.~Y. Cingiloglu, and M.~Frank,
\newblock Phys. Rev. D {\bf 107}, 036018 (2023).

\bibitem{Branco:2011iw}
G.~C. Branco et~al.,
\newblock Phys. Rept. {\bf 516}, 1 (2012).

\bibitem{Gunion:2002zf}
J.~F. Gunion and H.~E. Haber,
\newblock Phys. Rev. D {\bf 67}, 075019 (2003).

\bibitem{Coleppa:2013dya}
B.~Coleppa, F.~Kling, and S.~Su,
\newblock JHEP {\bf 01}, 161 (2014).

\bibitem{Altmannshofer:2012ar}
W.~Altmannshofer, S.~Gori, and G.~D. Kribs,
\newblock Phys. Rev. D {\bf 86}, 115009 (2012).

\bibitem{Ivanov:2006yq}
I.~P. Ivanov,
\newblock Phys. Rev. D {\bf 75}, 035001 (2007),
\newblock [Erratum: Phys.Rev.D 76, 039902 (2007)].

\bibitem{Ginzburg:2005dt}
I.~F. Ginzburg and I.~P. Ivanov,
\newblock Phys. Rev. D {\bf 72}, 115010 (2005).

\bibitem{Basler:2017nzu}
P.~Basler, P.~M. Ferreira, M.~M\"uhlleitner, and R.~Santos,
\newblock Phys. Rev. D {\bf 97}, 095024 (2018).

\bibitem{Accomando:2019jrb}
E.~Accomando, D.~Englert, C.~Byers, J.~Hays, and S.~Moretti,
\newblock 1905.07313  (2019).

\bibitem{Song:2022xts}
H.~Song, W.~Su, and M.~Zhang,
\newblock JHEP {\bf 10}, 048 (2022).

\bibitem{Cherchiglia:2017uwv}
A.~Cherchiglia, D.~St\"ockinger, and H.~St\"ockinger-Kim,
\newblock Phys. Rev. D {\bf 98}, 035001 (2018).

\bibitem{Haber:2015pua}
H.~E. Haber and O.~St\r{a}l,
\newblock Eur. Phys. J. C {\bf 75}, 491 (2015),
\newblock [Erratum: Eur.Phys.J.C 76, 312 (2016)].

\bibitem{BhupalDev:2014bir}
P.~S. Bhupal~Dev and A.~Pilaftsis,
\newblock JHEP {\bf 12}, 024 (2014),
\newblock [Erratum: JHEP 11, 147 (2015)].

\bibitem{Baglio:2014nea}
J.~Baglio, O.~Eberhardt, U.~Nierste, and M.~Wiebusch,
\newblock Phys. Rev. D {\bf 90}, 015008 (2014).

\bibitem{Eberhardt:2013uba}
O.~Eberhardt, U.~Nierste, and M.~Wiebusch,
\newblock JHEP {\bf 07}, 118 (2013).

\bibitem{Barroso:2013awa}
A.~Barroso, P.~M. Ferreira, I.~P. Ivanov, and R.~Santos,
\newblock JHEP {\bf 06}, 045 (2013).

\bibitem{Eriksson:2009ws}
D.~Eriksson, J.~Rathsman, and O.~Stal,
\newblock Comput. Phys. Commun. {\bf 181}, 189 (2010).

\bibitem{WahabElKaffas:2007xd}
A.~Wahab El~Kaffas, P.~Osland, and O.~M. Ogreid,
\newblock Phys. Rev. D {\bf 76}, 095001 (2007).

\bibitem{Maniatis:2006fs}
M.~Maniatis, A.~von Manteuffel, O.~Nachtmann, and F.~Nagel,
\newblock Eur. Phys. J. C {\bf 48}, 805 (2006).

\bibitem{Davidson:2005cw}
S.~Davidson and H.~E. Haber,
\newblock Phys. Rev. D {\bf 72}, 035004 (2005),
\newblock [Erratum: Phys.Rev.D 72, 099902 (2005)].

\bibitem{Kim:1986ax}
J.~E. Kim,
\newblock Phys. Rept. {\bf 150}, 1 (1987).

\bibitem{Peccei:1977hh}
R.~D. Peccei and H.~R. Quinn,
\newblock Phys. Rev. Lett. {\bf 38}, 1440 (1977).

\bibitem{Trodden:1998qg}
M.~Trodden,
\newblock {Electroweak baryogenesis: A Brief review},
\newblock in {\em {33rd Rencontres de Moriond: Electroweak Interactions and
  Unified Theories}}, pages 471--480, 1998.

\bibitem{Sirlin:1985ux}
A.~Sirlin and R.~Zucchini,
\newblock Nucl. Phys. B {\bf 266}, 389 (1986).

\bibitem{Hempfling:1994ar}
R.~Hempfling and B.~A. Kniehl,
\newblock Phys. Rev. D {\bf 51}, 1386 (1995).

\bibitem{Aguilar-Saavedra:2013qpa}
J.~A. Aguilar-Saavedra, R.~Benbrik, S.~Heinemeyer, and M.~P\'erez-Victoria,
\newblock Phys. Rev. D {\bf 88}, 094010 (2013).

\bibitem{Aguilar-Saavedra:2013wba}
J.~A. Aguilar-Saavedra,
\newblock EPJ Web Conf. {\bf 60}, 16012 (2013).

\bibitem{Ellis:2014dza}
S.~A.~R. Ellis, R.~M. Godbole, S.~Gopalakrishna, and J.~D. Wells,
\newblock JHEP {\bf 09}, 130 (2014).

\bibitem{Carmi:2012yp}
D.~Carmi, A.~Falkowski, E.~Kuflik, and T.~Volansky,
\newblock JHEP {\bf 07}, 136 (2012).

\bibitem{Fajfer:2013wca}
S.~Fajfer, A.~Greljo, J.~F. Kamenik, and I.~Mustac,
\newblock JHEP {\bf 07}, 155 (2013).

\bibitem{ATLAS:2015uhg}
G.~Aad et~al.,
\newblock JHEP {\bf 02}, 110 (2016).

\bibitem{ATLAS:2016scx}
G.~Aad et~al.,
\newblock Eur. Phys. J. C {\bf 76}, 442 (2016).

\bibitem{ATLAS:2015ktd}
G.~Aad et~al.,
\newblock JHEP {\bf 08}, 105 (2015).

\bibitem{ATLAS:2015vzd}
G.~Aad et~al.,
\newblock Phys. Rev. D {\bf 91}, 112011 (2015).

\bibitem{ATLAS:2017lvm}
ATLAS-CONF-2017-015, ATLAS-CONF-2017-015  (2017).

\bibitem{CMS:2017ked}
A.~M. Sirunyan et~al.,
\newblock JHEP {\bf 11}, 085 (2017).

\bibitem{CMS:2017gsh}
A.~M. Sirunyan et~al.,
\newblock JHEP {\bf 05}, 029 (2017).

\bibitem{CMS:2015hyy}
V.~Khachatryan et~al.,
\newblock Phys. Rev. D {\bf 93}, 112009 (2016).

\bibitem{CMS:2015jwh}
V.~Khachatryan et~al.,
\newblock JHEP {\bf 06}, 080 (2015).

\bibitem{CMS:2013wkd}
S.~Chatrchyan et~al.,
\newblock Phys. Rev. Lett. {\bf 112}, 171801 (2014).

\bibitem{CMS:2013hwy}
S.~Chatrchyan et~al.,
\newblock Phys. Lett. B {\bf 729}, 149 (2014).

\bibitem{ATLAS:2023bfh}
G.~Aad et~al.,
\newblock 2307.07584  (2023).

\bibitem{ATLAS:2022hnn}
G.~Aad et~al.,
\newblock Phys. Lett. B {\bf 843}, 138019 (2023).

\bibitem{ATLAS:2017vdo}
M.~Aaboud et~al.,
\newblock JHEP {\bf 08}, 052 (2017).

\bibitem{ATLAS:2018cye}
M.~Aaboud et~al.,
\newblock JHEP {\bf 07}, 089 (2018).

\bibitem{ATLAS:2023pja}
G.~Aad et~al.,
\newblock JHEP {\bf 08}, 153 (2023).

\bibitem{CDF:2022hxs}
T.~Aaltonen et~al.,
\newblock Science {\bf 376}, 170 (2022).

\bibitem{Lopez-Val:2014jva}
D.~L\'opez-Val and T.~Robens,
\newblock Phys. Rev. D {\bf 90}, 114018 (2014).

\bibitem{Xiao:2014kba}
M.-L. Xiao and J.-H. Yu,
\newblock Phys. Rev. D {\bf 90}, 014007 (2014),
\newblock [Addendum: Phys.Rev.D 90, 019901 (2014)].

\bibitem{Bahrami:2014ska}
S.~Bahrami and M.~Frank,
\newblock Phys. Rev. D {\bf 90}, 035017 (2014).

\bibitem{Shimizu:2023rvi}
Y.~Shimizu and S.~Takeshita,
\newblock Nucl. Phys. B {\bf 994}, 116290 (2023).

\bibitem{Aguilar-Saavedra:2006uim}
J.~A. Aguilar-Saavedra,
\newblock JHEP {\bf 12}, 033 (2006).

\bibitem{Dawson:2012di}
S.~Dawson and E.~Furlan,
\newblock Phys. Rev. D {\bf 86}, 015021 (2012).

\bibitem{Benbrik:2015fyz}
R.~Benbrik, C.-H. Chen, and T.~Nomura,
\newblock Phys. Rev. D {\bf 93}, 055034 (2016).

\bibitem{Arhrib:2016rlj}
A.~Arhrib et~al.,
\newblock Phys. Rev. D {\bf 97}, 095015 (2018).

\bibitem{Benbrik:2022kpo}
R.~Benbrik, M.~Boukidi, and S.~Moretti,
\newblock (2022).

\bibitem{Horejsi:2005da}
J.~Horejsi and M.~Kladiva,
\newblock Eur. Phys. J. C {\bf 46}, 81 (2006).

\bibitem{Tang:2013bz}
Y.~Tang,
\newblock Mod. Phys. Lett. A {\bf 28}, 1330002 (2013).

\bibitem{Hiller:2022rla}
G.~Hiller, T.~H\"ohne, D.~F. Litim, and T.~Steudtner,
\newblock Phys. Rev. D {\bf 106}, 115004 (2022).

\bibitem{Gopalakrishna:2018uxn}
S.~Gopalakrishna and A.~Velusamy,
\newblock Phys. Rev. D {\bf 99}, 115020 (2019).

\bibitem{ALEPH:2013htx}
G.~Abbiendi et~al.,
\newblock Eur. Phys. J. C {\bf 73}, 2463 (2013).

\bibitem{Kang:2022mdy}
S.~K. Kang, J.~Kim, S.~Lee, and J.~Song,
\newblock Phys. Rev. D {\bf 107}, 015025 (2023).

\bibitem{Kanemura:2019kjg}
S.~Kanemura, M.~Kikuchi, K.~Mawatari, K.~Sakurai, and K.~Yagyu,
\newblock Nucl. Phys. B {\bf 949}, 114791 (2019).

\bibitem{Hiller:2023bdb}
G.~Hiller, T.~H\"ohne, D.~F. Litim, and T.~Steudtner,
\newblock {Vacuum Stability as a Guide for Model Bulding},
\newblock in {\em {57th Rencontres de Moriond on Electroweak Interactions and
  Unified Theories}}, 2023.

\bibitem{ATLAS:2022ozf}
G.~Aad et~al.,
\newblock Phys. Rev. D {\bf 105}, 092012 (2022).

\bibitem{Heo:2022dey}
Y.~Heo, D.-W. Jung, and J.~S. Lee,
\newblock Phys. Lett. B {\bf 833}, 137274 (2022).

\bibitem{Xu:2017vpq}
X.-J. Xu,
\newblock Phys. Rev. D {\bf 95}, 115019 (2017).

\bibitem{Machacek:1983tz}
M.~E. Machacek and M.~T. Vaughn,
\newblock Nucl. Phys. B {\bf 222}, 83 (1983).

\bibitem{Machacek:1983fi}
M.~E. Machacek and M.~T. Vaughn,
\newblock Nucl. Phys. B {\bf 236}, 221 (1984).

\bibitem{Machacek:1984zw}
M.~E. Machacek and M.~T. Vaughn,
\newblock Nucl. Phys. B {\bf 249}, 70 (1985).

\bibitem{Peskin:1991sw}
M.~E. Peskin and T.~Takeuchi,
\newblock Phys. Rev. D {\bf 46}, 381 (1992).

\bibitem{ParticleDataGroup:2020ssz}
P.~A. Zyla et~al.,
\newblock PTEP {\bf 2020}, 083C01 (2020).

\bibitem{Hahn:1999mt}
T.~Hahn,
\newblock Acta Phys. Polon. B {\bf 30}, 3469 (1999).

\bibitem{Shtabovenko:2020gxv}
V.~Shtabovenko, R.~Mertig, and F.~Orellana,
\newblock Comput. Phys. Commun. {\bf 256}, 107478 (2020).

\bibitem{ATLAS:2023hyd}
G.~Aad et~al.,
\newblock CERN-EP-2022-228  (2023).

\bibitem{Kanemura:2015mxa}
S.~Kanemura, M.~Kikuchi, and K.~Yagyu,
\newblock Nucl. Phys. B {\bf 896}, 80 (2015).

\bibitem{He:2022zjz}
S.-P. He,
\newblock Chin. Phys. C {\bf 47}, 043102 (2023).

\bibitem{Lavoura:1992np}
L.~Lavoura and J.~P. Silva,
\newblock Phys. Rev. D {\bf 47}, 2046 (1993).

\bibitem{Chen:2017hak}
C.-Y. Chen, S.~Dawson, and E.~Furlan,
\newblock Phys. Rev. D {\bf 96}, 015006 (2017).

\bibitem{Cao:2022mif}
J.~Cao, L.~Meng, L.~Shang, S.~Wang, and B.~Yang,
\newblock Phys. Rev. D {\bf 106}, 055042 (2022).

\bibitem{Hagiwara:1994pw}
K.~Hagiwara, S.~Matsumoto, D.~Haidt, and C.~S. Kim,
\newblock Z. Phys. C {\bf 64}, 559 (1994),
\newblock [Erratum: Z.Phys.C 68, 352 (1995)].

\end{thebibliography}

\end{document}